\documentclass[a4paper,fleqn,usenatbib,useAMS]{mnras}
\usepackage{graphicx}	
\usepackage{amsmath}	
\usepackage{amssymb}	
\usepackage{multicol}        
\usepackage{url}
\usepackage{subfig}
\usepackage[T1]{fontenc}
\usepackage{ae,aecompl}
\usepackage{newtxtext,newtxmath}
\usepackage{pdfpages}

\title[Large-scale B-field in the First Galactic Quadrant]{The Complex Large-scale Magnetic Fields in the First Galactic Quadrant as Revealed by the Faraday Depth Profile Disparity}

\author[Y.\ K.\ Ma et al.]{Yik Ki Ma$^1$\thanks{Contact e-mail: \href{mailto:jackieac53@gmail.com}{jackieac53@gmail.com}},
S.~A.~Mao$^1$,
A.~Ordog$^2$, and
J.~C.~Brown$^2$
\\
$^1$Max-Planck-Institut f\"{u}r Radioastronomie, Auf dem H\"{u}gel 69, 53121 Bonn, Germany \\
$^2$Department of Physics and Astronomy, University of Calgary, Calgary, AB T2N 1N4, Canada
}

\date{Accepted 2020 July 15. Received 2020 July 9; in original form 2020 May 25}

\pubyear{2020}

\begin{document}
\label{firstpage}
\pagerange{\pageref{firstpage}--\pageref{lastpage}}
\maketitle

\begin{abstract}
The Milky Way is one of the very few spiral galaxies known to host large-scale magnetic field reversals. The existence of the field reversal in the first Galactic quadrant near the Sagittarius spiral arm has been well established, yet poorly characterised due to the insufficient number of reliable Faraday depths (FDs) from extragalactic radio sources (EGSs) through this reversal region. We have therefore performed broadband ($1$--$2\,{\rm GHz}$) spectro-polarimetric observations with the Karl G.\ Jansky Very Large Array (VLA) to determine the FD values of 194 EGSs in the Galactic longitude range of $20^\circ$--$52^\circ$ within $\pm 5^\circ$ from the Galactic mid-plane, covering the Sagittarius arm tangent. This factor of five increase in the EGS FD density has led to the discovery of a disparity in FD values across the Galactic mid-plane in the Galactic longitude range of $40^\circ$--$52^\circ$. Combined with existing pulsar FD measurements, we suggest that the Sagittarius arm can host an odd-parity disk field. We further compared our newly derived EGS FDs with the predictions of three major Galactic magnetic field models, and concluded that none of them can adequately reproduce our observational results. This has led to our development of new, improved models of the Milky Way disk magnetic field that will serve as an important step towards major future improvements in Galactic magnetic field models.
\end{abstract}

\begin{keywords}
ISM: magnetic fields -- Galaxy: structure
\end{keywords}

\section{INTRODUCTION} \label{sec:intro}
The magnetic field is an essential constituent of the interstellar medium. The $\sim \mu{\rm G}$ field present in galaxies is believed to have substantial effects on star formation, propagation of cosmic ray particles, galactic outflows, and evolution of galaxies \cite[e.g.,][]{beck13,beck15}. The study of the global magnetic field of the Milky Way is particularly interesting, since we have a unique perspective of its structure from within. It also allows us to attain a resolution in physical scale that is challenging to match with studies of external galaxies \citep[e.g.,][]{kierdorf20,lopez20}.

The magnetic field of galaxies, including that of the Milky Way, can be modelled as the sum of several components \citep[see, e.g.,][]{haverkorn15,beck15}. In terms of physical scales, the galactic magnetic field can be roughly divided into a large-scale field with coherence length of the order of the size of the galaxy ($\sim 1$--$10\,{\rm kpc}$), and a small-scale field with coherence length of $\lesssim 0.1\,{\rm kpc}$. Meanwhile, the field can also be separated into components occupying different spatial volumes: the disk component that dominates the galactic disk, and the halo component that fills the galactic halo. The distinction in magnetic field properties (such as the strength, geometry, and coherence length) between these components is primarily due to differences in their generation mechanisms.

The $\alpha$-$\Omega$ dynamo is the leading theory for the ordering process of the large-scale disk fields in galaxies \citep[e.g.,][]{ruzmaikin88book}, and was developed from pioneering works in the 1970s \citep[e.g.,][]{parker71,stix75,white78}. On the other hand, the small-scale disk fields can be generated by a small-scale dynamo \citep[][see also \citealt{beresnyak15}]{kazantsev68} or from the tangling of the large-scale field, both as the result of violent astrophysical phenomena such as supernova explosions \citep[e.g.,][]{norman96,maclow04,haverkorn08} or galactic spiral shocks \citep[e.g.,][]{kim06}. The origin of the halo field is still under active debate. It could have been generated by a dynamo in the galactic halo \citep{sokoloff90}, or it could have originally been the disk field and was subsequently transported by galactic outflows to the halo \citep[e.g.,][]{brandenburg93,heald12,krause19}.

One common way to measure the magnetic field of the Milky Way is by radio polarisation observations of background extragalactic radio sources \citep[EGSs; e.g.,][]{simard-normandin80,brown01,brown07,taylor09,stil11,vaneck11,mao12}. As the polarised emission propagates through the foreground magneto-ionic medium, it can experience the Faraday rotation effect, causing a change in the polarisation position angle (PA; [rad]) given by
\begin{equation}
\Delta {\rm PA} = \left[ 0.812 \int_l^0 n_e(s) B_\parallel (s)\,{\rm d}s \right] \cdot \lambda^2 \equiv {\rm FD} \cdot \lambda^2{\rm ,}
\end{equation}
where $l$ [pc] is the physical distance to the EGS, $n_e$ [${\rm cm}^{-3}$] and $B_\parallel$ [$\mu{\rm G}$] are the thermal electron number density and the strength of the magnetic field projected along the line of sight ($s$), respectively, $\lambda$ [m] is the wavelength of the polarised emission, and FD [${\rm rad\,m}^{-2}$] is the Faraday depth of the EGS. The obtained FD values carry information about the foreground magnetic field strength as well as its direction, with magnetic fields pointing towards or away from the observer leading to positive or negative FD values, respectively. Traditionally, the FD of a polarised source is obtained from a linear fit to the measured PA in $\lambda^2$ space, and for such cases the FD is often referred to as the rotation measure (RM)\footnote{To maintain a consistent notation, RM values from previous works are referred to as FD throughout this paper.} instead. This method implicitly assumes that the polarised source is Faraday simple and emits at a single FD only. With the advent of broadband capabilities of radio telescopes, algorithms that can uncover emission from multiple FDs such as RM-Synthesis \citep{brentjens05} and Stokes \textit{QU}-fitting \citep{farnsworth11,osullivan12} are becoming the new standard for determining FD values of polarised target sources.

By measuring the FD values of numerous polarised sources behind an astrophysical object of interest, one forms an FD grid that can be utilised to measure the magnetic field structure of the foreground object \citep[e.g.,][]{gaensler05,harvey-smith11,vaneck11,mao17,betti19,shanahan19}. Specifically for the Milky Way, the large-scale magnetic field can be revealed by spatial averaging of EGS FD values at an angular scale of $\sim 1^\circ$ \citep[e.g.,][]{sun08,mao10,vaneck11,mao12}. Similar studies of the Milky Way magnetic field can be performed using Galactic pulsars as the background polarised sources \citep[e.g.,][]{thomson80,noutsos08,han18,sobey19}, with the pulsars at different distances offering a tomographic view of the structure of the Galactic magnetic field. However, these studies are usually confined to the Milky Way disk where the pulsar number density is high, and are currently limited by the number of pulsars with both measured FD values and reliably determined distances \citep[e.g.,][]{han17}. There are also recent efforts in exploiting nearby H~{\sc ii} regions as depolarising screens to measure the magnetic field in the solar neighbourhood from diffuse Galactic emission \citep{thomson19}.

It is generally agreed that the large-scale disk field of the Milky Way is directed clockwise when viewed from the North Galactic Pole, with at least one reversal of the field direction near the Sagittarius spiral arm in the first Galactic quadrant \citep{simard-normandin80,thomson80,rand94,vallee05,sun08,vaneck11,jansson12,ordog17,han18}. Such a magnetic field configuration is rarely seen among spiral galaxies \citep[see][]{krause89,beck13,beck15,stein19}. It has been suggested by numerical simulations \citep{moss12,moss13} that such large-scale magnetic field reversals can emerge from $\alpha$-$\Omega$ dynamo and survive for $\sim$ Gyr, provided that the initial turbulent seed field is strong (close to equipartition) and the $\alpha$-$\Omega$ dynamo is efficient (e.g., from a strong differential rotation). An accurate knowledge of the large-scale field reversals of the Milky Way at the present epoch can allow us to trace back the physical conditions of the Milky Way to its infancy stages of the magnetic field evolution \citep[see, e.g.,][]{moss13}.

However, the Milky Way magnetic field models in the literature have not yet converged on the exact details of such large-scale field reversals, including the number, location, field strength, and magnetic pitch angle \citep[see, e.g.,][for summary]{haverkorn15}. This is at least partially due to the lack of reliable EGS FD measurements towards Galactic volumes hosting such complex magnetic field structures. In this study, we contribute to this problem by increasing the number of measured EGS FD values in a region of the Galactic disk.

We identified a sky region of $20^\circ$--$52^\circ$ in Galactic longitude ($\ell$) and within $\pm 5^\circ$ in Galactic latitude ($b$) that we will focus on in this study. Part of this region intercepts the large-scale magnetic field reversal region of the Sagittarius arm mentioned above. Our chosen region has only 43 reliable FD measurements (one per $7.3\,{\rm deg}^2$) from \cite{vaneck11}, determined from legacy Very Large Array (VLA) observations at 1.4\,GHz in spectral line mode that mitigated the $n\pi$-ambiguity issue. Although in this same sky area there are 106 reported FD values (one per $3.0\,{\rm deg}^2$) from the \cite{taylor09} catalogue, their FD values were deemed unreliable in this particular sky region due to the $n\pi$-ambiguity problem \citep{ma19a}. A new, deep EGS FD grid is clearly necessary to unveil the complex magnetic field structure in this sky region.

We performed broadband spectro-polarimetric observations in L-band ($1$--$2\,{\rm GHz}$) with the Karl G.\ Jansky Very Large Array (VLA) to determine the FD values of 194 EGSs in our region, resulting in an EGS FD density of one per $1.6\,{\rm deg}^2$. This is almost a factor of five increase from that of \cite{vaneck11}. Our goal is to carefully study the complex Milky Way magnetic field structure there. We present the EGS source selection criteria in Section~\ref{sec:srcsel}, and describe the details of the observations and data reduction procedures in Section~\ref{sec:obs}. In Section~\ref{sec:rmsynth}, we show our RM-Synthesis results, and interpret the newly derived EGS FD values in Section~\ref{sec:discussion}. We conclude our findings in Section~\ref{sec:conclusion}.

\section{TARGET SOURCE SELECTION CRITERIA}
\label{sec:srcsel}

In this study, we focus on the large-scale magnetic field near the Galactic mid-plane ($|b| \leq 5^\circ$) in the Galactic longitude range of $20^\circ \leq \ell \leq 52^\circ$. This chosen region intercepts the large-scale magnetic field reversal near the Sagittarius arm mentioned above in Section~\ref{sec:intro}. The lower limit in longitude was placed to exclude the complex Galactic centre region \citep[e.g.,][see also \citealt{haverkorn15,han17}]{roy08,law11,pare19}, while the upper limit joins the Canadian Galactic Plane Survey \citep[CGPS;][]{taylor03} that has been used to derive FD values for Galactic magnetism studies \citep{brown03,ordog17}. The imposed range of Galactic latitude ensures a complete coverage of the Milky Way disk field~-- at a distance of $28.5\,{\rm kpc}$ to the far side of our Galaxy, $|b| = 5^\circ$ corresponds to a Galactic height of $|z| = 2.5\,{\rm kpc}$, well covering the scale heights of thermal electrons \citep[$\approx 1.3$--$1.8\,{\rm kpc}$;][]{gaensler08,schnitzeler12} and the disk magnetic field \citep[$\approx 1$--$2\,{\rm kpc}$;][]{jansson09,kronberg11}.

The target EGSs\footnote{Known Galactic sources were identified and excluded from observations.} were selected using two criteria. Firstly, we chose sources from the original NVSS catalogue \citep{condon98} based on their listed polarisation properties: debiased polarised intensity \citep[PI;][]{wardle74} of more than $4\sigma$, and polarisation fraction ($p$) of more than 0.5 per cent. These ensure that an adequate signal-to-noise ($S/N$) ratio in polarisation can be reached within a reasonable per-source integration time ($\lesssim 5\,{\rm minutes}$), and that unpolarised sources that falsely appear as polarised due to residual off-axis instrumental polarisation are not included. These target sources will be referred to as the \emph{NVSS targets} from hereon. Secondly, we further included all EGSs in the \cite{taylor09} catalogue that were not already selected as NVSS targets, and will refer to them as the \emph{\cite{taylor09} targets}.

While both the original NVSS catalogue and the \cite{taylor09} catalogue were derived from the same NVSS data \citep{condon98}, they were processed differently and are therefore sensitive to different populations of polarised EGSs. The original NVSS catalogue combined the two 42\,MHz-wide sub-bands centred at 1364.9 and 1435.1\,MHz, while the \cite{taylor09} catalogue processed them independently. This makes the original NVSS more sensitive to low-PI EGSs than \cite{taylor09} in general. However, for the same reason the \cite{taylor09} catalogue is less susceptible to bandwidth depolarisation, with the observed PI reduced by more than 50 per cent for sources with $|{\rm FD}| \gtrsim 600\,{\rm rad\,m}^{-2}$. This is much greater than the $|{\rm FD}| \gtrsim 220\,{\rm rad\,m}^{-2}$ for the original NVSS catalogue. In other words, the \cite{taylor09} catalogue is much more sensitive to the high $|{\rm FD}|$ population, which is essential for our study since the $|{\rm FD}|$ values of EGSs within our region of interest can reach $\gtrsim 600\,{\rm rad\,m}^{-2}$ \citep[e.g.,][]{vaneck11}. We note the recent discovery of EGSs with extremely high $|{\rm FD}|$ (up to $4000\,{\rm rad\,m}^{-2}$) towards the Sagittarius arm tangent that are believed to trace the compressed warm ionised medium \citep{shanahan19}. This does not impact our study here focusing on the Galactic-scale magnetic field, but must be taken into account to obtain a complete picture of the magnetism of the Milky Way (see Section~\ref{sec:connection}).

Finally, we identified targets that lie within $2^\prime$ from each other. These close-by targets were grouped together and were then observed by a single pointing to optimise the granted observing time. We made sure that off-axis instrumental polarisation did not affect our final results (see Section~\ref{sec:rmsynth}).

\section{Observations and Data Reduction} 
\label{sec:obs}

We performed new broadband spectro-polarimetric observations of the 176 on-axis targets using the VLA in L-band (1--2\,GHz) in D-array configuration under project code 18A-332. The project was divided into seven observing blocks executed on 2018 September 1--10. In all observing blocks, 3C 286 was observed and used as the absolute flux, bandpass, and PA calibrators, while J1407$+$2827 was chosen as the unpolarised on-axis leakage calibrator. Depending on the observing block, either J1822$-$0938, J1859$+$1259, J1941$+$1026, or J1942$+$1026 was used as the phase calibrator. The observations totalled 15 hours, with an integration time of 3--5\,minutes on each on-axis target.

The Common Astronomy Software Applications (\texttt{CASA}) package \citep[version 5.3.0;][]{mcmullin07} was used for all data reduction procedures. Measurement sets from the seven observing blocks were independently calibrated. We first applied Hanning smoothing to all visibility data in frequency space to remove the Gibbs phenomenon. Corrupted data due to radio frequency interference (RFI) or phase instabilities were then flagged. Afterwards, we performed antenna position, delay, absolute flux, bandpass, complex gain, on-axis instrumental polarisation, and polarisation PA calibrations. Specifically, we followed the \cite{perley13a} and \cite{perley13b} scales for the flux density and PA of 3C 286, respectively. Finally, phase self calibration solutions were determined for each on-axis target for one iteration, but were applied only if they led to significant improvements in the image rms noise (reduction by more than 10 per cent).

The calibrated data from above were used to form channel images of the on-axis targets in Stokes \textit{I}, \textit{Q}, and \textit{U} across L-band. We binned visibility data in 4\,MHz channels to form the channel images instead of using the native 1\,MHz channel width. This improved the per-channel $S/N$ ratio (by a factor of two) without significant loss of information, as the Hanning smoothing procedure above had already reduced the effective spectral resolution by a factor of two. We formed the images using the \texttt{CASA} task \texttt{TCLEAN}, adopting a Briggs visibilities weighting of \texttt{robust} $= 0$ \citep{briggs95} to balance between angular resolution and image rms noise. Deconvolution of the dirty images was performed using the Clark algorithm, and no further smoothing was applied to the resulting channel images as we did not directly combine them. Lastly, primary beam corrections were performed to all images. The typical angular resolution of the images is $50^{\prime\prime} \times 42^{\prime\prime}$ at $1.5\,{\rm GHz}$, and the typical rms noise of the channel images near pointing centres in Stokes \textit{I}, \textit{Q}, and \textit{U} are 4.3, 1.4, and $1.5\,{\rm mJy\,beam}^{-1}$, respectively.

We further identified off-axis targets near our 176 on-axis targets by consulting the NVSS catalogue \citep{condon98}. These off-axis targets include, but are not limited to, the close-by targets described at the end of Section~\ref{sec:srcsel}. All sources that are within $5^\prime$ from the pointing centres and have reported NVSS flux densities of $\ge 20\,{\rm mJy}$ in Stokes \textit{I} are considered. The primary beam attenuation level is still close to unity at this $5^\prime$ distance (0.86 at 2\,GHz), and therefore the image rms noise level is still at an acceptable level after the primary beam correction. Moreover, we do not expect off-axis instrumental polarisation to significantly alter the astrophysical polarisation signals within this $5^\prime$ radius. This has been carefully assessed to make sure that our conclusions are not affected (see Section~\ref{sec:rmsynth}).

The images of each source were examined carefully, leading us to discard 15 of the target sources (both on- and off-axis). These sources were not confidently detected even in Stokes \textit{I}, either because they were too faint or they were affected by poor image fidelity due to bright neighbouring sources. Furthermore, some of the off-axis targets were not clearly spatially distinguished from the corresponding on-axis targets in our images. For such cases, we extracted their combined flux densities (in Stokes \textit{I}, \textit{Q}, and \textit{U}) below instead of separating them. We list both the discarded and spatially blended sources in Appendix~A in the Online Supporting Information.

\begin{table*}
\caption{RM-Synthesis Results for On-axis Targets}
\label{table:rm1}
\begin{tabular}{lcccccccc}
\hline
\multicolumn{1}{c}{Target Source} & $\ell$ & $b$ & ${\rm FD}_{\rm new}$ & ${\rm FD}_{\rm VE11}$ & ${\rm FD}_{\rm TSS09}$ & $p_{\rm new}$ & $p_{\rm TSS09}$ & $p_{\rm NVSS}$ \\
\multicolumn{1}{c}{(NVSS)} & ($^\circ$) & ($^\circ$) & (${\rm rad\,m}^{-2}$) & (${\rm rad\,m}^{-2}$) & (${\rm rad\,m}^{-2}$) & (\%) & (\%) & (\%) \\
\hline
J184415$-$131243 & $20.34$ & $-4.42$ & \phantom{0}$-96.0 \pm 2.4$\phantom{0} & --- & --- & \phantom{0}$3.35 \pm 0.13$ & --- & \phantom{0}$2.54 \pm 0.54$ \\
J181343$-$090743 & $20.49$ & $+4.11$ & \phantom{0}$+11.6 \pm 6.1$\phantom{0} & --- & --- & \phantom{0}$4.42 \pm 0.44$ & --- & \phantom{0}$9.67 \pm 2.19$ \\
J182038$-$094716 & $20.71$ & $+2.29$ & --- & --- & --- & ($0.02$) & --- & \phantom{0}$0.58 \pm 0.08$ \\
J183519$-$111559 & $21.08$ & $-1.59$ & \phantom{0}$-60.4 \pm 1.5$\phantom{0} & \phantom{0}$-66 \pm 12$ & \phantom{0}$-23.4 \pm 17.6$ & \phantom{0}$2.31 \pm 0.06$ & \phantom{0}$3.02 \pm 0.20$ & \phantom{0}$2.70 \pm 0.22$ \\
J181851$-$090659 & $21.10$ & $+3.00$ & $+237.6 \pm 2.5$\phantom{0} & --- & --- & \phantom{0}$4.07 \pm 0.16$ & --- & \phantom{0}$2.13 \pm 0.48$ \\
J181931$-$091059 & $21.12$ & $+2.82$ & $+186.5 \pm 2.6$\phantom{0} & --- & $+204.3 \pm 12.5$ & \phantom{0}$4.81 \pm 0.20$ & \phantom{0}$4.85 \pm 0.28$ & \phantom{0}$2.17 \pm 0.33$ \\
J183759$-$112627 & $21.23$ & $-2.25$ & \phantom{0}$-91.0 \pm 1.2$\phantom{0} & \phantom{0}$-83 \pm 4$\phantom{0} & \phantom{0}$-75.7 \pm 6.0$\phantom{0} & $11.49 \pm 0.23$ & $10.10 \pm 0.29$ & \phantom{0}$8.41 \pm 0.47$ \\
J183220$-$103510 & $21.35$ & $-0.63$ & --- & --- & \phantom{0}$-27.0 \pm 10.4$ & ($0.04$) & \phantom{0}$1.01 \pm 0.05$ & \phantom{0}$0.21 \pm 0.07$ \\
J182443$-$092933 & $21.44$ & $+1.54$ & \phantom{0}$+54.3 \pm 3.6$\phantom{0} & --- & --- & \phantom{0}$5.81 \pm 0.34$ & --- & \phantom{0}$3.52 \pm 0.46$ \\
J184606$-$115808 & $21.66$ & $-4.26$ & $-182.1 \pm 1.5$\phantom{0} & --- & $-206.1 \pm 11.9$ & \phantom{0}$2.23 \pm 0.05$ & \phantom{0}$1.72 \pm 0.09$ & \phantom{0}$1.36 \pm 0.08$ \\
J181419$-$073733 & $21.88$ & $+4.69$ & \phantom{0}$+69.4 \pm 2.6$\phantom{0} & --- & --- & $15.93 \pm 0.68$ & --- & $12.15 \pm 2.73$ \\
J184059$-$110139 & $21.93$ & $-2.72$ & \phantom{0}$-19.6 \pm 1.6$\phantom{0} & --- & --- & \phantom{0}$3.91 \pm 0.10$ & --- & \phantom{0}$2.85 \pm 0.51$ \\
J182503$-$085445 & $22.00$ & $+1.74$ & $+164.8 \pm 1.1$\phantom{0} & --- & $+148.0 \pm 9.1$\phantom{0} & \phantom{0}$9.88 \pm 0.18$ & \phantom{0}$9.37 \pm 0.41$ & \phantom{0}$6.19 \pm 0.61$ \\
J182542$-$083723 & $22.33$ & $+1.73$ & \phantom{0}$+42.9 \pm 1.9$\phantom{0} & \phantom{0}$+40 \pm 13$ & \phantom{0}$+56.3 \pm 11.9$ & \phantom{0}$3.25 \pm 0.10$ & \phantom{0}$2.49 \pm 0.15$ & \phantom{0}$2.58 \pm 0.19$ \\
J183942$-$101038 & $22.54$ & $-2.05$ & $+127.8 \pm 1.9$\phantom{0} & --- & $+146.9 \pm 14.1$ & \phantom{0}$8.87 \pm 0.27$ & \phantom{0}$7.97 \pm 0.52$ & \phantom{0}$6.96 \pm 0.82$ \\
J184750$-$110658 & $22.61$ & $-4.25$ & \phantom{0}$-47.2 \pm 1.3$\phantom{0} & --- & \phantom{0}$-71.2 \pm 7.7$\phantom{0} & \phantom{0}$7.32 \pm 0.15$ & \phantom{0}$6.80 \pm 0.25$ & \phantom{0}$5.87 \pm 0.33$ \\
J184911$-$111241 & $22.68$ & $-4.59$ & \phantom{00}$-0.7 \pm 2.6$\phantom{0} & --- & --- & \phantom{0}$2.24 \pm 0.09$ & --- & \phantom{0}$2.14 \pm 0.33$ \\
J182530$-$080945 & $22.71$ & $+1.99$ & $-150.1 \pm 4.4$\phantom{0} & --- & --- & \phantom{0}$3.30 \pm 0.23$ & --- & \phantom{0}$1.83 \pm 0.39$ \\
J184812$-$105133 & $22.88$ & $-4.22$ & \phantom{0}$+33.5 \pm 2.4$\phantom{0} & --- & --- & $16.61 \pm 0.65$ & --- & $10.79 \pm 2.47$ \\
J184552$-$103126 & $22.93$ & $-3.56$ & \phantom{00}$-1.2 \pm 1.7$\phantom{0} & --- & --- & $13.08 \pm 0.37$ & --- & \phantom{0}$4.23 \pm 0.93$ \\
J181949$-$065524 & $23.15$ & $+3.81$ & $+126.1 \pm 4.7$\phantom{0} & --- & \phantom{0}$+77.4 \pm 19.4$ & \phantom{0}$2.44 \pm 0.18$ & \phantom{0}$2.10 \pm 0.17$ & \phantom{0}$1.28 \pm 0.25$ \\
J182537$-$073729 & $23.20$ & $+2.21$ & \phantom{0}$-58.0 \pm 1.3$\phantom{0} & \phantom{0}$-62 \pm 13$ & \phantom{0}$-81.1 \pm 13.7$ & \phantom{0}$2.20 \pm 0.05$ & \phantom{0}$2.18 \pm 0.13$ & \phantom{0}$1.92 \pm 0.19$ \\
J182431$-$072714 & $23.23$ & $+2.53$ & \phantom{0}$+27.7 \pm 1.0$\phantom{0} & --- & \phantom{0}$+13.9 \pm 5.7$\phantom{0} & \phantom{0}$6.43 \pm 0.11$ & \phantom{0}$6.88 \pm 0.18$ & \phantom{0}$6.05 \pm 0.30$ \\
J182920$-$073400$^\odot$ & $23.68$ & $+1.42$ & $+360.4 \pm 3.0$\phantom{0} & --- & $+332.6 \pm 16.6$ & \phantom{0}$7.48 \pm 0.37$ & \phantom{0}$7.03 \pm 0.49$ & $-0.30 \pm 0.77$ \\
J184644$-$094654 & $23.68$ & $-3.41$ & \phantom{0}$+90.5 \pm 1.7$\phantom{0} & --- & \phantom{0}$+96.5 \pm 10.1$ & \phantom{0}$5.60 \pm 0.16$ & \phantom{0}$5.13 \pm 0.25$ & \phantom{0}$4.30 \pm 0.30$ \\
J184547$-$093821 & $23.70$ & $-3.14$ & $+190.4 \pm 6.3$\phantom{0} & --- & --- & \phantom{0}$5.60 \pm 0.57$ & --- & \phantom{0}$2.16 \pm 0.36$ \\
J183052$-$074402$^\odot$ & $23.71$ & $+1.01$ & $+517.8 \pm 1.8$\phantom{0} & --- & $+518.1 \pm 18.9$ & $11.37 \pm 0.33$ & \phantom{0}$6.90 \pm 0.60$ & \phantom{0}$3.61 \pm 1.08$ \\
J182043$-$062415 & $23.71$ & $+3.86$ & \phantom{0}$-14.3 \pm 2.3$\phantom{0} & --- & --- & \phantom{0}$6.41 \pm 0.24$ & --- & \phantom{0}$6.51 \pm 0.79$ \\
J185239$-$101324 & $23.95$ & $-4.91$ & \phantom{0}$+65.5 \pm 1.5$\phantom{0} & --- & \phantom{0}$+45.8 \pm 9.5$\phantom{0} & \phantom{0}$4.35 \pm 0.11$ & \phantom{0}$5.84 \pm 0.22$ & \phantom{0}$5.97 \pm 0.76$ \\
J183902$-$083023$^\dagger$$^\odot$ & $23.95$ & $-1.14$ & $+526.6 \pm 0.4$\phantom{0} & --- & $-119.6 \pm 5.5$\phantom{0} & $12.63 \pm 0.09$ & \phantom{0}$8.40 \pm 0.22$ & \phantom{0}$6.67 \pm 0.42$ \\
J182104$-$060915 & $23.98$ & $+3.90$ & \phantom{00}$-6.0 \pm 4.2$\phantom{0} & --- & --- & \phantom{0}$4.44 \pm 0.30$ & --- & \phantom{0}$3.71 \pm 0.54$ \\
J183321$-$073121$^\odot$ & $24.18$ & $+0.56$ & $+776.8 \pm 3.1$\phantom{0} & --- & --- & \phantom{0}$1.23 \pm 0.06$ & --- & \phantom{0}$0.69 \pm 0.13$ \\
J183409$-$071802$^\odot$ & $24.47$ & $+0.49$ & --- & --- & \phantom{0}$-10.0 \pm 4.4$\phantom{0} & ($0.02$) & \phantom{0}$0.85 \pm 0.02$ & \phantom{0}$0.16 \pm 0.05$ \\
J185030$-$090659 & $24.70$ & $-3.94$ & $+172.0 \pm 0.8$\phantom{0} & --- & $+151.3 \pm 5.7$\phantom{0} & \phantom{0}$6.53 \pm 0.09$ & \phantom{0}$5.65 \pm 0.15$ & \phantom{0}$3.15 \pm 0.19$ \\
J184249$-$075604$^\dagger$$^\odot$ & $24.89$ & $-1.71$ & $+935.2 \pm 2.0$\phantom{0} & --- & $+160.8 \pm 5.2$\phantom{0} & \phantom{0}$6.18 \pm 0.20$ & \phantom{0}$2.13 \pm 0.06$ & \phantom{0}$0.62 \pm 0.06$ \\
J182351$-$052429 & $24.96$ & $+3.63$ & \phantom{0}$+17.4 \pm 5.3$\phantom{0} & --- & --- & \phantom{0}$7.32 \pm 0.63$ & --- & $11.46 \pm 1.49$ \\
J182111$-$050219 & $24.98$ & $+4.39$ & $+186.8 \pm 6.2$\phantom{0} & --- & --- & \phantom{0}$2.02 \pm 0.20$ & --- & \phantom{0}$1.19 \pm 0.20$ \\
J184629$-$081333$^\odot$ & $25.05$ & $-2.65$ & $+476.1 \pm 1.0$\phantom{0} & $+491 \pm 8$\phantom{0} & --- & \phantom{0}$6.09 \pm 0.10$ & --- & \phantom{0}$2.13 \pm 0.40$ \\
J182013$-$042541 & $25.41$ & $+4.89$ & \phantom{0}$+68.9 \pm 1.0$\phantom{0} & --- & \phantom{0}$+59.9 \pm 5.5$\phantom{0} & \phantom{0}$5.04 \pm 0.08$ & \phantom{0}$4.19 \pm 0.11$ & \phantom{0}$4.35 \pm 0.18$ \\
J184511$-$060146$^\odot$ & $26.85$ & $-1.36$ & $+117.4 \pm 2.3$\phantom{0} & --- & --- & \phantom{0}$7.61 \pm 0.29$ & --- & \phantom{0}$4.67 \pm 0.84$ \\
J183253$-$042628$^\odot$ & $26.86$ & $+2.09$ & $+188.8 \pm 1.7$\phantom{0} & --- & --- & \phantom{0}$3.60 \pm 0.10$ & --- & \phantom{0}$2.06 \pm 0.40$ \\
J182634$-$030927 & $27.27$ & $+4.08$ & $+225.5 \pm 1.7$\phantom{0} & --- & $+164.7 \pm 9.5$\phantom{0} & \phantom{0}$3.96 \pm 0.11$ & \phantom{0}$3.42 \pm 0.14$ & \phantom{0}$1.84 \pm 0.18$ \\
J183847$-$040042$^\odot$ & $27.92$ & $+0.98$ & $+312.3 \pm 0.6$\phantom{0} & --- & $+287.1 \pm 8.2$\phantom{0} & \phantom{0}$4.08 \pm 0.04$ & \phantom{0}$2.82 \pm 0.08$ & \phantom{0}$0.54 \pm 0.09$ \\
J183400$-$030340$^\odot$ & $28.22$ & $+2.48$ & $+162.8 \pm 9.7$\phantom{0} & --- & --- & \phantom{0}$0.33 \pm 0.05$ & --- & \phantom{0}$0.71 \pm 0.16$ \\
J185054$-$050942$^\odot$ & $28.27$ & $-2.23$ & $+583.9 \pm 1.1$\phantom{0} & $+577 \pm 10$ & --- & \phantom{0}$4.48 \pm 0.08$ & --- & \phantom{0}$2.14 \pm 0.25$ \\
J184415$-$041757$^\odot$ & $28.29$ & $-0.36$ & \phantom{0}$+51.8 \pm 7.1$\phantom{0} & --- & --- & \phantom{0}$0.41 \pm 0.05$ & --- & \phantom{0}$0.83 \pm 0.17$ \\
J185523$-$053804$^\odot$ & $28.36$ & $-3.44$ & $+173.7 \pm 1.2$\phantom{0} & --- & --- & $12.03 \pm 0.23$ & --- & \phantom{0}$6.16 \pm 1.00$ \\
J183652$-$024606$^\odot$ & $28.81$ & $+1.97$ & $+571.5 \pm 2.8$\phantom{0} & --- & --- & \phantom{0}$1.06 \pm 0.05$ & --- & \phantom{0}$0.99 \pm 0.24$ \\
J185744$-$052527 & $28.81$ & $-3.87$ & $+232.7 \pm 1.5$\phantom{0} & --- & $+222.8 \pm 13.5$ & $11.12 \pm 0.27$ & \phantom{0}$9.14 \pm 0.68$ & \phantom{0}$3.64 \pm 0.62$ \\
J183939$-$030047$^\odot$ & $28.91$ & $+1.24$ & $+675.0 \pm 0.8$\phantom{0} & --- & $+639.4 \pm 10.7$ & \phantom{0}$9.59 \pm 0.13$ & \phantom{0}$4.41 \pm 0.22$ & \phantom{0}$5.03 \pm 0.47$ \\
J183717$-$015034 & $29.68$ & $+2.30$ & $+307.8 \pm 0.9$\phantom{0} & --- & $+284.2 \pm 9.1$\phantom{0} & \phantom{0}$3.53 \pm 0.05$ & \phantom{0}$1.90 \pm 0.08$ & \phantom{0}$0.39 \pm 0.15$ \\
J182900$-$002018 & $30.07$ & $+4.84$ & \phantom{0}$-56.8 \pm 3.3$\phantom{0} & --- & --- & \phantom{0}$2.00 \pm 0.11$ & --- & \phantom{0}$1.67 \pm 0.34$ \\
J184124$-$015255 & $30.11$ & $+1.37$ & \phantom{0}$+20.4 \pm 2.0$\phantom{0} & --- & $+338.7 \pm 10.7$ & \phantom{0}$0.87 \pm 0.03$ & \phantom{0}$1.10 \pm 0.06$ & \phantom{0}$0.22 \pm 0.14$ \\
J183840$-$012957 & $30.14$ & $+2.16$ & $+345.9 \pm 1.1$\phantom{0} & --- & $+326.6 \pm 7.1$\phantom{0} & \phantom{0}$6.56 \pm 0.11$ & \phantom{0}$8.33 \pm 0.25$ & \phantom{0}$0.38 \pm 0.38$ \\
J183551$-$005941 & $30.27$ & $+3.01$ & $+157.2 \pm 0.7$\phantom{0} & --- & $+152.5 \pm 11.6$ & \phantom{0}$6.94 \pm 0.08$ & \phantom{0}$5.65 \pm 0.28$ & \phantom{0}$3.54 \pm 0.35$ \\
J190014$-$033504 & $30.74$ & $-3.59$ & $+554.9 \pm 1.8$\phantom{0} & --- & --- & \phantom{0}$3.05 \pm 0.09$ & --- & \phantom{0}$1.42 \pm 0.34$ \\
J184959$-$013256 & $31.39$ & $-0.38$ & $+216.9 \pm 3.5$\phantom{0} & --- & \phantom{00}$-7.5 \pm 9.9$\phantom{0} & \phantom{0}$0.37 \pm 0.02$ & \phantom{0}$0.77 \pm 0.03$ & \phantom{0}$0.25 \pm 0.04$ \\
J183838$+$000858 & $31.60$ & $+2.92$ & $+115.3 \pm 1.8$\phantom{0} & --- & --- & \phantom{0}$3.56 \pm 0.10$ & --- & \phantom{0}$2.34 \pm 0.38$ \\
\hline
\multicolumn{9}{l}{\texttt{NOTE} -- ${\rm FD}_{\rm new}$ and $p_{\rm new}$ from this work; ${\rm FD}_{\rm VE11}$ from \cite{vaneck11}; ${\rm FD}_{\rm TSS09}$ and $p_{\rm TSS09}$ from \cite{taylor09}; and} \\
\multicolumn{9}{l}{\phantom{\texttt{NOTE} -- }$p_{\rm NVSS}$ from \cite{condon98}.} \\
\multicolumn{9}{l}{$^\odot$ Situated behind the prominent H {\sc ii} structure G26.5} \\
\multicolumn{9}{l}{$^\dagger$ Suffers from $n\pi$-ambiguity in \cite{taylor09}}
\end{tabular}
\end{table*}

\begin{table*}
\ContinuedFloat
\caption{(Continued) RM-Synthesis Results for On-axis Targets}
\begin{tabular}{lccccccccc}
\hline
\multicolumn{1}{c}{Target Source} & $\ell$ & $b$ & ${\rm FD}_{\rm new}$ & ${\rm FD}_{\rm VE11}$ & ${\rm FD}_{\rm TSS09}$ & $p_{\rm new}$ & $p_{\rm TSS09}$ & $p_{\rm NVSS}$ \\
\multicolumn{1}{c}{(NVSS)} & ($^\circ$) & ($^\circ$) & (${\rm rad\,m}^{-2}$) & (${\rm rad\,m}^{-2}$) & (${\rm rad\,m}^{-2}$) & (\%) & (\%) & (\%) \\
\hline
J183418$+$004852 & $31.70$ & $+4.18$ & \phantom{0}$+80.3 \pm 0.6$\phantom{0} & --- & \phantom{0}$+58.6 \pm 6.4$\phantom{0} & \phantom{0}$6.82 \pm 0.07$ & \phantom{0}$7.44 \pm 0.23$ & \phantom{0}$6.73 \pm 0.45$ \\
J183931$+$001447 & $31.79$ & $+2.76$ & $+216.7 \pm 1.8$\phantom{0} & --- & --- & \phantom{0}$8.97 \pm 0.26$ & --- & \phantom{0}$5.17 \pm 0.86$ \\
J183307$+$011535 & $31.97$ & $+4.65$ & $+372.7 \pm 1.3$\phantom{0} & --- & $+315.4 \pm 11.7$ & \phantom{0}$1.66 \pm 0.03$ & \phantom{0}$1.97 \pm 0.10$ & \phantom{0}$0.23 \pm 0.14$ \\
J183437$+$010519 & $31.98$ & $+4.24$ & \phantom{0}$+72.3 \pm 1.5$\phantom{0} & --- & --- & \phantom{0}$5.61 \pm 0.14$ & --- & \phantom{0}$3.59 \pm 0.70$ \\
J185822$-$013654 & $32.28$ & $-2.28$ & $+558.3 \pm 2.1$\phantom{0} & --- & --- & \phantom{0}$1.12 \pm 0.04$ & --- & \phantom{0}$0.68 \pm 0.15$ \\
J184704$-$000446 & $32.36$ & $+0.93$ & \phantom{0}$+64.3 \pm 1.8$\phantom{0} & --- & \phantom{0}$+54.1 \pm 15.3$ & \phantom{0}$2.34 \pm 0.07$ & \phantom{0}$2.11 \pm 0.13$ & \phantom{0}$2.21 \pm 0.24$ \\
J190833$-$023000 & $32.65$ & $-4.95$ & $+124.2 \pm 1.7$\phantom{0} & --- & --- & \phantom{0}$3.21 \pm 0.09$ & --- & \phantom{0}$1.92 \pm 0.39$ \\
J183511$+$014620 & $32.66$ & $+4.42$ & $+209.7 \pm 0.3$\phantom{0} & --- & $+196.9 \pm 4.4$\phantom{0} & \phantom{0}$9.74 \pm 0.05$ & \phantom{0}$6.29 \pm 0.14$ & \phantom{0}$3.53 \pm 0.18$ \\
J183337$+$020355 & $32.74$ & $+4.91$ & $+194.0 \pm 6.7$\phantom{0} & --- & --- & \phantom{0}$0.27 \pm 0.03$ & --- & \phantom{0}$0.87 \pm 0.15$ \\
J184821$+$001108 & $32.75$ & $+0.77$ & $-145.7 \pm 1.3$\phantom{0} & --- & $-107.2 \pm 5.2$\phantom{0} & \phantom{0}$5.17 \pm 0.11$ & $11.26 \pm 0.32$ & $11.32 \pm 0.64$ \\
J185351$-$002508$^\dagger$ & $32.84$ & $-0.73$ & $+374.7 \pm 0.8$\phantom{0} & --- & $-341.3 \pm 10.0$ & \phantom{0}$6.22 \pm 0.08$ & \phantom{0}$4.62 \pm 0.22$ & $-0.11 \pm 0.34$ \\
J185751$-$004817$^\dagger$ & $32.95$ & $-1.80$ & $+737.6 \pm 4.2$\phantom{0} & --- & \phantom{0}$-26.6 \pm 10.8$ & \phantom{0}$0.37 \pm 0.03$ & \phantom{0}$1.94 \pm 0.10$ & \phantom{0}$0.49 \pm 0.09$ \\
J190042$-$005151 & $33.22$ & $-2.46$ & $+411.3 \pm 1.3$\phantom{0} & --- & $+373.4 \pm 19.6$ & \phantom{0}$3.46 \pm 0.07$ & \phantom{0}$2.63 \pm 0.22$ & \phantom{0}$0.90 \pm 0.29$ \\
J190407$-$011342 & $33.29$ & $-3.38$ & $+278.0 \pm 1.1$\phantom{0} & --- & $+211.4 \pm 12.3$ & \phantom{0}$3.57 \pm 0.06$ & \phantom{0}$3.28 \pm 0.19$ & \phantom{0}$1.05 \pm 0.24$ \\
J185146$+$003532 & $33.50$ & $+0.19$ & $-274.0 \pm 0.6$\phantom{0} & --- & $-313.1 \pm 3.4$\phantom{0} & \phantom{0}$2.82 \pm 0.03$ & \phantom{0}$3.02 \pm 0.05$ & \phantom{0}$0.84 \pm 0.09$ \\
J190832$-$011929 & $33.70$ & $-4.41$ & $+188.2 \pm 1.8$\phantom{0} & --- & --- & \phantom{0}$4.86 \pm 0.14$ & --- & \phantom{0}$3.16 \pm 0.64$ \\
J184755$+$012221 & $33.75$ & $+1.41$ & $+138.0 \pm 3.8$\phantom{0} & --- & --- & \phantom{0}$3.93 \pm 0.24$ & --- & \phantom{0}$4.26 \pm 0.67$ \\
J185857$+$000727 & $33.90$ & $-1.61$ & $+268.3 \pm 1.4$\phantom{0} & --- & $+240.5 \pm 10.6$ & \phantom{0}$4.74 \pm 0.11$ & \phantom{0}$3.79 \pm 0.18$ & \phantom{0}$0.82 \pm 0.28$ \\
J190017$+$000355 & $34.00$ & $-1.94$ & $+547.1 \pm 1.1$\phantom{0} & --- & --- & \phantom{0}$2.08 \pm 0.04$ & --- & \phantom{0}$0.85 \pm 0.18$ \\
J184435$+$020933 & $34.07$ & $+2.51$ & \phantom{0}$+43.5 \pm 2.7$\phantom{0} & --- & --- & \phantom{0}$7.25 \pm 0.32$ & --- & \phantom{0}$7.14 \pm 1.13$ \\
J190831$-$004855 & $34.16$ & $-4.17$ & \phantom{0}$-14.9 \pm 2.5$\phantom{0} & --- & --- & \phantom{0}$7.77 \pm 0.31$ & --- & \phantom{0}$6.70 \pm 1.57$ \\
J191010$-$005622 & $34.23$ & $-4.60$ & \phantom{0}$-54.7 \pm 2.6$\phantom{0} & --- & --- & \phantom{0}$8.62 \pm 0.37$ & --- & \phantom{0}$8.44 \pm 2.11$ \\
J190532$-$000941 & $34.40$ & $-3.21$ & $+126.8 \pm 1.3$\phantom{0} & --- & $+114.7 \pm 11.9$ & \phantom{0}$4.70 \pm 0.10$ & \phantom{0}$5.30 \pm 0.31$ & \phantom{0}$4.47 \pm 0.47$ \\
J190559$+$000721 & $34.70$ & $-3.18$ & \phantom{0}$+20.6 \pm 1.7$\phantom{0} & --- & --- & \phantom{0}$2.97 \pm 0.08$ & --- & \phantom{0}$2.50 \pm 0.40$ \\
J190655$+$000339 & $34.75$ & $-3.42$ & \phantom{00}$+8.8 \pm 2.5$\phantom{0} & --- & --- & \phantom{0}$5.41 \pm 0.22$ & --- & \phantom{0}$9.41 \pm 2.13$ \\
J190741$+$000038 & $34.80$ & $-3.61$ & \phantom{0}$-73.1 \pm 0.8$\phantom{0} & --- & \phantom{0}$-74.2 \pm 13.1$ & \phantom{0}$5.37 \pm 0.07$ & \phantom{0}$4.37 \pm 0.33$ & \phantom{0}$4.17 \pm 0.32$ \\
J183848$+$040424 & $35.13$ & $+4.66$ & $+146.0 \pm 1.3$\phantom{0} & --- & $+129.8 \pm 16.1$ & \phantom{0}$2.57 \pm 0.05$ & \phantom{0}$1.15 \pm 0.09$ & \phantom{0}$0.57 \pm 0.10$ \\
J185515$+$021054$^\dagger$ & $35.31$ & $+0.15$ & \phantom{0}$+92.2 \pm 1.7$\phantom{0} & --- & $-553.2 \pm 12.5$ & $10.79 \pm 0.30$ & \phantom{0}$6.38 \pm 0.41$ & \phantom{0}$7.25 \pm 0.69$ \\
J191133$+$001449 & $35.45$ & $-4.36$ & \phantom{0}$-38.3 \pm 2.0$\phantom{0} & --- & --- & \phantom{0}$6.24 \pm 0.20$ & --- & \phantom{0}$5.37 \pm 0.82$ \\
J190426$+$011036 & $35.46$ & $-2.36$ & \phantom{0}$-68.4 \pm 2.4$\phantom{0} & --- & --- & \phantom{0}$4.87 \pm 0.19$ & --- & \phantom{0}$6.35 \pm 1.18$ \\
J185114$+$025939 & $35.57$ & $+1.41$ & $+177.3 \pm 1.4$\phantom{0} & $+175 \pm 19$ & \phantom{0}$+34.6 \pm 14.1$ & \phantom{0}$1.75 \pm 0.04$ & \phantom{0}$4.46 \pm 0.30$ & \phantom{0}$1.17 \pm 0.12$ \\
J184320$+$040256 & $35.62$ & $+3.65$ & \phantom{0}$+31.1 \pm 2.0$\phantom{0} & --- & --- & \phantom{0}$4.08 \pm 0.14$ & --- & \phantom{0}$3.36 \pm 0.59$ \\
J190944$+$005558 & $35.85$ & $-3.65$ & \phantom{0}$+35.7 \pm 3.3$\phantom{0} & --- & --- & \phantom{0}$3.56 \pm 0.19$ & --- & \phantom{0}$3.63 \pm 0.67$ \\
J191417$+$002421 & $35.91$ & $-4.90$ & \phantom{0}$-20.3 \pm 1.8$\phantom{0} & --- & --- & $16.25 \pm 0.46$ & --- & \phantom{0}$7.76 \pm 1.60$ \\
J190712$+$012709 & $36.02$ & $-2.84$ & $+193.6 \pm 0.7$\phantom{0} & --- & $+134.7 \pm 9.4$\phantom{0} & \phantom{0}$2.69 \pm 0.03$ & \phantom{0}$1.68 \pm 0.08$ & \phantom{0}$0.91 \pm 0.11$ \\
J185213$+$033255$^\dagger$ & $36.18$ & $+1.44$ & $+185.7 \pm 0.9$\phantom{0} & $+188 \pm 7$\phantom{0} & $-458.4 \pm 8.7$\phantom{0} & \phantom{0}$7.03 \pm 0.10$ & \phantom{0}$5.89 \pm 0.22$ & \phantom{0}$3.63 \pm 0.27$ \\
J185837$+$024518 & $36.20$ & $-0.34$ & $+161.3 \pm 2.8$\phantom{0} & $+132 \pm 11$ & --- & \phantom{0}$5.63 \pm 0.26$ & --- & \phantom{0}$3.21 \pm 0.67$ \\
J184500$+$043812 & $36.33$ & $+3.54$ & \phantom{00}$+2.9 \pm 2.0$\phantom{0} & --- & --- & $14.33 \pm 0.47$ & --- & \phantom{0}$9.34 \pm 1.60$ \\
J184604$+$043450 & $36.40$ & $+3.28$ & \phantom{0}$+34.8 \pm 3.0$\phantom{0} & --- & --- & \phantom{0}$6.92 \pm 0.33$ & --- & \phantom{0}$4.39 \pm 0.70$ \\
J185802$+$031316$^\dagger$ & $36.55$ & $+0.00$ & $+424.1 \pm 0.3$\phantom{0} & --- & $-241.0 \pm 3.6$\phantom{0} & \phantom{0}$4.75 \pm 0.03$ & \phantom{0}$2.17 \pm 0.03$ & \phantom{0}$0.79 \pm 0.05$ \\
J185306$+$044052$^\dagger$ & $37.29$ & $+1.76$ & $+294.2 \pm 0.8$\phantom{0} & --- & $-367.1 \pm 10.6$ & $14.53 \pm 0.20$ & $11.22 \pm 0.61$ & \phantom{0}$3.24 \pm 1.19$ \\
J184718$+$055022 & $37.67$ & $+3.57$ & $+123.4 \pm 6.4$\phantom{0} & --- & --- & \phantom{0}$3.09 \pm 0.32$ & --- & \phantom{0}$5.43 \pm 1.18$ \\
J184438$+$062651 & $37.91$ & $+4.44$ & $+200.6 \pm 2.1$\phantom{0} & --- & $+179.0 \pm 16.1$ & \phantom{0}$5.32 \pm 0.18$ & \phantom{0}$4.34 \pm 0.31$ & \phantom{0}$2.16 \pm 0.34$ \\
J191406$+$025549 & $38.13$ & $-3.70$ & $+545.8 \pm 3.1$\phantom{0} & --- & --- & \phantom{0}$5.30 \pm 0.27$ & --- & \phantom{0}$3.09 \pm 0.78$ \\
J184432$+$064257 & $38.14$ & $+4.58$ & $+213.7 \pm 0.5$\phantom{0} & --- & $+183.7 \pm 6.3$\phantom{0} & \phantom{0}$2.75 \pm 0.02$ & \phantom{0}$2.47 \pm 0.07$ & \phantom{0}$1.35 \pm 0.08$ \\
J185513$+$052158$^\dagger$ & $38.14$ & $+1.60$ & $+369.9 \pm 0.9$\phantom{0} & --- & $-325.5 \pm 10.5$ & \phantom{0}$7.68 \pm 0.12$ & \phantom{0}$6.29 \pm 0.31$ & $-0.16 \pm 0.52$ \\
J184919$+$063211 & $38.52$ & $+3.44$ & \phantom{0}$+21.9 \pm 2.4$\phantom{0} & --- & --- & $12.72 \pm 0.49$ & --- & \phantom{0}$9.67 \pm 2.35$ \\
J191325$+$034308$^\dagger$ & $38.76$ & $-3.18$ & $+348.1 \pm 1.2$\phantom{0} & --- & $-331.6 \pm 5.3$\phantom{0} & \phantom{0}$4.84 \pm 0.09$ & \phantom{0}$7.45 \pm 0.18$ & $-0.12 \pm 0.27$ \\
J191849$+$030442 & $38.81$ & $-4.67$ & \phantom{0}$+70.3 \pm 2.0$\phantom{0} & --- & \phantom{0}$-66.1 \pm 18.0$ & \phantom{0}$1.48 \pm 0.05$ & \phantom{0}$2.01 \pm 0.17$ & \phantom{0}$0.29 \pm 0.13$ \\
J184753$+$071538 & $39.00$ & $+4.09$ & $+181.0 \pm 1.9$\phantom{0} & --- & $+169.6 \pm 15.7$ & \phantom{0}$3.95 \pm 0.12$ & \phantom{0}$4.75 \pm 0.33$ & \phantom{0}$3.86 \pm 0.47$ \\
J190343$+$055256 & $39.56$ & $-0.04$ & $+445.6 \pm 2.9$\phantom{0} & --- & --- & \phantom{0}$1.10 \pm 0.05$ & --- & \phantom{0}$0.76 \pm 0.12$ \\
J190043$+$064546$^\dagger$ & $40.01$ & $+1.03$ & $+384.4 \pm 1.1$\phantom{0} & $+380 \pm 5$\phantom{0} & $-252.3 \pm 13.4$ & \phantom{0}$7.69 \pm 0.13$ & \phantom{0}$6.30 \pm 0.39$ & \phantom{0}$1.70 \pm 0.46$ \\
J191725$+$044236 & $40.10$ & $-3.61$ & \phantom{0}$+83.5 \pm 1.5$\phantom{0} & $+107 \pm 14$ & --- & \phantom{0}$2.94 \pm 0.07$ & --- & \phantom{0}$1.97 \pm 0.33$ \\
J192049$+$042052 & $40.17$ & $-4.52$ & \phantom{0}$+77.3 \pm 1.2$\phantom{0} & --- & --- & \phantom{0}$5.68 \pm 0.11$ & --- & \phantom{0}$3.41 \pm 0.34$ \\
J190734$+$060446$^\dagger$ & $40.18$ & $-0.80$ & $+409.6 \pm 0.9$\phantom{0} & --- & $-226.7 \pm 15.2$ & \phantom{0}$4.11 \pm 0.06$ & \phantom{0}$2.84 \pm 0.19$ & \phantom{0}$0.79 \pm 0.22$ \\
J191840$+$043932 & $40.20$ & $-3.91$ & \phantom{0}$-18.9 \pm 5.1$\phantom{0} & --- & --- & \phantom{0}$6.32 \pm 0.52$ & --- & \phantom{0}$9.23 \pm 2.24$ \\
J184731$+$090047 & $40.53$ & $+4.96$ & $+228.0 \pm 1.0$\phantom{0} & --- & $+211.7 \pm 8.2$\phantom{0} & $10.07 \pm 0.16$ & $10.71 \pm 0.40$ & \phantom{0}$3.54 \pm 0.68$ \\
J192258$+$044354 & $40.76$ & $-4.82$ & $+125.2 \pm 2.2$\phantom{0} & --- & --- & \phantom{0}$6.65 \pm 0.24$ & --- & \phantom{0}$2.89 \pm 0.72$ \\
\hline
\multicolumn{9}{l}{\texttt{NOTE} -- ${\rm FD}_{\rm new}$ and $p_{\rm new}$ from this work; ${\rm FD}_{\rm VE11}$ from \cite{vaneck11}; ${\rm FD}_{\rm TSS09}$ and $p_{\rm TSS09}$ from \cite{taylor09}; and} \\
\multicolumn{9}{l}{\phantom{\texttt{NOTE} -- }$p_{\rm NVSS}$ from \cite{condon98}.} \\
\multicolumn{9}{l}{$^\odot$ Situated behind the prominent H {\sc ii} structure G26.5} \\
\multicolumn{9}{l}{$^\dagger$ Suffers from $n\pi$-ambiguity in \cite{taylor09}}
\end{tabular}
\end{table*}

\begin{table*}
\ContinuedFloat
\caption{(Continued) RM-Synthesis Results for On-axis Targets}
\begin{tabular}{lccccccccc}
\hline
\multicolumn{1}{c}{Target Source} & $\ell$ & $b$ & ${\rm FD}_{\rm new}$ & ${\rm FD}_{\rm VE11}$ & ${\rm FD}_{\rm TSS09}$ & $p_{\rm new}$ & $p_{\rm TSS09}$ & $p_{\rm NVSS}$ \\
\multicolumn{1}{c}{(NVSS)} & ($^\circ$) & ($^\circ$) & (${\rm rad\,m}^{-2}$) & (${\rm rad\,m}^{-2}$) & (${\rm rad\,m}^{-2}$) & (\%) & (\%) & (\%) \\
\hline
J192243$+$045126 & $40.84$ & $-4.71$ & $+129.7 \pm 1.7$\phantom{0} & --- & --- & \phantom{0}$0.83 \pm 0.02$ & --- & \phantom{0}$0.59 \pm 0.08$ \\
J191310$+$064158 & $41.37$ & $-1.75$ & \phantom{0}$+17.5 \pm 3.3$\phantom{0} & --- & --- & \phantom{0}$4.30 \pm 0.23$ & --- & \phantom{0}$3.34 \pm 0.75$ \\
J184951$+$094850 & $41.51$ & $+4.81$ & $+567.5 \pm 1.4$\phantom{0} & --- & --- & \phantom{0}$7.45 \pm 0.17$ & --- & \phantom{0}$2.74 \pm 0.62$ \\
J191917$+$061942 & $41.75$ & $-3.27$ & \phantom{0}$+35.6 \pm 2.6$\phantom{0} & \phantom{0}$+53 \pm 32$ & --- & \phantom{0}$1.48 \pm 0.06$ & --- & \phantom{0}$1.20 \pm 0.23$ \\
J190614$+$084226 & $42.36$ & $+0.70$ & $+125.6 \pm 2.8$\phantom{0} & --- & --- & \phantom{0}$7.03 \pm 0.31$ & --- & \phantom{0}$7.16 \pm 1.39$ \\
J185557$+$102011 & $42.66$ & $+3.70$ & $+570.2 \pm 2.0$\phantom{0} & --- & --- & \phantom{0}$1.13 \pm 0.04$ & --- & \phantom{0}$0.72 \pm 0.16$ \\
J192233$+$071048 & $42.88$ & $-3.59$ & $+196.8 \pm 4.0$\phantom{0} & \phantom{0}$+78 \pm 20$ & --- & \phantom{0}$1.67 \pm 0.11$ & --- & \phantom{0}$2.14 \pm 0.45$ \\
J190741$+$090717 & $42.90$ & $+0.57$ & $+706.5 \pm 0.9$\phantom{0} & $+703 \pm 12$ & --- & \phantom{0}$3.52 \pm 0.05$ & --- & \phantom{0}$1.51 \pm 0.13$ \\
J192245$+$073933 & $43.33$ & $-3.40$ & $+231.1 \pm 2.1$\phantom{0} & --- & $+187.9 \pm 16.3$ & \phantom{0}$5.10 \pm 0.18$ & \phantom{0}$6.46 \pm 0.42$ & \phantom{0}$2.96 \pm 1.04$ \\
J192820$+$070355 & $43.46$ & $-4.91$ & \phantom{0}$+45.6 \pm 3.5$\phantom{0} & --- & --- & \phantom{0}$2.21 \pm 0.13$ & --- & \phantom{0}$1.97 \pm 0.35$ \\
J191906$+$081920 & $43.49$ & $-2.30$ & $+220.5 \pm 0.9$\phantom{0} & $+229 \pm 6$\phantom{0} & $+191.5 \pm 15.7$ & \phantom{0}$9.85 \pm 0.15$ & \phantom{0}$8.82 \pm 0.62$ & \phantom{0}$4.31 \pm 0.69$ \\
J185728$+$111021 & $43.57$ & $+3.75$ & $+524.1 \pm 0.6$\phantom{0} & --- & --- & $11.04 \pm 0.11$ & --- & \phantom{0}$3.12 \pm 0.53$ \\
J191641$+$090147 & $43.84$ & $-1.44$ & $+509.9 \pm 2.1$\phantom{0} & $+505 \pm 6$\phantom{0} & --- & $14.29 \pm 0.49$ & --- & \phantom{0}$7.99 \pm 1.70$ \\
J185952$+$112514 & $44.06$ & $+3.34$ & $+660.8 \pm 2.6$\phantom{0} & $+655 \pm 7$\phantom{0} & --- & \phantom{0}$6.17 \pm 0.26$ & --- & \phantom{0}$2.55 \pm 0.57$ \\
J190323$+$112905$^\dagger$ & $44.51$ & $+2.60$ & $+833.6 \pm 0.3$\phantom{0} & $+831 \pm 3$\phantom{0} & $+141.3 \pm 7.4$\phantom{0} & $17.21 \pm 0.08$ & \phantom{0}$4.52 \pm 0.19$ & \phantom{0}$3.56 \pm 0.26$ \\
J192840$+$084849$^\dagger$ & $45.04$ & $-4.15$ & $+528.1 \pm 1.0$\phantom{0} & --- & $-153.4 \pm 16.1$ & \phantom{0}$2.63 \pm 0.04$ & \phantom{0}$1.59 \pm 0.10$ & \phantom{0}$1.56 \pm 0.15$ \\
J192355$+$094424 & $45.31$ & $-2.68$ & $+298.3 \pm 1.5$\phantom{0} & --- & $+282.2 \pm 8.7$\phantom{0} & \phantom{0}$4.20 \pm 0.10$ & \phantom{0}$4.29 \pm 0.19$ & \phantom{0}$0.69 \pm 0.23$ \\
J185923$+$125912 & $45.41$ & $+4.15$ & $+263.6 \pm 0.4$\phantom{0} & --- & $+154.3 \pm 2.6$\phantom{0} & \phantom{0}$1.88 \pm 0.01$ & \phantom{0}$1.46 \pm 0.02$ & \phantom{0}$0.57 \pm 0.03$ \\
J191005$+$114748$^\dagger$ & $45.54$ & $+1.28$ & $+844.7 \pm 0.9$\phantom{0} & --- & $+149.5 \pm 11.4$ & \phantom{0}$4.41 \pm 0.07$ & \phantom{0}$1.46 \pm 0.07$ & \phantom{0}$0.67 \pm 0.18$ \\
J191000$+$122524 & $46.09$ & $+1.59$ & $+771.1 \pm 2.4$\phantom{0} & $+783 \pm 11$ & --- & \phantom{0}$3.60 \pm 0.14$ & --- & \phantom{0}$1.31 \pm 0.30$ \\
J192922$+$095808$^\dagger$ & $46.14$ & $-3.76$ & \phantom{0}$+19.6 \pm 0.6$\phantom{0} & \phantom{0}$+24 \pm 6$\phantom{0} & $+686.6 \pm 7.0$\phantom{0} & \phantom{0}$4.59 \pm 0.04$ & \phantom{0}$3.73 \pm 0.12$ & \phantom{0}$5.40 \pm 0.21$ \\
J191733$+$114215$^\dagger$ & $46.31$ & $-0.38$ & $-123.8 \pm 1.5$\phantom{0} & $-117 \pm 8$\phantom{0} & $+529.2 \pm 11.6$ & \phantom{0}$6.67 \pm 0.16$ & \phantom{0}$7.98 \pm 0.43$ & \phantom{0}$7.75 \pm 0.72$ \\
J190501$+$132047 & $46.35$ & $+3.09$ & $+588.0 \pm 1.0$\phantom{0} & $+575 \pm 10$ & --- & \phantom{0}$4.93 \pm 0.08$ & --- & \phantom{0}$1.90 \pm 0.28$ \\
J193434$+$104340 & $47.43$ & $-4.52$ & $-288.8 \pm 4.5$\phantom{0} & --- & --- & \phantom{0}$1.31 \pm 0.10$ & --- & \phantom{0}$1.62 \pm 0.40$ \\
J190247$+$145137$^\dagger$ & $47.46$ & $+4.26$ & $+540.8 \pm 1.5$\phantom{0} & --- & $-143.1 \pm 14.7$ & \phantom{0}$4.51 \pm 0.11$ & \phantom{0}$4.15 \pm 0.25$ & \phantom{0}$2.64 \pm 0.35$ \\
J193357$+$105642 & $47.54$ & $-4.28$ & $-158.6 \pm 1.2$\phantom{0} & --- & --- & \phantom{0}$4.31 \pm 0.08$ & --- & \phantom{0}$2.56 \pm 0.37$ \\
J191025$+$140125$^\dagger$ & $47.56$ & $+2.24$ & $+541.3 \pm 1.5$\phantom{0} & --- & $-125.1 \pm 5.9$\phantom{0} & \phantom{0}$3.54 \pm 0.08$ & \phantom{0}$9.29 \pm 0.23$ & \phantom{0}$8.24 \pm 0.65$ \\
J192540$+$122738 & $47.91$ & $-1.77$ & \phantom{0}$+65.4 \pm 2.6$\phantom{0} & \phantom{0}$+63 \pm 12$ & --- & \phantom{0}$1.80 \pm 0.08$ & --- & \phantom{0}$2.23 \pm 0.44$ \\
J190451$+$152148$^\dagger$ & $48.13$ & $+4.05$ & $+545.6 \pm 0.4$\phantom{0} & --- & $-117.1 \pm 5.1$\phantom{0} & \phantom{0}$4.14 \pm 0.03$ & \phantom{0}$2.10 \pm 0.05$ & \phantom{0}$1.65 \pm 0.09$ \\
J190414$+$153638 & $48.28$ & $+4.29$ & $+662.7 \pm 1.8$\phantom{0} & --- & --- & \phantom{0}$2.09 \pm 0.06$ & --- & \phantom{0}$1.03 \pm 0.18$ \\
J192458$+$130033 & $48.31$ & $-1.36$ & $+524.5 \pm 6.3$\phantom{0} & $+435 \pm 8$\phantom{0} & --- & \phantom{0}$1.86 \pm 0.19$ & --- & \phantom{0}$1.78 \pm 0.43$ \\
J190655$+$152342$^\dagger$ & $48.39$ & $+3.62$ & $+636.8 \pm 0.5$\phantom{0} & $+629 \pm 4$\phantom{0} & \phantom{0}$-37.5 \pm 10.1$ & $12.52 \pm 0.10$ & \phantom{0}$7.27 \pm 0.34$ & \phantom{0}$6.66 \pm 0.46$ \\
J193335$+$120844 & $48.56$ & $-3.62$ & \phantom{0}$-34.5 \pm 2.5$\phantom{0} & --- & --- & \phantom{0}$7.90 \pm 0.32$ & --- & \phantom{0}$4.91 \pm 1.10$ \\
J190355$+$160147$^\dagger$ & $48.62$ & $+4.55$ & $+439.8 \pm 0.6$\phantom{0} & --- & $-240.1 \pm 12.9$ & \phantom{0}$5.98 \pm 0.06$ & \phantom{0}$3.89 \pm 0.21$ & \phantom{0}$1.75 \pm 0.27$ \\
J191644$+$150349$^\dagger$ & $49.19$ & $+1.37$ & $+534.5 \pm 0.9$\phantom{0} & $+541 \pm 10$ & $-145.5 \pm 8.8$\phantom{0} & \phantom{0}$3.17 \pm 0.05$ & \phantom{0}$2.36 \pm 0.10$ & \phantom{0}$1.85 \pm 0.13$ \\
J192517$+$135919 & $49.21$ & $-0.97$ & $+450.9 \pm 2.1$\phantom{0} & $+470 \pm 7$\phantom{0} & $+442.5 \pm 3.6$\phantom{0} & \phantom{0}$3.70 \pm 0.13$ & \phantom{0}$3.78 \pm 0.06$ & \phantom{0}$1.19 \pm 0.09$ \\
J190516$+$163706$^\dagger$ & $49.30$ & $+4.53$ & $+489.8 \pm 0.9$\phantom{0} & --- & $-219.6 \pm 9.2$\phantom{0} & \phantom{0}$3.72 \pm 0.05$ & \phantom{0}$2.19 \pm 0.11$ & \phantom{0}$1.55 \pm 0.13$ \\
J193302$+$131335 & $49.44$ & $-2.98$ & \phantom{0}$-74.3 \pm 0.7$\phantom{0} & --- & \phantom{0}$-76.7 \pm 4.9$\phantom{0} & \phantom{0}$4.20 \pm 0.05$ & \phantom{0}$3.03 \pm 0.07$ & \phantom{0}$2.98 \pm 0.19$ \\
J191133$+$161431$^\dagger$ & $49.65$ & $+3.02$ & $+616.0 \pm 1.4$\phantom{0} & --- & \phantom{00}$+7.4 \pm 14.7$ & \phantom{0}$1.67 \pm 0.04$ & \phantom{0}$1.03 \pm 0.06$ & \phantom{0}$0.37 \pm 0.09$ \\
J191158$+$161147 & $49.66$ & $+2.91$ & $+711.8 \pm 1.3$\phantom{0} & --- & --- & \phantom{0}$1.31 \pm 0.03$ & --- & \phantom{0}$0.86 \pm 0.13$ \\
J190901$+$163944$^\dagger$ & $49.75$ & $+3.75$ & $+446.5 \pm 0.5$\phantom{0} & $+451 \pm 10$ & $-235.4 \pm 7.9$\phantom{0} & \phantom{0}$3.49 \pm 0.03$ & \phantom{0}$2.71 \pm 0.10$ & \phantom{0}$1.28 \pm 0.12$ \\
J191219$+$161628$^\dagger$ & $49.77$ & $+2.87$ & $+751.2 \pm 0.6$\phantom{0} & $+751 \pm 8$\phantom{0} & \phantom{0}$+41.7 \pm 8.2$\phantom{0} & \phantom{0}$5.45 \pm 0.05$ & \phantom{0}$2.23 \pm 0.09$ & \phantom{0}$1.86 \pm 0.13$ \\
J192835$+$142156 & $49.92$ & $-1.49$ & $+127.7 \pm 3.6$\phantom{0} & --- & --- & $12.67 \pm 0.74$ & --- & $14.25 \pm 2.27$ \\
J192910$+$141952 & $49.96$ & $-1.63$ & $+125.0 \pm 1.6$\phantom{0} & --- & --- & \phantom{0}$8.73 \pm 0.23$ & --- & \phantom{0}$6.06 \pm 0.73$ \\
J191649$+$155836$^\dagger$ & $50.00$ & $+1.77$ & $+393.6 \pm 0.8$\phantom{0} & --- & $-284.9 \pm 6.7$\phantom{0} & \phantom{0}$6.04 \pm 0.08$ & \phantom{0}$5.21 \pm 0.15$ & \phantom{0}$0.76 \pm 0.23$ \\
J191549$+$160834 & $50.04$ & $+2.06$ & $+471.6 \pm 1.2$\phantom{0} & $+482 \pm 7$\phantom{0} & --- & \phantom{0}$5.81 \pm 0.11$ & --- & \phantom{0}$2.44 \pm 0.42$ \\
J194012$+$125809 & $50.06$ & $-4.64$ & $-157.2 \pm 2.3$\phantom{0} & --- & $-142.3 \pm 11.5$ & \phantom{0}$2.65 \pm 0.10$ & \phantom{0}$4.42 \pm 0.24$ & \phantom{0}$3.04 \pm 0.41$ \\
J191414$+$163640$^\dagger$ & $50.28$ & $+2.62$ & $+556.9 \pm 0.8$\phantom{0} & $+556 \pm 8$\phantom{0} & $-123.7 \pm 15.9$ & \phantom{0}$4.80 \pm 0.07$ & \phantom{0}$3.16 \pm 0.25$ & \phantom{0}$2.85 \pm 0.25$ \\
J192439$+$154043$^\dagger$ & $50.63$ & $-0.03$ & $+420.1 \pm 0.6$\phantom{0} & --- & $-178.8 \pm 14.9$ & \phantom{0}$5.27 \pm 0.05$ & \phantom{0}$1.74 \pm 0.10$ & \phantom{0}$0.29 \pm 0.16$ \\
J193939$+$134604 & $50.70$ & $-4.13$ & $-186.4 \pm 3.1$\phantom{0} & --- & --- & \phantom{0}$3.43 \pm 0.17$ & --- & \phantom{0}$3.28 \pm 0.72$ \\
J192032$+$162557 & $50.82$ & $+1.20$ & $+533.9 \pm 1.3$\phantom{0} & $+543 \pm 6$\phantom{0} & --- & \phantom{0}$9.79 \pm 0.21$ & --- & \phantom{0}$5.14 \pm 0.81$ \\
J192203$+$162243$^\dagger$ & $50.95$ & $+0.85$ & $+457.2 \pm 1.2$\phantom{0} & $+466 \pm 9$\phantom{0} & $-228.1 \pm 10.7$ & \phantom{0}$3.78 \pm 0.07$ & \phantom{0}$2.89 \pm 0.16$ & \phantom{0}$1.30 \pm 0.24$ \\
J193306$+$145624 & $50.95$ & $-2.17$ & $+146.5 \pm 1.9$\phantom{0} & $+186 \pm 15$ & $+137.0 \pm 11.8$ & \phantom{0}$2.29 \pm 0.06$ & \phantom{0}$2.10 \pm 0.11$ & \phantom{0}$1.68 \pm 0.16$ \\
J193321$+$150446$^\dagger$ & $51.10$ & $-2.16$ & $+353.4 \pm 2.6$\phantom{0} & --- & $-238.9 \pm 14.3$ & \phantom{0}$1.15 \pm 0.05$ & \phantom{0}$1.39 \pm 0.09$ & --- \\
J193052$+$153235 & $51.22$ & $-1.41$ & $+139.6 \pm 1.5$\phantom{0} & $+196 \pm 22$ & --- & \phantom{0}$1.44 \pm 0.03$ & --- & \phantom{0}$1.30 \pm 0.09$ \\
\hline
\multicolumn{9}{l}{\texttt{NOTE} -- ${\rm FD}_{\rm new}$ and $p_{\rm new}$ from this work; ${\rm FD}_{\rm VE11}$ from \cite{vaneck11}; ${\rm FD}_{\rm TSS09}$ and $p_{\rm TSS09}$ from \cite{taylor09}; and} \\
\multicolumn{9}{l}{\phantom{\texttt{NOTE} -- }$p_{\rm NVSS}$ from \cite{condon98}.} \\
\multicolumn{9}{l}{$^\odot$ Situated behind the prominent H {\sc ii} structure G26.5} \\
\multicolumn{9}{l}{$^\dagger$ Suffers from $n\pi$-ambiguity in \cite{taylor09}}
\end{tabular}
\end{table*}

\begin{table*}
\caption{RM-Synthesis Results for Off-axis Targets}
\label{table:rm2}
\begin{tabular}{lccccc}
\hline
\multicolumn{1}{c}{Target Source} & $\ell$ & $b$ & ${\rm FD}_{\rm new}$ & $p_{\rm new}$ & $p_{\rm NVSS}$ \\
\multicolumn{1}{c}{(NVSS)} & ($^\circ$) & ($^\circ$) & (${\rm rad\,m}^{-2}$) & (\%) & (\%) \\
\hline
J183756$-$112202 & $21.28$ & $-2.21$ & --- & ($0.77$) & \phantom{0}$2.11 \pm 2.65$ \\
J184555$-$115813 & $21.64$ & $-4.22$ & $-126.1 \pm 2.7$\phantom{0} & $13.74 \pm 0.60$ & \phantom{0}$2.32 \pm 2.64$ \\
J182535$-$083948 & $22.28$ & $+1.74$ & \phantom{00}$+4.8 \pm 7.3$\phantom{0} & \phantom{0}$9.39 \pm 1.12$ & \phantom{0}$0.91 \pm 2.26$ \\
J183931$-$101336 & $22.48$ & $-2.03$ & $+141.2 \pm 7.5$\phantom{0} & \phantom{0}$2.50 \pm 0.31$ & $-0.57 \pm 0.90$ \\
J184906$-$111430 & $22.64$ & $-4.59$ & \phantom{0}$+42.2 \pm 4.3$\phantom{0} & \phantom{0}$5.40 \pm 0.38$ & $-0.56 \pm 1.49$ \\
J184808$-$105535 & $22.82$ & $-4.23$ & \phantom{0}$+16.9 \pm 4.0$\phantom{0} & \phantom{0}$7.08 \pm 0.46$ & \phantom{0}$2.28 \pm 2.72$ \\
J184541$-$093643 & $23.72$ & $-3.10$ & $+190.3 \pm 1.9$\phantom{0} & \phantom{0}$2.74 \pm 0.08$ & \phantom{0}$1.14 \pm 2.13$ \\
J185027$-$091037 & $24.64$ & $-3.96$ & \phantom{0}$+98.0 \pm 5.9$\phantom{0} & \phantom{0}$0.70 \pm 0.07$ & \phantom{0}$0.62 \pm 0.21$ \\
J182058$-$050223 & $24.95$ & $+4.44$ & \phantom{0}$+67.4 \pm 3.4$\phantom{0} & \phantom{0}$4.79 \pm 0.26$ & \phantom{0}$0.53 \pm 0.28$ \\
J184617$-$081126$^\odot$ & $25.05$ & $-2.59$ & $+620.3 \pm 6.1$\phantom{0} & \phantom{0}$4.46 \pm 0.44$ & \phantom{0}$1.65 \pm 1.48$ \\
J182644$-$030952 & $27.29$ & $+4.04$ & $+250.3 \pm 8.7$\phantom{0} & \phantom{0}$4.85 \pm 0.69$ & $-0.75 \pm 1.64$ \\
J183414$-$030119$^\odot$ & $28.28$ & $+2.44$ & $+481.0 \pm 0.9$\phantom{0} & \phantom{0}$4.28 \pm 0.06$ & \phantom{0}$0.69 \pm 0.23$ \\
J183701$-$015140 & $29.63$ & $+2.36$ & $+307.9 \pm 6.8$\phantom{0} & \phantom{0}$1.48 \pm 0.16$ & \phantom{0}$0.32 \pm 0.48$ \\
J183827$-$013111 & $30.10$ & $+2.19$ & $+290.1 \pm 3.2$\phantom{0} & $10.03 \pm 0.53$ & $-0.21 \pm 2.53$ \\
J183603$-$005747 & $30.32$ & $+2.98$ & \phantom{0}$+34.3 \pm 4.4$\phantom{0} & \phantom{0}$1.84 \pm 0.13$ & \phantom{0}$1.58 \pm 0.71$ \\
J183415$+$004451 & $31.64$ & $+4.16$ & \phantom{0}$+52.0 \pm 6.9$\phantom{0} & \phantom{0}$2.51 \pm 0.28$ & \phantom{0}$1.85 \pm 0.93$ \\
J183935$+$001547 & $31.81$ & $+2.76$ & $+167.0 \pm 2.7$\phantom{0} & \phantom{0}$6.37 \pm 0.28$ & \phantom{0}$2.02 \pm 0.81$ \\
J183433$+$010127 & $31.92$ & $+4.22$ & --- & ($0.49$) & \phantom{0}$0.33 \pm 2.33$ \\
J185807$-$004834 & $32.97$ & $-1.86$ & --- & ($0.19$) & \phantom{0}$0.77 \pm 2.25$ \\
J190832$-$005319 & $34.09$ & $-4.21$ & \phantom{00}$+7.0 \pm 3.8$\phantom{0} & $11.89 \pm 0.73$ & $-1.15 \pm 9.39$ \\
J190721$+$012341 & $35.99$ & $-2.90$ & $+106.3 \pm 7.5$\phantom{0} & \phantom{0}$2.06 \pm 0.22$ & $-0.52 \pm 1.83$ \\
J185222$+$033347 & $36.21$ & $+1.42$ & $+162.9 \pm 7.3$\phantom{0} & \phantom{0}$5.46 \pm 0.65$ & $-0.09 \pm 2.03$ \\
J191833$+$043928 & $40.18$ & $-3.88$ & \phantom{0}$+88.6 \pm 4.8$\phantom{0} & \phantom{0}$6.45 \pm 0.50$ & \phantom{0}$0.73 \pm 1.99$ \\
J190616$+$083858 & $42.31$ & $+0.67$ & --- & ($0.25$) & \phantom{0}$0.74 \pm 1.18$ \\
J192802$+$070219 & $43.40$ & $-4.85$ & $+131.3 \pm 2.7$\phantom{0} & \phantom{0}$8.07 \pm 0.36$ & --- \\
J191630$+$090223 & $43.83$ & $-1.39$ & $+544.0 \pm 2.6$\phantom{0} & \phantom{0}$6.63 \pm 0.28$ & \phantom{0}$2.71 \pm 1.06$ \\
J190319$+$112950 & $44.51$ & $+2.62$ & $+787.0 \pm 2.6$\phantom{0} & \phantom{0}$2.15 \pm 0.09$ & \phantom{0}$0.65 \pm 0.24$ \\
J190235$+$145023 & $47.41$ & $+4.30$ & $+549.9 \pm 3.8$\phantom{0} & \phantom{0}$3.94 \pm 0.24$ & \phantom{0}$1.46 \pm 1.01$ \\
J190653$+$152650 & $48.43$ & $+3.65$ & $+506.9 \pm 4.9$\phantom{0} & \phantom{0}$1.99 \pm 0.16$ & $-0.10 \pm 0.66$ \\
J193328$+$120953 & $48.56$ & $-3.59$ & \phantom{0}$-96.8 \pm 3.2$\phantom{0} & $12.87 \pm 0.67$ & \phantom{0}$2.16 \pm 2.40$ \\
J192030$+$162333 & $50.78$ & $+1.19$ & --- & ($0.51$) & \phantom{0}$1.72 \pm 2.66$ \\
J192032$+$162429 & $50.80$ & $+1.19$ & --- & ($0.52$) & \phantom{0}$1.57 \pm 1.92$ \\
J192157$+$162501 & $50.97$ & $+0.89$ & --- & ($0.08$) & $-0.22 \pm 0.39$ \\
\hline
\multicolumn{6}{l}{\texttt{NOTE} -- None of these sources are listed in \cite{taylor09} or \cite{vaneck11}.}\\
\multicolumn{6}{l}{\phantom{\texttt{NOTE} -- }${\rm FD}_{\rm new}$ and $p_{\rm new}$ from this work; and $p_{\rm NVSS}$ from \cite{condon98}.} \\
\multicolumn{6}{l}{$^\odot$ Situated behind the prominent H {\sc ii} structure G26.5}
\end{tabular}
\end{table*}

The flux densities of our target sources in Stokes \textit{I}, \textit{Q}, and \textit{U} were extracted by two different methods, depending on whether they were spatially unresolved or extended. For unresolved sources, we used the \texttt{CASA} task \texttt{IMFIT} to obtain the integrated flux densities. Specifically, we used a 2D Gaussian function for each target and frequency channel, with its size and orientation fixed to that of the corresponding image's synthesised beam. The source's position in each channel image in Stokes \textit{I} was then fitted for, and this position (along with size and orientation) was subsequently fixed to obtain the integrated flux densities in Stokes \textit{I}, \textit{Q}, and \textit{U}. For extended sources, we first formed Stokes \textit{I} images of them using the entire usable L-band with the multi-frequency synthesis (MFS) algorithm. Contours of $3\sigma$ level enclosing each of the extended sources were then defined using the MFS images, and finally the flux densities of the extended sources were extracted with these contours using the \texttt{CASA} task \texttt{IMSTAT}. The Stokes \textit{I} radio spectra of our target sources are reported in Appendix~B in the Online Supporting Information.

\section{Rotation Measure Synthesis Results}
\label{sec:rmsynth}

Using the lists of Stokes \textit{I}, \textit{Q}, and \textit{U} values across frequency for our 204 sources (171 on-axis plus 33 off-axis) described in Section~\ref{sec:obs}, we performed RM-Synthesis \citep{brentjens05} to determine the FD values of our target EGSs. A \texttt{python} implementation of RM-Synthesis in \texttt{CIRADA-Tools}\footnote{Available on \href{https://github.com/CIRADA-Tools/RM}{https://github.com/CIRADA-Tools/RM}.} was used. We used $q = Q/I$ and $u = U/I$ as the inputs to remove the effect of spectral index \citep[this implicitly assumes that the total intensity and linear polarisation originates from the same emission volume; see][]{schnitzeler17,schnitzeler18}, and adopted a normalised inverse noise variance weighting function \citep[e.g.,][]{schnitzeler17} to produce dirty Faraday spectra within $|{\rm FD}| \leq 2000\,{\rm rad\,m}^{-2}$ at steps of $2\,{\rm rad\,m}^{-2}$. Deconvolution of the dirty spectra were subsequently performed with the RM-Clean algorithm \citep[e.g.,][]{heald09} until the residual spectra fell below $6\sigma$. With our observational setup, the resolution of the Faraday spectrum, the maximum detectable scale, and the maximum detectable FD are \citep[equations 61--63 in][]{brentjens05}
\begin{gather}
\delta {\rm FD}_0 \approx \frac{2\sqrt{3}}{\Delta \lambda^2} \approx 123\,{\rm rad\,m}^{-2}{\rm ,} \\
\text{max-scale} \approx \frac{\pi}{\lambda_{\rm min}^2} \approx 144\,{\rm rad\,m}^{-2}{\rm ,~and} \\
||{\rm FD}_{\rm max}|| \approx \frac{\sqrt{3}}{\delta \lambda^2} \approx (6\text{--}20) \times 10^3\,{\rm rad\,m}^{-2}{\rm ,}
\end{gather}
respectively. The quoted range for $||{\rm FD}_{\rm max}||$ represents the difference in widths of the $4\,{\rm MHz}$ channels in $\lambda^2$ space at the two ends of the usable L-band. As an additional check, we formed another set of Faraday spectra within $|{\rm FD}| \leq 20000\,{\rm rad\,m}^{-2}$ to make sure that we did not miss any polarised components with $|{\rm FD}| > 2000\,{\rm rad\,m}^{-2}$.

\begin{figure*}
\includegraphics[width=480pt]{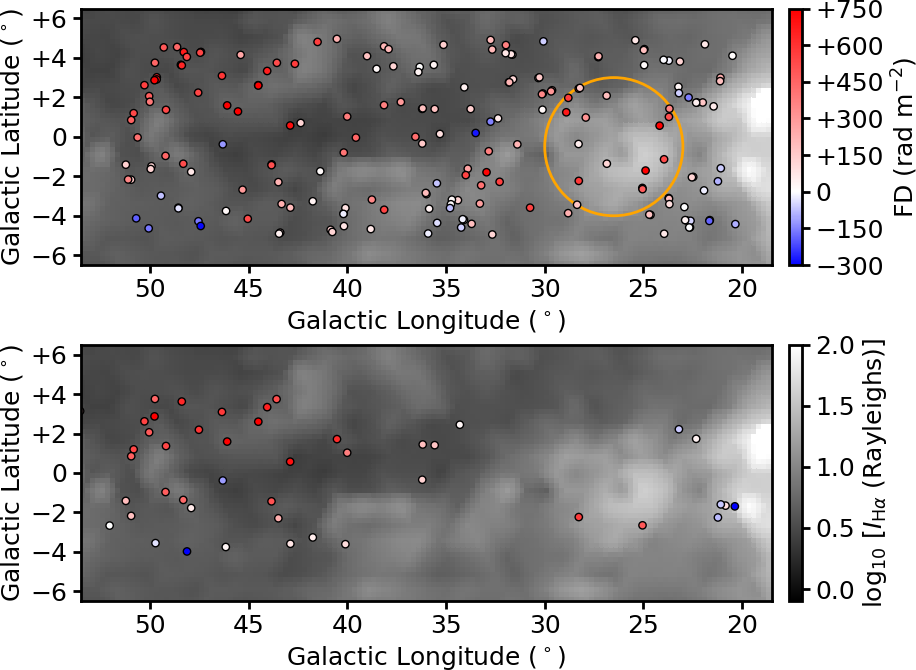}
\caption{EGS FD values from \textbf{(Top)} our new VLA observations and \textbf{(Bottom)} \citet{vaneck11} observations, both plotted as colour dots. The background greyscale map represents the WHAMSS H$\alpha$ map \citep{haffner03,haffner10}. The typical uncertainty of our new FD is about $2\,{\rm rad\,m}^{-2}$, while that of the \citet{vaneck11} is about $10\,{\rm rad\,m}^{-2}$. The orange circle in the top panel outlines a prominent H\,{\sc ii} structure, G26.5, and EGSs within this circle were not considered in our analysis (see Section~\ref{sec:hii_cont}).}
\label{fig:new_ha}
\end{figure*}

The Faraday spectra (amplitudes; $||\mathbf{F}||$) are presented in Appendix C in the Online Supporting Information. We only considered polarised components that are above $6\sigma$ in polarisation fraction, and disregarded signals below this cutoff as manifestations of polarisation bias \citep[see, e.g.,][]{george12}. The FD and $p$ values of our target sources were obtained by fitting a second-order polynomial to the highest peak in $||\mathbf{F}||$ of each source \citep[e.g.,][]{heald09,mao10,betti19}. Specifically, we fitted to the seven data points nearest to the highest peak, with the FD uncertainty calculated by $\displaystyle \frac{\delta {\rm FD}_0}{2 \cdot (S/N)}$ \citep[e.g.,][]{mao10,iacobelli13}. We did not correct for the Ricean polarisation bias since we do not expect it to have significant effects for our case with $S/N > 6$ \citep{wardle74,george12}. All these results from our new VLA observations (${\rm FD}_{\rm new}$ and $p_{\rm new}$), along with their counterparts in the original NVSS catalogue ($p_{\rm NVSS}$), the \cite{taylor09} catalogue (${\rm FD}_{\rm TSS09}$ and $p_{\rm TSS09}$), and \cite{vaneck11} observations (${\rm FD}_{\rm VE11}$) are listed in Tables~\ref{table:rm1} and \ref{table:rm2} for on- and off-axis targets, respectively. We found that three on-axis and seven off-axis sources are unpolarised in our new observations (i.e., $p_{\rm new}$ below our $6\sigma$ cutoff). The $1\sigma$ values in $p_{\rm new}$ of these sources are reported within parentheses in the two Tables. In total, we derived the FD values of 194 polarised EGSs, with 168 and 26 on- and off-axis targets, respectively. The sky distribution of our EGS FD is plotted in Figure~\ref{fig:new_ha} top panel. 

In order to assess the residual leakage signal level present in our data, we formed Faraday spectra of our unpolarised on-axis leakage calibrator (J1407$+$2827) using data from each of the seven observing blocks. We find that the highest peak out of the seven Faraday spectra is at a value of $p = 0.024 \pm 0.007$\,per~cent, much lower than that of all of our polarised target sources. Note that the weak polarisation signal we see from J1407$+$2827 here is likely due to polarisation bias stemming from random noise fluctuations \citep[e.g.,][]{george12}, and therefore serves as an upper limit to the actual residual on-axis leakage level.

Finally, we verified that our off-axis targets have not been significantly affected by the off-axis instrumental polarisation of the VLA. All these sources are within $5^\prime$ from their respective pointing centres, with a mean distance of $3\farcm2$. For the specific case of the VLA in L-band, off-axis instrumental polarisation can reach $\approx 5$\,per cent at the half-power point of the primary beam ($\approx 15^\prime$ away from the pointing centre), and is expected to manifest as an instrumental polarised component with ${\rm FD} \approx 0\,{\rm rad\,m}^{-2}$ \citep{jagannathan17}. If we approximate this off-axis instrumental polarisation beam pattern as a second-order polynomial, the expected off-axis leakage level at $5^\prime$ from the pointing centre would be $\approx 0.5$\,per cent. For our polarised off-axis targets (Table~\ref{table:rm2}), we noted that six of them have $|{\rm FD}_{\rm new}| \leq 0.5 \cdot \delta {\rm FD}_0 \approx 60\,{\rm rad\,m}^{-2}$, of which NVSS J183603$-$005747 has the lowest $p_{\rm new}$ of $1.84 \pm 0.13$\,per cent. This is much higher than the 0.5\,per cent we estimated above. Therefore, we conclude that the FD values of our off-axis targets are reliable.

\begin{figure}
\centering
\includegraphics[width=215pt]{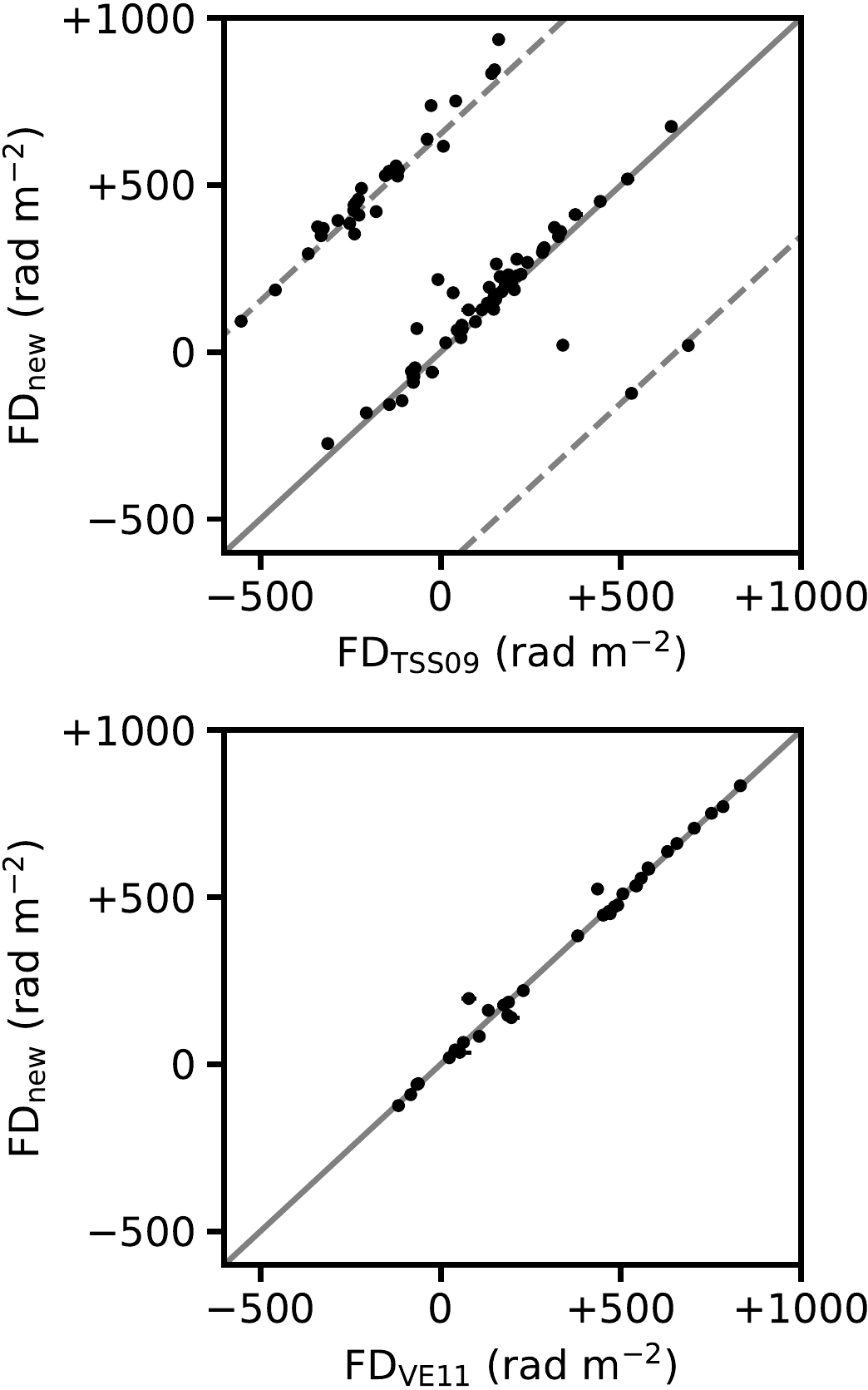}
\caption{Comparisons of EGS FD values between our new observations and \textbf{(Top)} \citet{taylor09} and \textbf{(Bottom)} \citet{vaneck11}. The error bars are shown but in almost all cases are too small to be noticeable. The grey solid lines are the lines of equality, and the dashed lines in the top panel correspond to FD offsets by $\pm 652.9\,{\rm rad\,m}^{-2}$ due to $n\pi$-ambiguity in \citet{taylor09}.}
\label{fig:rm_compare}
\end{figure}

\section{Discussion}
\label{sec:discussion}

\subsection{Comparisons with Existing Faraday Depth Measurements}
\label{sec:compare_old}
Out of our 168 polarised on-axis targets, 85 were also listed in the \cite{taylor09} catalogue. Upon comparison between our derived ${\rm FD}_{\rm new}$ and their listed ${\rm FD}_{\rm TSS09}$ (Figure~\ref{fig:rm_compare} top panel), we found that 32 out of the 85 sources (almost 40 per cent) have the two values differ by more than $500\,{\rm rad\,m}^{-2}$. This is most likely due to the $n\pi$-ambiguity issue of the \cite{taylor09} catalogue \cite[see][]{ma19a}. Furthermore, we noted two sources (NVSS J183220$-$103510 and NVSS J183409$-$071802) that, despite being listed as polarised in the \cite{taylor09} catalogue at $1.01 \pm 0.05$ and $0.85 \pm 0.02$~per~cent, respectively, were unpolarised in our new VLA observations (with $p_{\rm new}$ lower than $6\sigma$ cutoffs of 0.24 and 0.12~per~cent, respectively). Such differences in fractional polarisation can be attributed to the off-axis instrumental polarisation of the NVSS observations \citep[see][]{ma19b}. We conclude that if one relies solely on the \cite{taylor09} FD values to study the Galactic magnetic field in this particular sky area, the reliability of the results will likely be affected.

Furthermore, we compared our new FD values with those from \cite{vaneck11} observations (${\rm FD}_{\rm VE11}$) for the 35 cross-matched sources, as shown in Figure~\ref{fig:new_ha} and the bottom panel of Figure~\ref{fig:rm_compare}. The two sets of measurements agree with each other within error bars in general, except for two sources for which we found significant differences (at $> 3\sigma$): NVSS J192233$+$071048 (${\rm FD}_{\rm new} = +196.8 \pm 4.0\,{\rm rad\,m}^{-2}$; ${\rm FD}_{\rm VE11} = +78.0 \pm 20.0\,{\rm rad\,m}^{-2}$) and NVSS J192458$+$130033 (${\rm FD}_{\rm new} = +524.5 \pm 6.3\,{\rm rad\,m}^{-2}$; ${\rm FD}_{\rm VE11} = +435.0 \pm 8.0\,{\rm rad\,m}^{-2}$). Both these sources were found to exhibit Faraday complexities in our new data (see Appendix C of Online Supporting Information), which we attribute as the cause of the discrepancy between ${\rm FD}_{\rm new}$ and ${\rm FD}_{\rm VE11}$.

As can be seen in Figure~\ref{fig:new_ha}, our new observations have led to a much higher polarised EGS source density than that of \cite{vaneck11}. Specifically, the source density over the entire region ($20^\circ \leq \ell \leq 52^\circ$ and $|b| \leq 5^\circ$) has increased by almost a factor of five (from their one source per $7.3\,{\rm deg}^2$ to our one source per $1.6\,{\rm deg}^2$), with the longitude range of $20^\circ$--$40^\circ$ seeing the largest improvement from one source per $16.6\,{\rm deg}^2$ (total of 12 sources) to one source per $1.6\,{\rm rad\,m}^{-2}$ (total of 125 sources). This increase in polarised EGS count enables our study of the complex large-scale magnetic fields in the Milky Way disk, especially in the latitude dependence of FD.

Finally, we compared our FD values with those recently published by \cite{shanahan19} as part of The H\,{\sc i} / OH / Recombination line (THOR) survey, conducted in L-band with the VLA in C-array configuration. Out of their 127 polarised compact sources in $39^\circ < \ell < 52^\circ$ and $|b| < 1\fdg25$, we found a total of 10 cross-matches, with most of them showing consistency in the two sets of FD values. The two sources with significant differences in FD are NVSS J190741$+$090717 (${\rm FD}_{\rm new} = +706.5 \pm 0.9\,{\rm rad\,m}^{-2}$; ${\rm FD}_{\rm THOR} = +695 \pm 1\,{\rm rad\,m}^{-2}$) and NVSS J192517$+$135919 (${\rm FD}_{\rm new} = +450.9 \pm 2.1\,{\rm rad\,m}^{-2}$; ${\rm FD}_{\rm THOR} = +424 \pm 1\,{\rm rad\,m}^{-2}$). Within the region where they found EGSs with extremely high $|{\rm FD}|$ values (up to $4000\,{\rm rad\,m}^{-2}$; at $47^\circ < \ell < 49^\circ$ and $|b| < 1\fdg25$), we found no cross-matches because none of our target sources reside there, likely due to a bias from our source selection caused by bandwidth depolarisation (see Section~\ref{sec:srcsel}). However, this does not affect our study of the Galactic-scale magnetic field here (see Section~\ref{sec:connection}).

\begin{figure}
\centering
\includegraphics[width=230pt]{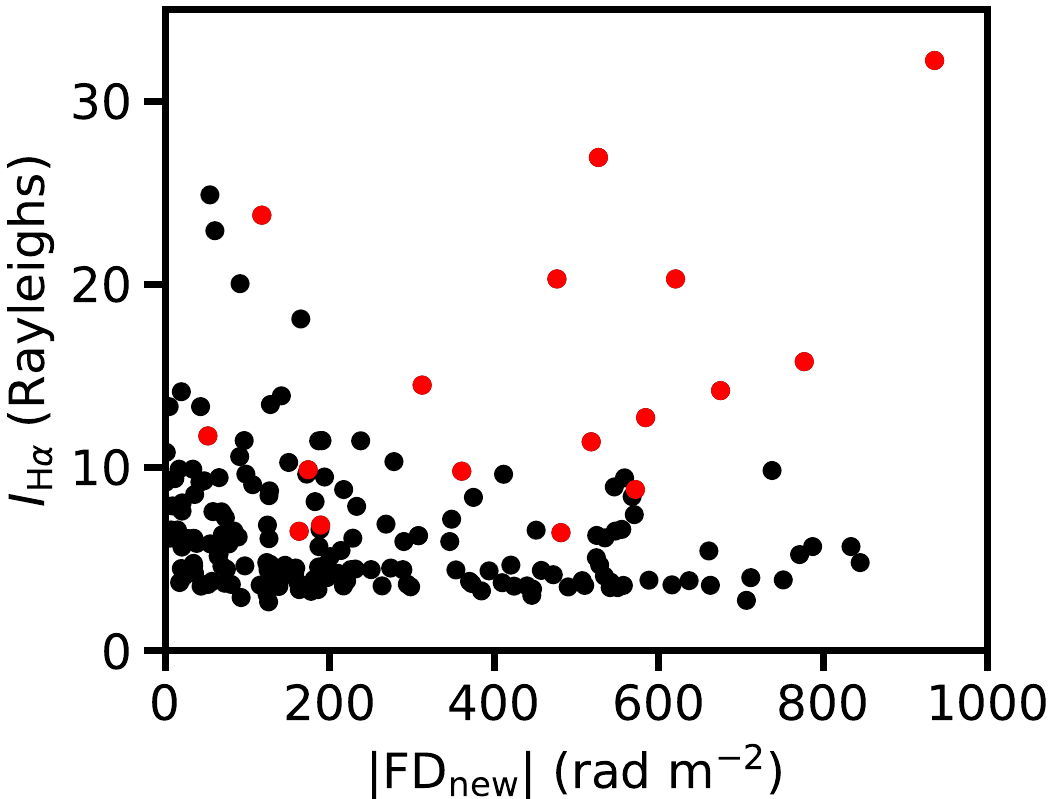}
\caption{WHAM velocity-integrated H$\alpha$ intensities \citep{haffner03,haffner10} against our newly derived EGS FD values. The 17 polarised EGSs discarded due to their positioning behind the H\,{\sc ii} structure G26.5 are marked as red.}
\label{fig:halpha_fd}
\end{figure}

\begin{figure*}
\includegraphics[width=450pt]{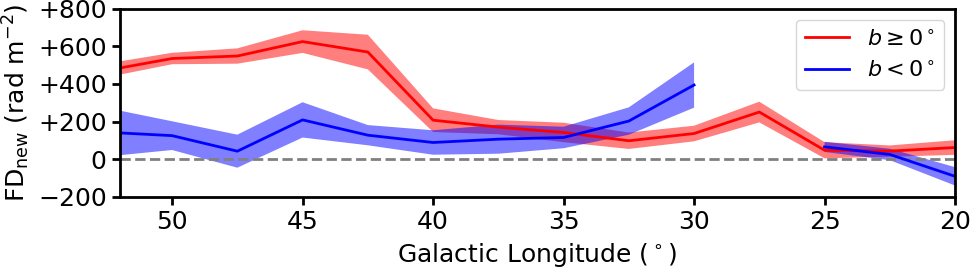}
\caption{Boxcar-binned FD profiles of our target EGSs across $\ell$, with the sources separated into above (red) and below (blue) the Galactic plane. A bin width of $5^\circ$ along $\ell$ was adopted, with the profile sampled at a $2\fdg5$ interval. The shaded area represents the standard error of median (SEM) of FD values in each bin.}
\label{fig:rm_l_bin}
\end{figure*}

\begin{figure*}
\includegraphics[width=435pt]{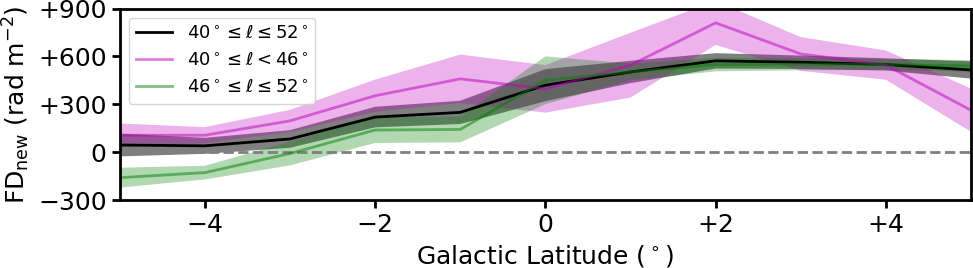}
\caption{Boxcar-binned FD profile of our target EGSs across $b$ for sources in $40^\circ \leq \ell \leq 52^\circ$ (black). We also considered two smaller sub-regions with longitude ranges of $40^\circ$--$46^\circ$ (magenta) and $46^\circ$--$52^\circ$ (green). A bin width of $2\fdg5$ along $b$ was adopted, with the profile sampled at a $1^\circ$ interval. The shaded area represents the standard error of median (SEM) of FD values in each bin.}
\label{fig:rm_b_bin2}
\end{figure*}

\subsection{Contamination by Galactic H\,{\sc ii} Structures}
\label{sec:hii_cont}
Upon inspection of the Wisconsin H-Alpha Mapper Sky Survey (WHAMSS) H$\alpha$ map in Figure~\ref{fig:new_ha}, we identified a large (diameter $\approx 7^\circ$) H\,{\sc ii} structure centred at $(\ell, b) = (26\fdg5, -0\fdg5)$ that contains smaller H\,{\sc ii} regions such as Sh~2-59 and Sh~2-60. This H\,{\sc ii} structure, which we call G26.5, appears to lead to an excess FD of $\approx +300\,{\rm rad\,m}^{-2}$ for EGSs behind it compared to those in the immediate surroundings. Galactic H\,{\sc ii} structures are known to lead to FD enhancements by $\sim 100\,{\rm rad\,m}^{-2}$ in magnitude for background EGSs \citep[e.g.,][]{harvey-smith11,purcell15}, which we believe to be the case as well for G26.5. Since the focus of our study is the Galactic-scale magnetic field of the Milky Way, we decided to discard the 17 polarised EGSs (15 on-axis plus two off-axis) situated behind G26.5 as the FD values of these sources are likely contaminated by this H\,{\sc ii} structure. As reference, we plotted the H$\alpha$ intensity against $|{\rm FD}_{\rm new}|$ for our polarised target sources in Figure~\ref{fig:halpha_fd}. This results in a final list of 177 EGS FD values (153 on-axis plus 24 off-axis) that we use for our study below. 

\subsection{Faraday Depth Disparity across Galactic Latitude}
\label{sec:fd_disparity}

\subsubsection{Identification from Newly Derived Faraday Depths}
An obvious feature in the spatial FD distribution can be identified from Figure~\ref{fig:new_ha} top panel: a disparity\footnote{In this work, we mean by disparity a great difference, and is not directly related to the technical terms of even/odd parity that we introduce below.} of FD across the Galactic mid-plane within $40^\circ \lesssim \ell \lesssim 52^\circ$. Within this longitude range, the median FD values for sources above and below the Galactic plane are $+550 \pm 40\,{\rm rad\,m}^{-2}$ and $+130 \pm 50\,{\rm rad\,m}^{-2}$, respectively. We further performed a Kolmogorov-Smirnov test (KS-test) with the null hypothesis being that these two samples have the same FD distribution. The resulting $p$-value is $4 \times 10^{-9}$, strongly supporting that the FD distributions on either side of the Galactic mid-plane are different. We note that hints of the same structure can already be seen in the \cite{vaneck11} data (Figure~\ref{fig:new_ha} bottom panel), but was not explicitly pointed out in their paper. 

\begin{figure*}
\includegraphics[width=450pt]{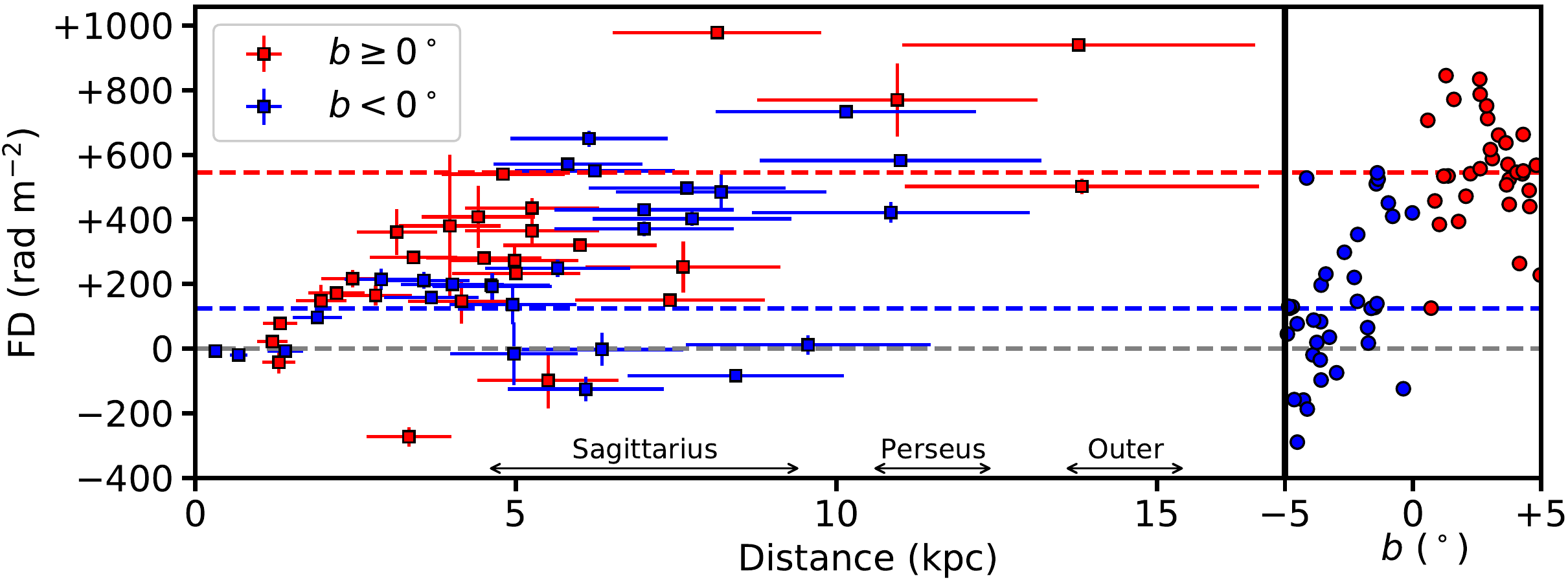}
\caption{FD values of pulsars within $40^\circ \leq \ell \leq 52^\circ$ and $|b| \leq 5^\circ$ across distance, obtained from the ATNF Pulsar Catalogue \citep{manchester05}. A typical uncertainty of pulsar distances of 20~per~cent has been assumed \citep[e.g.,][]{han17}. The approximate locations of the Sagittarius, Perseus, and Outer spiral arms along the line of sight are indicated. Our newly derived EGS FDs in the same sky region are plotted on the right against Galactic latitude, with the median FD for above ($+550\,{\rm rad\,m}^{-2}$) and below ($+130\,{\rm rad\,m}^{-2}$) the Galactic mid-plane shown as the red and blue dashed lines, respectively.}
\label{fig:psr_fd_dist}
\end{figure*}

\begin{figure*}
\includegraphics[width=450pt]{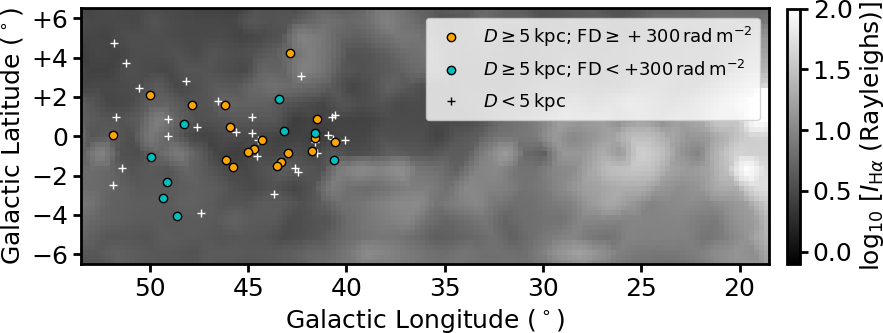}
\caption{Spatial distribution of the pulsars considered in Figure~\ref{fig:psr_fd_dist}. In particular, pulsars at distances of more than $5\,{\rm kpc}$ are shown as colour dots, and the rest are shown as crosses. The background map shows H$\alpha$ intensity from WHAMSS \citep{haffner03,haffner10}.}
\label{fig:fd_2color}
\end{figure*}

The boxcar-binned EGS FD profiles across $\ell$ are shown in Figure~\ref{fig:rm_l_bin}, with the sources separated into above and below the Galactic plane. By performing such spatial averaging of FD values, the FD profiles represent the large-scale magnetic field of the Milky Way, since we expect the FD contaminations from various sources (see below) to be smoothed out by the spatial binning. We adopted a bin size along Galactic longitude of $5^\circ$, chosen as the smallest bin size with which smooth FD trends along longitude could be seen, meaning that in most bins there are enough data points for robust statistics. We verified that choosing slightly larger bin sizes ($< 10^\circ$) would still give consistent results. The FD profiles are sampled at a $2\fdg5$ interval. The solid lines show the median FD within the moving $5^\circ$ bin, while the shaded areas represent the FD uncertainty. We calculated the FD uncertainty as the standard error of median (SEM) of each individual bin:
\begin{equation}
{\rm SEM} = \frac{1.2533 \cdot \sigma}{\sqrt{N}}{\rm ,} \label{eq:sem}
\end{equation}
where $\sigma$ and $N$ are the standard deviation and the number of FD values in each bin, respectively. For our case here, $\sigma$ accounts for the contamination by the small-scale Galactic magnetic field \citep[$\sim 100\,{\rm rad\,m}^{-2}$ over $\sim 1^\circ$; e.g.,][]{haverkorn08,stil11}, the intrinsic FD of the EGSs \citep[$\sim 10$--$100\,{\rm rad\,m}^{-2}$; e.g.,][]{schnitzeler10,oppermann15,anderson19}, magnetic fields in the intergalactic medium \citep[$\lesssim 10\,{\rm rad\,m}^{-2}$; e.g.,][]{vernstrom19,osullivan20}, and the uncertainty of our measurements ($\approx 2\,{\rm rad\,m}^{-2}$). The use of SEM as our FD uncertainty implicitly assumes that the above sources of FD contaminations are not spatially correlated, which is not strictly the case for small-scale Galactic magnetic field \citep[see][]{haverkorn08}. We therefore warn that our FD uncertainty can be slightly underestimated. Finally, we mask out the FD profile of $b < 0^\circ$ in the Galactic longitude range of $25^\circ$--$30^\circ$, since within this range there is only one EGS remaining in the $5^\circ$ bin after we discarded EGSs situated behind G26.5 (Section~\ref{sec:hii_cont}), meaning that the uncertainty of the FD profile diverges there.

We can see a clear disparity in the two FD profiles in the Galactic longitude range of $40^\circ$--$52^\circ$, but not in $20^\circ$--$40^\circ$. This immediately shows that the distributions of the large-scale magnetic field and/or the Galactic free electron number density are not symmetric on the two sides of the Galactic mid-plane within the longitude range of $40^\circ$--$52^\circ$. We further investigated this by plotting in Figure~\ref{fig:rm_b_bin2} the FD profile along Galactic latitude, considering sources in the longitude range of $40^\circ$--$52^\circ$ only (black line). Here, we used a boxcar bin width of $2\fdg5$ along latitude, and sampled the FD profile at a $1^\circ$ interval. If the FD disparity occurs only beyond a certain Galactic latitude, say at $|b| > b_0$, we would expect the FD profile here to be symmetric about $b = 0^\circ$ for $|b| < b_0$, which is not what we found. Instead, we see a steady increase in FD from $\approx +100\,{\rm rad\,m}^{-2}$ at $b = -3^\circ$ to $\approx +600\,{\rm rad\,m}^{-2}$ at $b = +2^\circ$, without signs of symmetry about $b = 0^\circ$. We further plotted the FD profiles in smaller longitude ranges of $40^\circ$--$46^\circ$ (magenta) and $46^\circ$--$52^\circ$ (green), and found that they are consistent with the picture above. We therefore conclude that the FD disparity must begin at latitude of very close to $0^\circ$.

\subsubsection{Distance Estimate from Existing Pulsar Measurements}
\label{sec:psr_fd}
From the ATNF Pulsar Catalogue \citep[version 1.60;][]{manchester05}\footnote{Available on \href{http://www.atnf.csiro.au/research/pulsar/psrcat/}{http://www.atnf.csiro.au/research/pulsar/psrcat/}.}, we obtained the FD values and distances of Galactic pulsars within $40^\circ \leq \ell \leq 52^\circ$ and $|b| \leq 5^\circ$. These measurements allow us to trace how FD changes across physical distances \citep[e.g.,][]{noutsos08,han18}, and allow us to constrain where along our line of sight the FD disparity occurs. We considered a total of 55 pulsars, out of which 10 have independent distance estimates \citep[e.g., parallax or H\,{\sc i} measurements; see][]{lorimer12,han17}. The remaining 45 have their distances inferred from their dispersion measure (DM) values by assuming the Galactic thermal electron distribution model of YMW16 \citep{ymw16}, which they showed gives more accurate pulsar distance estimates than by assuming the NE2001 model \citep{ne2001}.

We plotted in Figure~\ref{fig:psr_fd_dist} the FD against distance of these pulsars, similar to the figures in \cite{han18} except that we have separated the sources into above and below the Galactic mid-plane. From this, we see that pulsars both above and below $b = 0^\circ$ follow the same trend of increasing FD with distance up to $\approx 5\,{\rm kpc}$, beyond which the FD trends deviate. For pulsars above the mid-plane, the FD continues to rise with increasing distance and eventually reaches the median EGS FD there of $+550\,{\rm rad\,m}^{-2}$. Meanwhile, pulsars below the mid-plane show a large spread of FD values from $-120\,{\rm rad\,m}^{-2}$ up to $+730\,{\rm rad\,m}^{-2}$. This can either be interpreted as a genuine increase in FD spread due to a highly turbulent magneto-ionic medium in that sky region, or that the pulsars are composed of two populations with a divide at $\approx +300\,{\rm rad\,m}^{-2}$. We favour the latter option for two reasons. Firstly, the population with ${\rm FD} < +300\,{\rm rad\,m}^{-2}$ shows a steadily decreasing FD with increasing distance and eventually roughly matches the median EGS FD of $+130\,{\rm rad\,m}^{-2}$ there. This means that the ${\rm FD} < +300\,{\rm rad\,m}^{-2}$ population could be representative of the diffuse warm ionised medium towards this sky region, while the pulsars with ${\rm FD} \geq +300\,{\rm rad\,m}^{-2}$ can be regarded as a ``peculiar'' population (see below). Secondly, we note that the population with ${\rm FD} \geq +300\,{\rm rad\,m}^{-2}$ shows a spatial clustering at $40^\circ \lesssim \ell \lesssim 46^\circ$ and $-1\fdg5 \lesssim b \lesssim 0^\circ$ (Figure~\ref{fig:fd_2color}), which is unexpected for the former option of a genuine FD spread. We speculate from such spatial clustering that this ``peculiar'' pulsar population could be a manifestation of longitudinal variations in FD, or these pulsars could be situated behind some localised magneto-ionic medium.

Assuming the two-population option above, we ignored the ${\rm FD} \geq +300\,{\rm rad\,m}^{-2}$ population below the mid-plane. This led us to the identification of the split in FD trends for pulsars above versus below the Galactic plane at a distance of $\approx 5\,{\rm kpc}$ away from us, hinting that the EGS FD disparity we discovered occurs in the Sagittarius spiral arm. Additionally, the increasing / decreasing FD trends with distance for pulsars above / below the mid-plane would mean that the plane-parallel magnetic field direction changes across the Galactic plane. However, we acknowledge the high uncertainty in our interpretation here, as we are limited by the number of pulsars with high accuracy FD and distance estimates beyond $5\,{\rm kpc}$ in this sky region.

\begin{figure*}
\includegraphics[width=410pt]{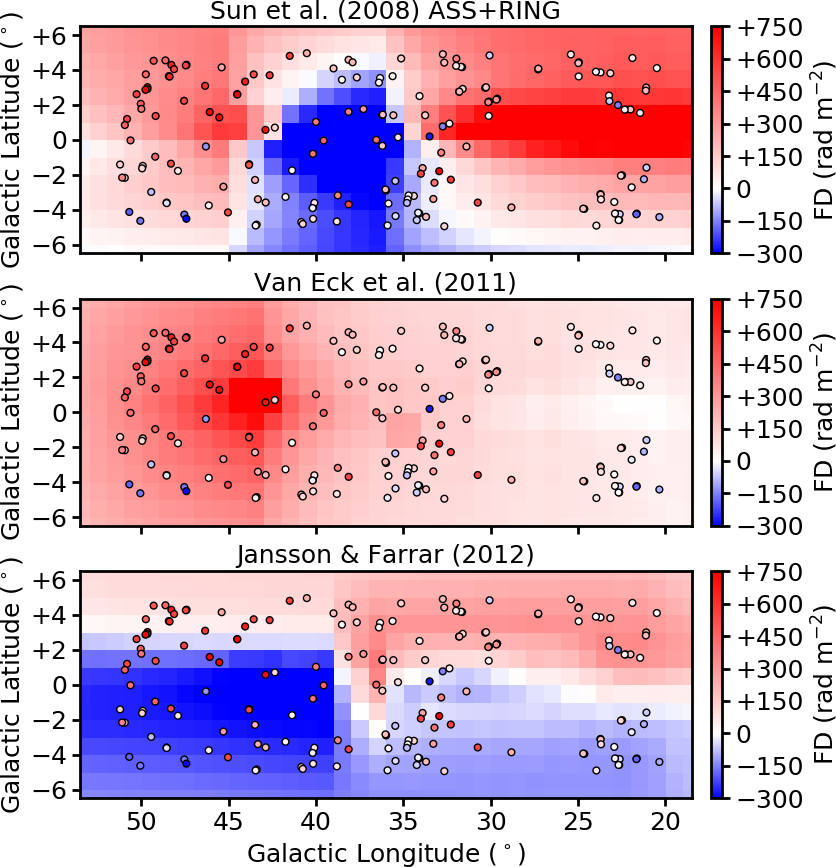}
\caption{Predicted FD maps of the large-scale magnetic field models of the Milky Way of \textbf{(Top)} \citet{sun08}, \textbf{(Middle)} \citet{vaneck11}, and \textbf{(Bottom)} \citet{jansson12}. The thermal electron number density model of NE2001 \citep{ne2001} has been adopted. The colour dots represent our newly derived EGS FD values.}
\label{fig:ne2001_rm}
\end{figure*}

\subsection{Performance of Existing Magnetic Field Models}
\label{sec:field_comparisons}

We now proceed to assess the performance of three recent major Milky Way magnetic field models, namely \cite{sun08}, \cite{vaneck11}, and \cite{jansson12}, by comparing their FD predictions with our newly derived EGS FD values. The three models were combined with the thermal electron number density model of NE2001\footnote{We repeated our investigation using the YMW16 model \citep{ymw16} instead, and found that the conclusions are unchanged.} \citep{ne2001} to generate the FD predictions. We first review these Galactic magnetic field models below.

\subsubsection{A Brief Review on the Galactic Magnetic Field Models}
\label{sec:fieldmodelreview}

Firstly, the \cite{sun08} model\footnote{We consider their ASS+RING model, since it has been shown to better fit to observations compared to their ASS+ARM and BSS models \citep{sun08,vaneck11}.} (shortened as Sun08) was developed using EGS FD measurements from the CGPS \citep{brown03} and the Southern Galactic Plane Survey \citep[SGPS;][]{gaensler01,brown07}. In particular, they used the NE2001 thermal electron number density model, and adjusted the free parameters of their large-scale magnetic field model to fit the predicted FD to the observed EGS FD values. A large-scale magnetic field reversal has been placed in a ring at Galacto-centric radius of $6$--$7.5\,{\rm kpc}$, with the strength of the disk field diminishing exponentially at increasing Galactic height. They have assumed that the disk field has an even parity, meaning the plane-parallel magnetic field direction is the same on either side of the Galactic plane. Meanwhile, their toroidal halo field has an odd-parity, meaning the plane-parallel magnetic field flips in direction across the Galactic mid-plane.

Next, the \cite{vaneck11} model (shortened as VE11) was based on their new FD measurements of 194 EGSs in the Galactic plane, in addition to EGS FD values from both the CGPS \citep{brown03} and the SGPS \citep{brown07}. They have also adopted the NE2001 model to fit to the EGS FD. Their study only focused on the Milky Way disk field (i.e., no halo field component), which was assumed to have an even parity and a constant field strength along Galactic height out to $\pm 1.5\,{\rm kpc}$ where the model has been truncated. The field model is composed of three independent sectors with different geometries. The region of interest in this paper resides in their Sector C, within which they found a large-scale magnetic field reversal ring at Galacto-centric radius of $5.8$--$8.4\,{\rm kpc}$ from their best-fit results.

Finally, the \cite{jansson12} model (shortened as JF12) is unique among the three models as it is the only one that has implemented a vertical field component. It is comprised of the disk, the toroidal halo, and the X-shaped halo components. Moreover, their field model is more physically motivated than the others, as they have implemented the divergence-free condition of magnetic fields, as well as the X-shaped halo field as motivated by the observational results from external edge-on galaxy studies \citep[see, e.g.,][]{krause09}. To determine the best-fit parameters of the large-scale magnetic field model, they combined different information, namely (1) a substantial list of FD measurements from the literature covering the entire sky, (2) the K-band ($22\,{\rm GHz}$) polarised synchrotron map of the Galactic foreground from \textit{WMAP} \citep{gold11}, (3) the NE2001 thermal electron number density model, and (4) the Galactic cosmic ray density models from GALPROP \citep{strong09} and \textit{WMAP} \citep{page07}. The details of their disk field component, which is the focus of our study here, were mainly constrained by the EGS FD values from the CGPS \citep{brown03}, SGPS \citep{brown07}, \cite{vaneck11} observations, and the \cite{taylor09} catalogue.

\begin{table}
\caption{$\chi^2$ Values of the Tested Galactic Magnetic Field Models}
\label{table:model_chi2}
\centering
\begin{tabular}{lccc}
\hline
Galactic & \multicolumn{3}{c}{Magnetic Field Models} \\
Longitude & Sun08 & VE11 & JF12 \\
\hline
\multicolumn{4}{c}{\textbf{All EGSs}} \\
\hline
$20^\circ$--$25^\circ$ & 178.79 & \phantom{0}\textbf{2.36} & 28.83 \\
$25^\circ$--$30^\circ$ & \phantom{0}21.74 & 10.87 & \phantom{0}\textbf{0.03} \\
$30^\circ$--$35^\circ$ & \phantom{0}16.41 & \phantom{0}\textbf{0.01} & 12.92 \\
$35^\circ$--$40^\circ$ & 107.74 & \phantom{0}\textbf{0.09} & \phantom{0}0.68 \\
$40^\circ$--$45^\circ$ & \phantom{0}18.38 & \phantom{0}\textbf{6.54} & 52.09 \\
$45^\circ$--$50^\circ$ & \phantom{00}8.11 & \phantom{0}\textbf{3.53} & 94.14 \\
(Average) & \phantom{0}58.53 & \phantom{0}\textbf{3.90} & 31.45 \\
\hline
\multicolumn{4}{c}{$\mathbf{b \geq 0^\circ}$} \\
\hline
$20^\circ$--$25^\circ$ & 271.32 & \phantom{0}\textbf{0.66} & \phantom{0}84.58 \\
$25^\circ$--$30^\circ$ & \phantom{0}15.28 & \phantom{0}8.12 & \phantom{00}\textbf{0.13} \\
$30^\circ$--$35^\circ$ & \phantom{0}33.76 & \phantom{0}\textbf{0.50} & \phantom{0}15.80 \\
$35^\circ$--$40^\circ$ & \phantom{0}61.63 & \phantom{0}\textbf{0.01} & \phantom{00}5.82 \\
$40^\circ$--$45^\circ$ & \phantom{0}21.61 & \phantom{0}\textbf{0.87} & \phantom{0}37.29 \\
$45^\circ$--$50^\circ$ & \phantom{0}28.07 & \textbf{19.51} & 147.40 \\
(Average) & \phantom{0}71.95 & \phantom{0}\textbf{4.95} & \phantom{0}48.50 \\
\hline
\multicolumn{4}{c}{$\mathbf{b < 0^\circ}$} \\
\hline
$20^\circ$--$25^\circ$ & 16.91 & \phantom{0}\textbf{1.83} & 22.17 \\
$25^\circ$--$30^\circ$ & --- & --- & --- \\
$30^\circ$--$35^\circ$ & 12.73 & \phantom{0}\textbf{1.02} & 16.29 \\
$35^\circ$--$40^\circ$ & 24.22 & \phantom{0}\textbf{0.23} & \phantom{0}7.41 \\
$40^\circ$--$45^\circ$ & 28.67 & \textbf{13.22} & 48.61 \\
$45^\circ$--$50^\circ$ & \phantom{0}\textbf{1.88} & 14.08 & 11.17 \\
(Average) & 16.88 & \phantom{0}\textbf{6.08} & 21.13 \\
\hline
\multicolumn{4}{l}{\texttt{NOTE} -- The lowest $\chi^2$ value of each row} \\
\multicolumn{4}{l}{\phantom{\texttt{NOTE} -- }is bold-faced.}
\end{tabular}
\end{table}

\begin{figure*}
\includegraphics[width=440pt]{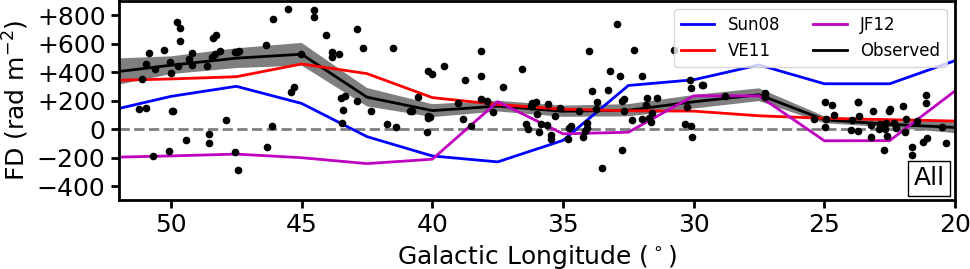}
\includegraphics[width=440pt]{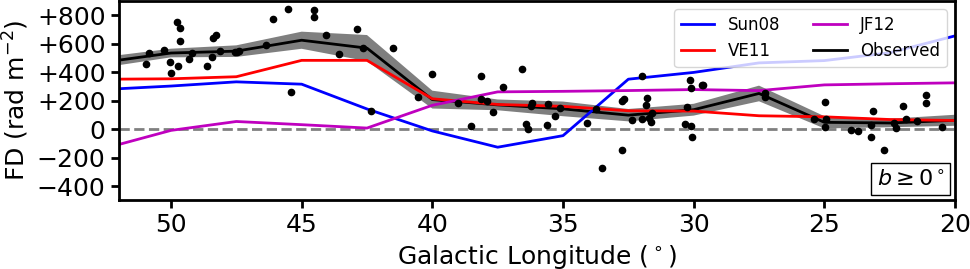}
\includegraphics[width=440pt]{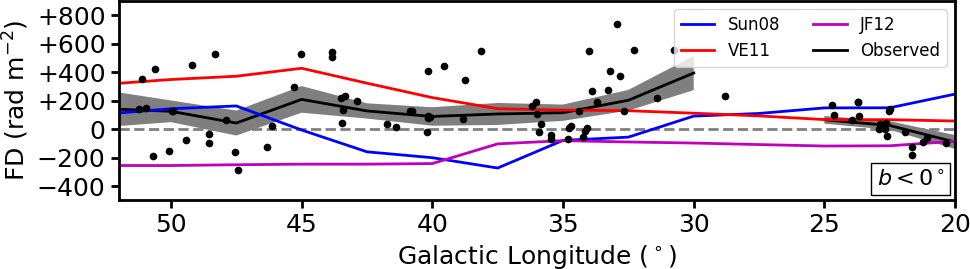}
\caption{Comparisons between the boxcar-binned FD profiles along $\ell$ of our observed values (black lines) and the predictions of the \citet{sun08} (Sun08; blue lines), \citet{vaneck11} (VE11; red lines), and \citet{jansson12} (JF12; magenta lines) models. The three panels show the results from considering \textbf{(Top)} all EGSs, \textbf{(Middle)} $b \geq 0^\circ$ only, and \textbf{(Bottom)} $b < 0^\circ$ only. The shaded area represents the standard error of median (SEM) of the observed EGS FD values in each bin. Our newly derived FD values of each individual EGS are marked as the black data points.}
\label{fig:ne2001_losbin}
\end{figure*}

\subsubsection{Performance of the Models}
\label{chap:bfield_chi2}
We present the predicted FD maps of the three Milky Way magnetic field models in Figure~\ref{fig:ne2001_rm}. It is immediately apparent that both the \cite{sun08} and \cite{jansson12} predictions exhibit significant asymmetries across the Galactic mid-plane, since both these models have implemented odd-parity halo field components. The \cite{vaneck11} FD prediction is reasonably symmetric about the Galactic plane because they only considered an even-parity disk field.

To facilitate comparisons between our newly derived EGS FD values with the predictions by the three models, we collapsed the Galactic latitude axis to generate boxcar-binned FD profiles along Galactic longitude, as shown in Figure~\ref{fig:ne2001_losbin}. This is similar to what we performed in Section~\ref{sec:fd_disparity}. Specifically here, we used the sky positions of our 177 polarised EGSs, and calculated the predicted FD values at those exact locations according to the three magnetic field models. This mitigates the possibility of sampling biases imposed by the particular positions where our polarised EGSs were located. Next, we evaluated the median FD values in the moving $5^\circ$ longitude bin sampled at a $2\fdg5$ interval, for our observed EGSs as well as the model predictions. Note that the overlapping bins were only used for plotting the smooth FD trends across longitude, and the model evaluations below were performed with independent bins. The FD profiles were generated considering (1) all EGSs, (2) $b \geq 0^\circ$ only, and (3) $b < 0^\circ$ only. We calculated the uncertainties of the observed FD profiles as the SEM as in Equation~\ref{eq:sem}.

With the boxcar-binned FD profiles, we performed a quantitative evaluation of the three Galactic magnetic field models. We obtained the median FD in independent bins by re-sampling the FD profiles at $5^\circ$ interval from $22\fdg5$ to $47\fdg5$. For each bin, we compared our observations with the three model predictions by evaluating
\begin{equation}
\chi^2 = \frac{(\overline{{\rm FD}}_{\rm obs} - \overline{{\rm FD}}_{\rm model})^2}{\sigma_{\rm FD}^2}{\rm ,} \label{eq:chi}
\end{equation}
where $\overline{{\rm FD}}_{\rm obs}$ is the observed FD median, $\overline{{\rm FD}}_{\rm model}$ is the model FD median, and $\sigma_{\rm FD}$ is the SEM of the observed FD. Note that even if a magnetic field model performs satisfactorily in a specific longitude bin, its $\chi^2$ value can still deviate from unity because of random fluctuations. However, one can compare the order of magnitude of $\chi^2$ between the three models to assess their relative performance in each longitude bin. Moreover, we calculated the average $\chi^2$ values for each model and latitude range combination over the six independent longitude bins. This averaged $\chi^2$ should converge to unity for a well performing model and thus allows an evaluation of the absolute performance of each model. We listed the results in Table~\ref{table:model_chi2}. It is obvious that the \cite{vaneck11} model performs the best overall, especially for the case where we considered the full Galactic latitude range ($|b| \leq 5^\circ$; averaged $\chi^2 = 3.90$). For the cases where we considered the two sides of the Galactic mid-plane separately, however, the performance of the \cite{vaneck11} model deteriorated slightly to averaged $\chi^2 = 4.95$ and $6.08$ for above and below the plane, respectively. This is because their model has only considered an even-parity disk field, and therefore it failed to capture the FD disparity that we identified. 

\subsection{Explanations to the Faraday Depth Disparity} \label{sec:scenario}

In light of the unsatisfactory performance of the three Milky Way magnetic field models in reproducing our newly derived EGS FD values (Section~\ref{sec:field_comparisons}), especially in the Galactic longitude range of $40^\circ$--$52^\circ$ where we discovered the FD disparity, we explored alternative astrophysical scenarios.

\subsubsection{Scenario I: Odd-Parity Large-scale Galactic Disk Field}
\label{sec:scenario1}
The first scenario that can explain the EGS FD disparity is that some regions in the Galactic disk host a large-scale magnetic field with odd parity. Both odd- and even-parity magnetic fields can be generated in galaxies according to the $\alpha$-$\Omega$ dynamo theory \citep[e.g.,][]{sokoloff90,brandenburg92,beck96,moss10}. For the case of the Galactic disk, an even-parity magnetic field is expected from the dynamo theory \citep[e.g.,][]{ruzmaikin88book} and has been observationally found to be the case for the local Galactic volume \citep{frick01} and the Perseus spiral arm \citep{mao12}. These have led to the common assumption of an even-parity disk field in Milky Way magnetic field modelling efforts \citep[e.g.,][]{sun08,vaneck11,jansson12}. Nonetheless, an odd-parity disk field can be generated under certain conditions, such as a sufficiently thick galactic disk or a weak galactic differential rotation \citep{ferriere05}. It can also be generated in the outskirts of galaxies as the result of turbulent pumping, galactic wind, and/or flaring of the galactic disk \citep{gressel13}.

\begin{figure*}
\includegraphics[width=430pt]{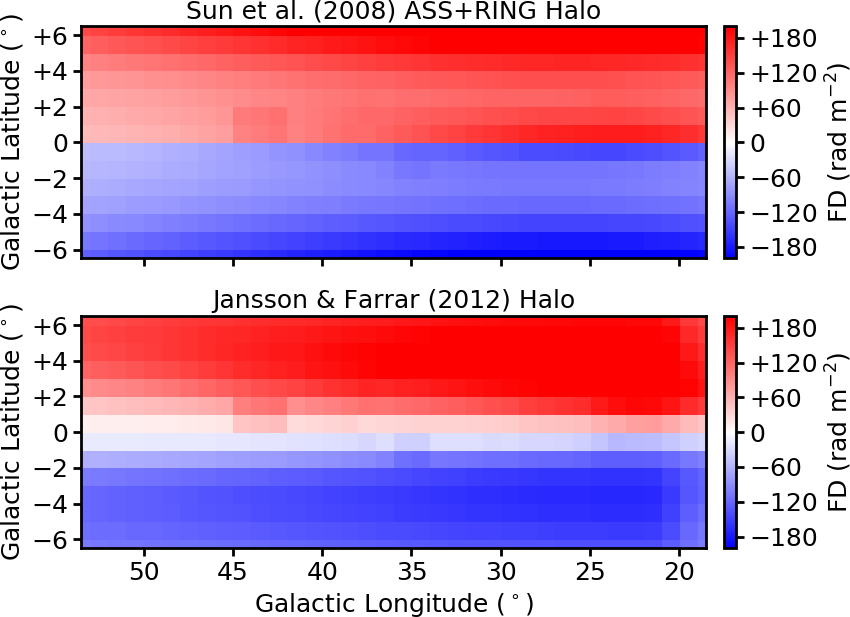}
\caption{Predicted FD contribution maps of the halo field component of \textbf{(Top)} \citet{sun08} and \textbf{(Bottom)} \citet{jansson12} models.}
\label{fig:halo_fd_map}
\end{figure*}

\begin{figure*}
\includegraphics[width=430pt]{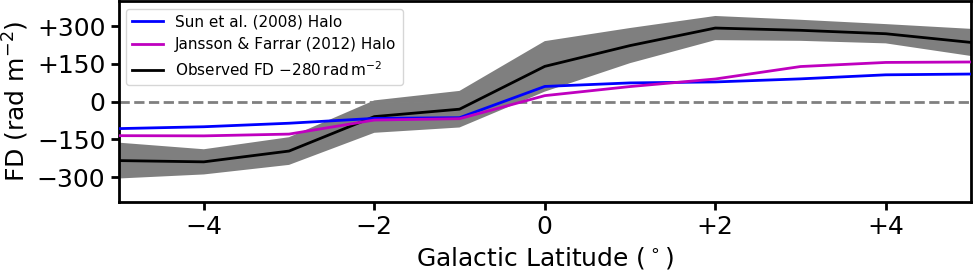}
\caption{Comparisons between the boxcar-binned FD profiles along $b$ of our observed values (black) and the predictions of the halo field component of \citet{sun08} (blue) and \citet{jansson12} (magenta) models. Only sources in $40^\circ \leq \ell \leq 52^\circ$ were considered. A $y$-offset of $-280\,{\rm rad\,m}^{-2}$ has been applied to the observed FD profile. The shaded area represents the standard error of median (SEM) of FD values in each bin.}
\label{fig:rm_b_bin}
\end{figure*}

The FD disparity can be caused by the change in magnetic field direction across the Galactic mid-plane of an odd-parity field, assuming that the thermal electron distribution is symmetric about $b = 0^\circ$. Such odd-parity field can either be the dominant component occupying a Galactic volume, or it can be in superposition with a stronger even-parity field. As revealed by the pulsar FD values increasing (decreasing) with distance above (below) the mid-plane in the longitude range of $40^\circ$--$52^\circ$ (Section~\ref{sec:psr_fd}), the Sagittarius arm could be hosting a dominant odd-parity field. The magnetic field direction for above and below the mid-plane is pointing towards and away from us, respectively. This suggests that the well-known large-scale field reversal of the Sagittarius arm occurs on one side of the Galactic mid-plane only, as the result of an odd-parity magnetic field there. Given the information that we have, this is our favoured scenario over the other two described below.

\subsubsection{Scenario II: Contributions from the Odd-parity Galactic Halo Field}
\label{sec:scenario2}
The second scenario is that the FD disparity is caused by the odd-parity magnetic field in the Milky Way halo. Such magnetic field structure is the preferred configuration from $\alpha$-$\Omega$ dynamo of spherical objects such as galactic halos \citep[e.g.,][]{sokoloff90,moss10}, and has indeed been suggested observationally to be the case for the Milky Way \citep[e.g.,][]{han97,sun08,taylor09}. The change in magnetic field direction of the halo field on either side of the Galactic plane can then lead to the observed FD disparity.

We first investigated the possibility of this scenario by considering at what Galactic height the halo field would become dominant. As noted from the FD profile across Galactic latitude (Figure~\ref{fig:rm_b_bin2}), the FD disparity begins at very close to $b = 0^\circ$. By adopting a generous upper limit of $|b| = 1^\circ$ where the halo field starts to show appreciable effects on our FD profile, and a distance of $5\,{\rm kpc}$ away from us where the FD disparity occurs (Secrion~\ref{sec:psr_fd}), this scenario requires the halo field to emerge at a Galactic height of no more than $\approx 90\,{\rm pc}$. It is challenging to reconcile this with the case study of the Perseus spiral arm \citep{mao12}, which showed from their EGS FD profile along Galactic latitude that the magnetic disk-halo transition occurs at a much higher Galactic height of $\sim 540\,{\rm pc}$ in the Perseus arm. Furthermore, as we do not see significant differences in the FD profile in the Galactic longitude range of $20^\circ$--$40^\circ$, this scenario would further require the halo field to have negligible FD contributions there, or an even-parity halo field in this longitude range.

In addition, we looked into whether the halo field prescriptions of the \cite{sun08} and \cite{jansson12} models can explain the FD disparity. Both these models include odd-parity halo fields that fill the entire Galactic volume. For reference, at a Galacto-centric radius of $8.5\,{\rm kpc}$ and Galactic height of $500\,{\rm pc}$ (i.e., in the solar neighbourhood near the magnetic disk-halo transition region), the \cite{sun08} and \cite{jansson12} halo fields have magnetic field strengths of $0.3$ and $0.9\,\mu{\rm G}$, respectively. We plot in Figure~\ref{fig:halo_fd_map} the predicted FD maps using only the halo field components of the two magnetic field models (i.e., the disk field components have been removed). Although both halo field models do indeed predict FD disparities in Galactic longitude range of $40^\circ$--$52^\circ$, the same is also expected to occur in the longitude range of $20^\circ$--$40^\circ$. This latter FD disparity is not seen in our newly derived FD values.

With only the halo component of the two models, we further generated the boxcar-binned FD profiles along Galactic latitude in the longitude range of $40^\circ$--$52^\circ$, as shown in Figure~\ref{fig:rm_b_bin}. We again chose a bin width of $2\fdg5$ and a sampling step size of $1^\circ$, and applied a $y$-offset of $-280\,{\rm rad\,m}^{-2}$ to the observed FD profile to centre it at ${\rm FD} \approx 0\,{\rm rad\,m}^{-2}$. This is because we are interested in comparing the profile shapes and amplitudes of the observed and predicted FD disparities. From the similarities in the functional forms between the observed and predicted FD profiles, we suggest that the Galactic halo field remains a plausible explanation to the observed FD disparity. However, we argue that the exact implementations of the two halo field models investigated here are insufficient, since (1) they cannot reproduce the amplitude of the FD disparity in $40^\circ \leq \ell \leq 52^\circ$, and (2) they cannot explain the absence of FD disparity in $20^\circ \leq \ell \leq 40^\circ$.

\subsubsection{Scenario III: Contamination by Ionised Structures}
\label{sec:scenario3}
In the above two scenarios, we attributed the FD disparity to asymmetries in the magnetic field across the Galactic mid-plane. In this final scenario, we consider the case where the FD disparity is caused by differences in thermal electron number density on either side of the Galactic plane due to discrete ionised structures. Such structures will have an angular extent of $\gtrsim 10^\circ$ along the Galactic longitude, which we could not identify from careful inspections of the WHAMSS H$\alpha$ map \citep{haffner03,haffner10}, as well as the extinction-corrected H$\alpha$ map of \cite{finkbeiner03}. The mean extinction-corrected H$\alpha$ intensities in $40^\circ \leq \ell \leq 52^\circ$ are $4.40$ and $4.35\,{\rm Rayleighs}$ for above and below the Galactic mid-plane, respectively. In addition, we looked into the H\,{\sc i} map of the Effelsberg-Bonn H\,{\sc i} Survey \citep[EBHIS;][]{winkel16}, the Stokes \textit{I} map of the Sino-German 6\,cm Polarization Survey \citep{sun11}, and the \textit{WISE} H\,{\sc ii} region catalogue \citep{anderson14}, but could not locate any corresponding structures of interest.

Finally, we noted a nearby H\,{\sc i} bubble centred at $(\ell, b) \approx (45^\circ, 25^\circ)$ with an angular diameter of $\approx 40^\circ$ that was found to have FD contribution of $10$--$100\,{\rm rad\,m}^{-2}$ in the Global Magneto-Ionic Medium Survey \citep[GMIMS;][]{wolleben10}. It is unlikely that the FD disparity we found is due to contaminations by this H\,{\sc i} bubble, as the \cite{wolleben10} FD map suggests a minimal FD contribution of $\lesssim 10\,{\rm rad\,m}^{-2}$ in magnitude within our region of $40^\circ \leq \ell \leq 52^\circ$ and $|b| \leq 5^\circ$.

\subsection{Refining the \citet{vaneck11} Model} \label{sec:newvaneck}

Building upon the idea of an odd-parity disk field in the Sagittarius arm (Section~\ref{sec:scenario1}), we attempt to improve the \cite{vaneck11} model by allowing the possibility of an odd-parity disk field and re-fitting it to newly available data. The \cite{vaneck11} model is chosen here because it was found to perform the best overall within the region that we studied (Section~\ref{sec:field_comparisons}). We only further develop their Sector C that covered Galactic longitude range of $20^\circ$--$100^\circ$.

\begin{figure}
\centering
\includegraphics[clip, trim=1.0cm 1.5cm 1.0cm 1.2cm, width=240pt]{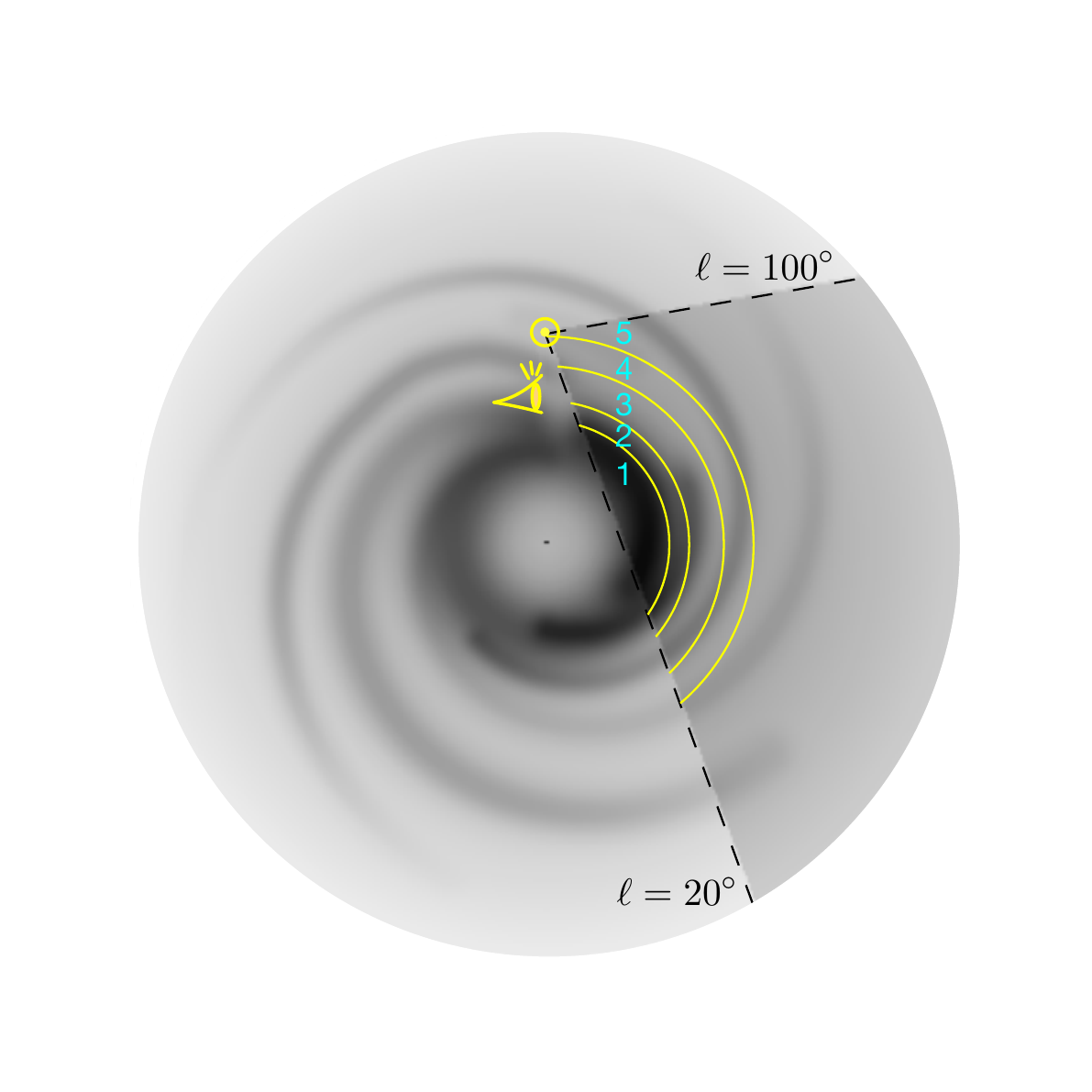}
\caption{Ring regions definition of our models following Sector C of \citet{vaneck11}, looking down onto the Galactic disk from the North Galactic Pole. The regions 1 through 5 are labelled, with the location of the Sun in the Milky Way marked by the symbol $\odot$. The eye symbol shows the vantage point of the edge-on illustration of Figure~\ref{fig:diagonal}. The background greyscale map shows the smooth component of NE2001 model.}
\label{fig:topdown}
\end{figure}

Most aspects of the skeleton of their model have been preserved, which we list in this paragraph as a summary. Starting from a Galacto-centric radius ($R$) of $3\,{\rm kpc}$, the Galactic volume is divided into five ring-shaped regions numbered 1 through 5 out to $R = 20\,{\rm kpc}$, with boundaries between these regions located at $R = 5.0$, $5.8$, $7.2$, and $8.4\,{\rm kpc}$. There is no modulation of field strength along Galactic height, and there is also no radial dependence of field strength within each region except for region 5, where it falls off by $R^{-1}$. A magnetic pitch angle of $11\fdg5$ has been adopted for regions 2--4, while for regions 1 and 5 the pitch angle is $0^\circ$. We show in Figure~\ref{fig:topdown} a schematic picture of this model.

Meanwhile, we made several modifications to their model. Firstly, while they truncated their magnetic field model at Galactic height of $\pm 1.5\,{\rm kpc}$, we extended this cutoff height slightly to $\pm 2.0\,{\rm kpc}$. This is to accommodate the few EGSs at the lowest Galactic longitude ($\ell \approx 20^\circ$) and the highest latitude ($b \approx \pm 5^\circ$), since otherwise the sight lines towards these sources will reach the cutoff height before passing the outer limit of $R = 20\,{\rm kpc}$. Secondly, we incorporated all components of the NE2001 model \citep{ne2001} instead of only the smooth components. Finally, while \cite{vaneck11} has assumed even-parity in all five regions, we only did so for regions 1, 2, and 5. The symmetry of magnetic field across the Galactic mid-plane for regions 3 and 4 are chosen differently for the different models investigated (see below).

\begin{table*}
\caption{Best-fit Results for the Tested New Models}
\label{table:newvaneck}
\centering
\begin{tabular}{cccccccccc}
\hline
& \multicolumn{7}{c}{Magnetic Field Strength ($\mu {\rm G}$)} \\
Model Name & Region 1 & Region 2 & Region 3a & Region 3b & Region 4a & Region 4b & Region 5 & dof & $\chi_{\rm red}^2$ \\
\hline
VE11 & $-0.15 \pm 0.04$ & $-0.40 \pm 0.01$ & $+2.23 \pm 0.13$ & $+2.23 \pm 0.13$ & $+0.09 \pm 0.05$ & $+0.09 \pm 0.05$ & $-0.86 \pm 0.09$ & 5 & \phantom{0}6.18 \\
Odd 3 & $+0.04 \pm 0.03$ & $+0.35 \pm 0.10$ & $+0.83 \pm 0.08$ & $-0.83 \pm 0.08$ & $+0.84 \pm 0.11$ & $+0.84 \pm 0.11$ & $-0.96 \pm 0.10$ & 5 & 10.35 \\
Odd 4 & $-0.02 \pm 0.03$ & $-0.37 \pm 0.12$ & $+1.56 \pm 0.10$ & $+1.56 \pm 0.10$ & $+0.61 \pm 0.08$ & $-0.61 \pm 0.08$ & $-0.68 \pm 0.07$ & 5 & \phantom{0}4.82 \\
Odd 3+4 & $+0.04 \pm 0.03$ & $+0.39 \pm 0.10$ & $+0.66 \pm 0.09$ & $-0.66 \pm 0.09$ & $+0.56 \pm 0.08$ & $-0.56 \pm 0.08$ & $-0.46 \pm 0.06$ & 5 & 10.30 \\
Free 3 & $-0.02 \pm 0.03$ & $-0.41 \pm 0.12$ & $+1.96 \pm 0.12$ & $+0.82 \pm 0.15$ & $+0.52 \pm 0.11$ & $+0.52 \pm 0.11$ & $-0.98 \pm 0.09$ & 6 & \phantom{0}5.12 \\
Free 4 & $-0.02 \pm 0.03$ & $-0.33 \pm 0.11$ & $+1.47 \pm 0.11$ & $+1.47 \pm 0.11$ & $+1.03 \pm 0.13$ & $-0.12 \pm 0.15$ & $-0.95 \pm 0.10$ & 6 & \phantom{0}4.51 \\
Free 3+4 & $-0.01 \pm 0.03$ & $-0.37 \pm 0.12$ & $+1.76 \pm 0.12$ & $+0.97 \pm 0.15$ & $+0.90 \pm 0.13$ & \phantom{+}$0.00 \pm 0.15$ & $-0.96 \pm 0.10$ & 7 & \phantom{0}3.87 \\
\hline
\multicolumn{10}{l}{\texttt{NOTE} -- For region 5, the magnetic field strength is modulated by $R^{-1}$, with the listed field strength above being that at $R = 8.5\,{\rm kpc}$.}
\end{tabular}
\end{table*}

We supplemented our 177 newly derived EGS FD values in the Galactic longitude range of $20^\circ$--$52^\circ$ with the EGS FDs from the 2020 data release of the CGPS compact source catalogue (CGPS2020; Van Eck et al.\ in prep.) that covered $53^\circ \leq \ell \leq 193^\circ$ and $-3^\circ \leq b \leq +5^\circ$. For our purpose, we only included the 622 CGPS EGSs within the longitude range of $53^\circ$--$100^\circ$. The CGPS2020 data were obtained from observations with the Dominion Radio Astrophysical Observatory (DRAO) Synthesis Telescope in four frequency bands centred at $1407.2$, $1414.1$, $1427.7$, and $1434.6\,{\rm MHz}$, with bandwidths of $7.5\,{\rm MHz}$ each \citep{landecker10}. Our new combined data set thus contains 799 EGS FD values, which is a significant improvement from the 378 used by \cite{vaneck11} in modelling the same Sector C. However, we did not incorporate pulsar FD measurements in our fitting procedures.

We investigated a total of six different models. The first three are ``odd-parity'' models, which have odd-parity magnetic fields in specified regions and even-parity magnetic fields in the remaining regions. Specifically, the model ``Odd~3'' has odd-parity field in region 3 (i.e., the magnetic field in this region has the same magnitude but opposite direction across the mid-plane), the model ``Odd~4'' has odd-parity field in region 4, and the model ``Odd~3+4'' has odd-parity fields in both regions 3 and 4. We further relaxed the symmetry constraint in our next three ``free'' models, with the magnetic field strength and direction on either side of the Galactic mid-plane in the specified ``free'' regions fitted independently. This can be thought of as a superposition of an even-parity field and an odd-parity field in a region. The model ``Free~3'' has region 3 set as such ``free'' region (while regions 1, 2, 4, and 5 have even-parity fields), and similarly for the ``Free~4'' and ``Free~3+4'' models. Note that regions 3 and 4 are situated at the Sagittarius arm, which is the reason that we modified their magnetic field symmetries in the models explored here.

For each model, we determined the best-fit values of the free parameters (namely, the magnetic field strength and direction in each region) by the Markov chain Monte Carlo (MCMC) method. Specifically, we used a \texttt{python} implementation of \texttt{emcee} \citep{emcee} here. We first binned our 799 EGS FD values across Galactic longitude in $5^\circ$ independent bins, centred at $\ell = 22\fdg5$, $27\fdg5$, ..., $97\fdg5$. The median of each bin was taken as the binned value, with the SEM (Equation~\ref{eq:sem}) adopted as the uncertainty. Such binning was performed independently for either side of the Galactic mid-plane. Next, we adjusted the free parameters and calculated the resulting predicted FD value of each bin. The performance of the set of free parameters is then evaluated by comparing the predicted with the actual FD values, simultaneously on the two sides of the Galactic mid-plane. Specifically, we adopted a likelihood function assuming Gaussian measurement uncertainties. We further chose a prior of uniform distribution within $\pm 10\,\mu{\rm G}$ for all free parameters (i.e., constraining the magnetic field strength in all regions to within $10\,\mu{\rm G}$). We initiated the runs for each model with 16 walkers randomly positioned in the parameter space (by uniform distribution within the constraint set by the prior), and proceeded for a variable number of steps depending on the complexity of each model. The auto-correlation time for each case was determined\footnote{We determined the auto-correlation time using the \texttt{emcee} function \texttt{get\_autocorr\_time()}, following the \texttt{emcee} tutorial \href{https://emcee.readthedocs.io/en/stable/tutorials/line/}{https://emcee.readthedocs.io/en/stable/tutorials/line/}.}, and we discarded 10 times that as the initial burn-in steps. In all cases, we are left with usable number of steps of more than 100 times the auto-correlation time, and more than 10,000 steps times 16 walkers from which we determine the best-fit results. We noted that all the best-fit values have highly symmetric uncertainties, and therefore we do not report asymmetric error bars (see Appendix~D in the Online Supporting Information).

\begin{figure*}
\includegraphics[width=450pt]{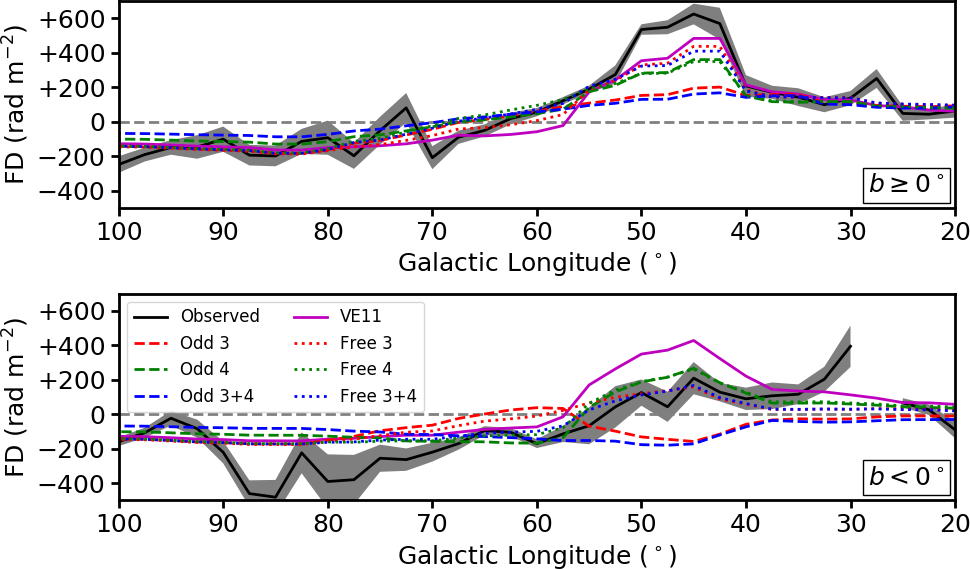}
\caption{Predicted boxcar-binned FD profiles along $\ell$ for the different new models for \textbf{(Top)} above and \textbf{(Bottom)} below the Galactic mid-plane. The observed EGS FD profiles and the predicted profiles of \citet{vaneck11} model are similarly plotted.}
\label{fig:newvaneck}
\end{figure*}

We list the best-fit values of all parameters for each model, alongside with those from the \cite{vaneck11} model as comparison, in Table~\ref{table:newvaneck}. Regions 3 and 4 are further divided into sub-regions a and b, representing above and below the Galactic mid-plane, respectively. As mentioned above, the magnetic field strength in region 5 is modulated by $R^{-1}$, and thus we choose to report the field strength in this region at $R = 8.5\,{\rm kpc}$. We follow the convention defined by \cite{vaneck11} that, positive and negative magnetic field strengths denote counter-clockwise (CCW) and clockwise (CW) field directions, respectively, when viewed from the North Galactic Pole. In the same Table, we also list the degrees of freedom (dof) of the models, as well as $\chi_{\rm red}^2$ defined as
\begin{equation}
\chi_{\rm red}^2 = \sum_i^N \frac{(\overline{\rm FD}_{{\rm obs}, i} - \overline{\rm FD}_{{\rm model}, i})^2}{(N-{\rm dof}) \cdot \sigma_{{\rm FD}, i}^2}{\rm ,}
\end{equation}
where $i$ is the index for the $N=51$ independent bins, and the remaining parameters as defined in Equation~\ref{eq:chi}. A smaller $\chi_{\rm red}^2$ value indicates a better performing model. Note that $\chi_{\rm red}^2$ values of much larger than unity can either signify inadequacies in the large-scale magnetic field in our models, or it can be attributed to the turbulent interstellar medium. Although we attempted to account for the latter through $\sigma_{{\rm FD}, i}$, we did not model the power-law nature of such turbulent interstellar medium \citep{haverkorn08,stil11} in our study here. We further plot the predicted FD profiles along Galactic longitude for all models in Figure~\ref{fig:newvaneck} for visual comparisons.

\begin{figure*}
\includegraphics[width=440pt]{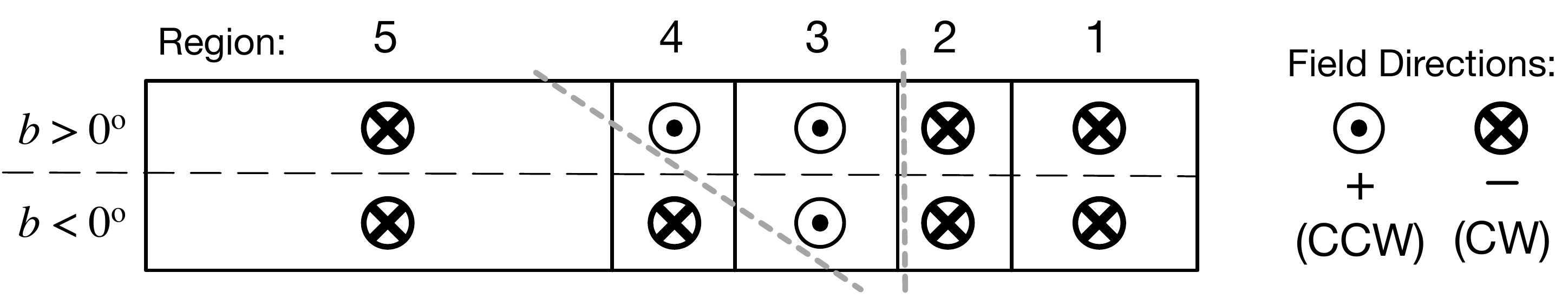}
\caption{Edge-on view through our Odd~4 and Free~4 models from the vantage point marked in Figure~\ref{fig:topdown}. In each region, the line-of-sight magnetic field direction is marked (see legend on the right). Interfaces of large-scale field reversals are denoted by the two grey dashed lines. Both these models depict a diagonal interface of field reversal, similar to the conclusion of \citet{ordog17}.}
\label{fig:diagonal}
\end{figure*}

It is evident that models Odd~4, Free~3, Free~4, and Free~3+4 show better fit to the data than that of \cite{vaneck11}, suggesting that our hypothesis of an odd-parity disk field in the Sagittarius arm is indeed improving the model of the Milky Way magnetic field. However, the models are still not deemed a satisfactory fit to our data, given the high $\chi_{\rm red}^2 > 3.5$ for all cases. In particular, we point out that none of the models can capture the ${\rm FD} \approx +600\,{\rm rad\,m}^{-2}$ peak at $\ell \approx 40^\circ$--$50^\circ$ above the Galactic mid-plane. Nonetheless, the results here serve as an important step towards a major future improvement in the model of the Milky Way magnetic field. Upon inspection of the FD profiles of the model Odd~3, we find that it does not only predict FD disparity over $\ell \approx 40^\circ$--$52^\circ$, but also over a wide range of $\ell \approx 20^\circ$--$60^\circ$. This clearly shows that with the geometry of region 3 defined by \cite{vaneck11}, one cannot simply impose an odd-parity field there to obtain FD disparity that starts from $\ell = 40^\circ$. Similarly for model Odd~4, we see FD disparity in longitude range of about $55^\circ$--$80^\circ$. These strongly suggest that if one were to further improve the \cite{vaneck11} model in the future, its geometry (namely, the shape of the individual regions, the locations of the region boundaries, and/or the magnetic pitch angles) must be modified to obtain better fit to data. Another possibility is to rebuild the \cite{vaneck11} model using the YMW16 thermal electron distribution model instead, which is beyond the scope of this paper.

\subsection{Connections to Other FD Grid Experiments}
\label{sec:connection}
In the past few years, there have been significant efforts in shedding new light on the complex large-scale magnetic fields in the first Galactic quadrant. \cite{ordog17} have shown, from CGPS FD data of both Galactic diffuse emission and EGSs, that there is a clear FD gradient across a diagonal line from $(\ell, b) = (67^\circ, +4^\circ)$ to $(56^\circ, -2^\circ)$. This is similar to the FD disparity we identified in this paper, which can be seen as an FD gradient across the Galactic mid-plane. Given the spatial proximity and similar nature of these two structures, it is possible that the two are physically linked. It has been shown by magnetohydrodynamics (MHD) simulations of the global Galactic disk magnetic field \citep{gressel13} that, magneto-rotational instabilities (MRI) can cause the interface of an odd-parity field to rise and fall through the Galactic mid-plane across Galactic azimuthal angle (``undulations''; see their figure 10). This serves as a possible connection between the two FD gradients. Moreover, an alternative interpretation of the \cite{ordog17} structure is that it traces the large-scale magnetic field reversal in the Sagittarius arm. This echoes the Odd~4 and Free~4 models that we presented in Section~\ref{sec:newvaneck}, which suggests a similar diagonal interface for the large-scale magnetic field reversal (see Figure~\ref{fig:diagonal}). Future dedicated simulation efforts are required to gain a full, accurate understanding in the physical nature of and connection between the \cite{ordog17} FD gradient and our FD disparity.

In addition, the THOR survey has recently uncovered an unexpectedly high FD (up to $\approx 4000\,{\rm rad\,m}^{-2}$ in magnitude) through the tangent of the Sagittarius arm within $|b| < 1\fdg25$ \citep{shanahan19}, likely tracing a compressed warm ionised medium in that spiral arm. Meanwhile, our work probing up to $|b| = 5^\circ$ towards the same spiral arm has discovered the FD disparity. These two complementary studies together paint a vibrant picture of the complex magneto-ionic medium in the Sagittarius arm. An investigation in how these two regimes are connected would require a much denser FD grid than is currently available. This could be achieved by on-going polarisation surveys such as POSSUM \citep{gaensler10} and VLASS \citep{myers14,lacy20}.

\section{Conclusion}
\label{sec:conclusion}
In this paper, we have conducted new broadband spectro-polarimetric observations with the VLA to study the large-scale magnetic field near the Milky Way mid-plane ($|b| \leq 5^\circ$) in the Galactic longitude range of $20^\circ$--$52^\circ$. The FD values of a total of 194 EGSs (168 on-axis plus 26 off-axis) have been determined, out of which 177 (153 on-axis plus 24 off-axis) were used for this study. Our effort has led to a significant increase in the number of reliable FD values by a factor of five in this complex Galactic region, leading to our discovery of a clear disparity in FD values across the Galactic mid-plane in the longitude range of $40^\circ$--$52^\circ$. We do not see similar FD disparities in the longitude range of $20^\circ$--$40^\circ$. From existing pulsar FD measurements, we found hints that the FD disparity occurs at a distance of $\approx 5\,{\rm kpc}$ away from us, corresponding to the Sagittarius spiral arm that has been known to host a large-scale magnetic field reversal. 

We further performed rigorous comparisons between our newly derived EGS FD values with the predictions of three major large-scale magnetic field models of the Milky Way -- \cite{sun08} ASS+RING, \cite{vaneck11}, and \cite{jansson12}, combined with the thermal electron number density model of NE2001 \citep{ne2001}. Our conclusion is that the \cite{vaneck11} model can best match our measured FD values overall. However, we also noted a short-coming of this model, namely it has assumed a-priori that the large-scale Galactic disk magnetic field has an even parity everywhere. It therefore could not adequately fit to the observed FD values in the longitude range of $40^\circ$--$52^\circ$ when we considered the two sides of the Galactic mid-plane separately.

Given the unsatisfactory performance of the above magnetic field models, we considered three astrophysical scenarios that could have led to this newly discovered FD disparity:
\begin{itemize}
\item Scenario I: The large-scale disk field in the Sagittarius arm can have an odd parity, either as the dominant component or in superposition with a stronger even-parity field;
\item Scenario II: An odd-parity halo field contributes significantly to our EGS FD values, causing the FD disparity; or
\item Scenario III: Some Galactic ionised structure contaminates the FD values of our target EGSs either above or below the Galactic plane.
\end{itemize}
We favour Scenario I given the currently available information, since Scenario II would require the odd-parity halo field to show appreciable effects at a very low Galactic height of $\lesssim 90\,{\rm pc}$ in $40^\circ \leq \ell \leq 52^\circ$ only. We could not identify notable structures in H$\alpha$, H\,{\sc i}, or 6\,cm radio continuum maps, or in the \textit{WISE} H\,{\sc ii} region catalogue that would support Scenario III.

Finally, we pursued an improved \cite{vaneck11} model by relaxing the even-parity field constraint. From this, we developed new models that showed better fit to the observed EGS FD values than the \cite{vaneck11} model. This will serve as an important step towards major future improvements in magnetic field models of the Milky Way.

Our study adds to the recent rapid progress in our understanding of the Galactic-scale magnetic fields in the first quadrant of the Milky Way, prompting the development of a vastly improved magnetic field model. On-going and future radio polarisation surveys will certainly further shed light on the complex magnetic field structure of the Galaxy. As the next step, we will repurpose this same dataset to study the small-scale Galactic magnetic field in this same sky region in a future publication (Ma et al.\ in prep.).

\section*{Acknowledgements}
We thank the manuscript referee, Tess Jaffe, for her careful reading and insightful comments that have improved this paper. We thank Aristeidis Noutsos for his valuable suggestions and comments as the MPIfR internal referee, and Rainer Beck and Marita Krause for fruitful discussions on this work. We further thank \mbox{Russell} Shanahan and Jeroen Stil for discussions on the FD measurements from the THOR survey, Cameron Van Eck for discussions about the CGPS data, and Jennifer West for her assistance on a \texttt{python} implementation of the NE2001 model. This publication is adapted from part of the PhD thesis of the lead author \citep[chapter 5 of][]{ma20thesis}. Y.K.M.\ was supported for this research by the International Max Planck Research School (IMPRS) for Astronomy and Astrophysics at the Universities of Bonn and Cologne. Y.K.M.\ acknowledges partial support through the Bonn-Cologne Graduate School of Physics and Astronomy. The National Radio Astronomy Observatory is a facility of the National Science Foundation operated under cooperative agreement by Associated Universities, Inc. The Canadian Galactic Plane Survey is a Canadian project with international partners. It was supported by the Natural Sciences and Engineering Research Council. The Dominion Radio Astrophysical Observatory is a national facility operated by the National Research Council Canada. This research has made use of the NASA/IPAC Extragalactic Database (NED) which is operated by the Jet Propulsion Laboratory, California Institute of Technology, under contract with the National Aeronautics and Space Administration. The Wisconsin H-Alpha Mapper and its Sky Survey have been funded primarily through awards from the U.S.\ National Science Foundation.

\section*{DATA AVAILABILITY}
The data underlying this article will be shared on reasonable request to the corresponding author.

\bsp

\bibliography{ms}
\label{lastpage}



\section*{Supplementary material}

# The Complex Large-scale Magnetic Fields in the First Galactic Quadrant as Revealed by the Faraday Depth Profile Disparity

Yik Ki Ma[1], S. A. Mao[1], A. Ordog[2], and J. C. Brown[2]

[1]*Max-Planck-Institut für Radioastronomie, Auf dem Hügel 69, 53121 Bonn, Germany*
[2]*Department of Physics and Astronomy, University of Calgary, Calgary, AB T2N 1N4, Canada*

## APPENDIX A: BLENDED AND DISCARDED SOURCES

As pointed out in Section 3 of the main text, 29 of the off-axis targets are spatially blended with their corresponding on-axis targets, and for such cases their combined flux densities were extracted instead of attempting to separate them. We list in Table A1 such spatially blended sources. In addition, a total of 15 target sources have been discarded because they were not confidently detected even in total intensity images. These sources are listed in Table A2.

## APPENDIX B: TOTAL INTENSITIES OF TARGET SOURCES

We also include here the total intensity information of the targets derived from our new observations. For each source, we fitted a power law to the radio spectrum:

$$S_\nu = S_{1.4\,\mathrm{GHz}} \cdot \left(\frac{\nu}{1.4\,\mathrm{GHz}}\right)^\alpha, \qquad (B1)$$

where $S_\nu$ [mJy] is the measured flux density at frequency $\nu$ [GHz], and $\alpha$ is the spectral index. The values of $\alpha$ and $S_{1.4\,\mathrm{GHz}}$, along with the reported integrated radio flux densities at also 1.4 GHz from the NVSS ($S_{\mathrm{NVSS}}$), are listed in Tables B1 and B2 for our on- and off-axis targets, respectively. Note that the discrepancies between $S_{1.4\,\mathrm{GHz}}$ and $S_{\mathrm{NVSS}}$ for some of the sources can be attributed to the spatial blending of multiple NVSS sources in our new observations (marked by the $\star$ symbol in the Tables; see Section 3 and Appendix A).

## APPENDIX C: FARADAY SPECTRA OF TARGET SOURCES

We present here the Faraday spectra (at $|\mathrm{FD}| < 1000\,\mathrm{rad\,m^{-2}}$) of the 204 target sources (194 polarised plus 10 unpolarised) observed for our study, as well as that of the on-axis polarisation leakage calibrator (J1407+2827) in the seven observing blocks (OBs). The spectra of on-axis targets (including the leakage calibrator) are shown in Figure C1, and those of off-axis targets are shown in Figure C2. The results from this RM-Synthesis analysis are listed in Tables 1 and 2 of the main text.

## APPENDIX D: MCMC RESULTS OF NEW MODELS

Finally, we present the corner plots of our Markov chain Monte Carlo (MCMC) fitting results of our new models of the Milky Way magnetic fields (see Section 5.6 of main text). The results of our models Odd 3, Odd 4, Odd 3+4, Free 3, Free 4, and Free 3+4 are shown in Figures D1, D2, D3, D4, D5, and D6, respectively. For all cases, the subscript denotes the region number (e.g., $B_2$ means the magnetic field strength in region 2), and for "odd-parity" models

**Table A1.** List of Spatially Blended Sources

| On-axis Source (NVSS) | Blended Off-axis Source (NVSS) |
|---|---|
| J181931−091059 | J181937−090914 |
| J183409−071802 | J183408−071904 |
|  | J183414−071628 |
| J183418+004852 | J183417+004939 |
| J183511+014620 | J183515+014611 |
| J183840−012957 | J183842−013105 |
| J184124−015255 | J184130−015257 |
|  | J184134−015252 |
|  | J184142−015246 |
| J184249−075604 | J184245−075613 |
| J184415−041757 | J184413−041938 |
|  | J184422−041746 |
|  | J184422−041848 |
| J184718+055022 | J184711+054948 |
| J184821+001108 | J184824+001158 |
| J185030−090659 | J185037−090633 |
| J185239−101324 | J185230−101410 |
| J185728+111021 | J185731+111053 |
| J185744−052527 | J185748−052631 |
| J190655+000339 | J190651+000113 |
|  | J190653+000213 |
| J190944+005558 | J190950+005550 |
| J191025+140125 | J191028+140019 |
|  | J191031+135810 |
|  | J191033+135653 |
| J191325+034308 | J191331+034218 |
| J192458+130033 | J192501+130143 |
| J192922+095808 | J192922+095901 |
| J193939+134604 | J193943+134652 |

**Table A2.** List of Discarded Sources

| Discarded On-axis Targets | | |
|---|---|---|
| J184500−015838 | J184622−015654 | J184655−021535 |
| J184809−005748 | J185521+011951 | |

| Discarded Off-axis Targets | | |
|---|---|---|
| J183205−103701 | J183207−103648 | J185529+052003 |
| J190352+054926 | J190359+055501 | J190402+055410 |
| J190743+090552 | J192508+135801 | J192511+135653 |
| J192519+135819 | | |

we only show the fitted field strengths above the mid-plane for odd-parity regions, since they would be identical to those below the mid-plane (apart from the obvious flip in sign).

This paper has been typeset from a TEX/LATEX file prepared by the author.

**Table B1.** Total Intensities and Spectral Indices of On-axis Targets

| Target Source (NVSS) | $\ell$ (°) | $b$ (°) | $\alpha$ | $S_{1.4\,\mathrm{GHz}}$ (mJy) | $S_{\mathrm{NVSS}}$ (mJy) |
|---|---|---|---|---|---|
| J184415−131243 | 20.34 | −4.42 | −1.05 ± 0.02 | 113.4 ± 0.3 | 121.4 ± 4.4 |
| J181343−090743 | 20.49 | +4.11 | −0.81 ± 0.05 | 36.4 ± 0.4 | 36.3 ± 1.5 |
| J182038−094716 | 20.71 | +2.29 | −0.79 ± 0.01 | 949.4 ± 1.5 | 901.8 ± 27.1 |
| J183519−111559 | 21.08 | −1.59 | −0.77 ± 0.01 | 324.1 ± 0.7 | 291.3 ± 8.8 |
| J181851−090659 | 21.10 | +3.00 | −0.82 ± 0.02 | 397.5 ± 1.5 | 411.6 ± 14.1 |
| J181931−091059★ | 21.12 | +2.82 | −0.92 ± 0.02 | 420.5 ± 1.3 | 274.6 ± 9.3 |
| J183759−112627 | 21.23 | −2.25 | −0.87 ± 0.02 | 155.2 ± 0.7 | 155.7 ± 4.7 |
| J183220−103510 | 21.35 | −0.63 | −0.01 ± 0.02 | 1051.7 ± 4.2 | 867.4 ± 26.0 |
| J182443−092933 | 21.44 | +1.54 | −0.80 ± 0.03 | 157.0 ± 0.9 | 154.4 ± 5.5 |
| J184606−115808 | 21.66 | −4.26 | −0.87 ± 0.01 | 783.3 ± 0.7 | 792.9 ± 23.8 |
| J181419−073733 | 21.88 | +4.69 | −0.75 ± 0.06 | 27.2 ± 0.3 | 25.6 ± 0.9 |
| J184059−110139 | 21.93 | −2.72 | −0.63 ± 0.01 | 119.4 ± 0.3 | 115.6 ± 3.5 |
| J182503−085445 | 22.00 | +1.74 | −1.16 ± 0.02 | 117.1 ± 0.5 | 118.8 ± 4.3 |
| J182542−083723 | 22.33 | +1.73 | −0.72 ± 0.01 | 445.1 ± 1.1 | 431.1 ± 15.0 |
| J183942−101038 | 22.54 | −2.05 | −0.85 ± 0.02 | 108.0 ± 0.5 | 109.9 ± 3.8 |
| J184750−110658 | 22.61 | −4.25 | −0.75 ± 0.01 | 250.8 ± 0.6 | 254.7 ± 9.0 |
| J184911−111241 | 22.68 | −4.59 | −1.04 ± 0.01 | 218.0 ± 0.3 | 215.0 ± 6.5 |
| J182530−080945 | 22.71 | +1.99 | −0.57 ± 0.02 | 167.4 ± 0.7 | 169.8 ± 5.1 |
| J184812−105133 | 22.88 | −4.22 | −0.84 ± 0.05 | 23.9 ± 0.2 | 22.9 ± 0.8 |
| J184552−103126 | 22.93 | −3.56 | −0.43 ± 0.03 | 38.9 ± 0.2 | 66.6 ± 2.1 |
| J181949−065524 | 23.15 | +3.81 | −0.98 ± 0.01 | 328.0 ± 0.8 | 326.2 ± 11.2 |
| J182537−073729 | 23.20 | +2.21 | −0.20 ± 0.01 | 400.7 ± 0.7 | 364.5 ± 10.9 |
| J182431−072714 | 23.23 | +2.53 | −0.64 ± 0.01 | 308.7 ± 0.7 | 312.6 ± 11.1 |
| J182920−073400 | 23.68 | +1.42 | −0.51 ± 0.04 | 115.9 ± 1.1 | 92.2 ± 3.3 |
| J184644−094654 | 23.68 | −3.41 | −0.94 ± 0.01 | 186.7 ± 0.4 | 198.5 ± 6.0 |
| J184547−093821 | 23.70 | −3.14 | −0.75 ± 0.06 | 27.0 ± 0.3 | 158.0 ± 4.8 |
| J183052−074402 | 23.71 | +1.01 | −0.33 ± 0.05 | 89.1 ± 0.9 | 83.1 ± 2.9 |
| J182043−062415 | 23.71 | +3.86 | −0.78 ± 0.02 | 87.1 ± 0.4 | 82.6 ± 2.5 |
| J185239−101324★ | 23.95 | −4.91 | −0.61 ± 0.01 | 579.9 ± 0.9 | 425.8 ± 14.0 |
| J183902−083023 | 23.95 | −1.14 | −0.97 ± 0.02 | 214.3 ± 0.8 | 221.2 ± 8.5 |
| J182104−060915 | 23.98 | +3.90 | −0.90 ± 0.02 | 127.1 ± 0.6 | 123.9 ± 3.7 |
| J183321−073121 | 24.18 | +0.56 | −0.64 ± 0.03 | 538.9 ± 3.9 | 489.0 ± 17.3 |
| J183409−071802★ | 24.47 | +0.49 | +0.29 ± 0.02 | 5282.3 ± 27.5 | 4091.2 ± 135.3 |
| J185030−090659★ | 24.70 | −3.94 | −0.64 ± 0.01 | 538.9 ± 0.7 | 470.6 ± 16.5 |
| J184249−075604★ | 24.89 | −1.71 | −0.63 ± 0.00 | 1541.7 ± 1.2 | 1074.8 ± 38.0 |
| J182351−052429 | 24.96 | +3.63 | −0.59 ± 0.04 | 73.9 ± 0.6 | 75.6 ± 2.7 |
| J182111−050219 | 24.98 | +4.39 | −0.47 ± 0.01 | 448.7 ± 0.8 | 404.8 ± 14.2 |
| J184629−081333 | 25.05 | −2.65 | −0.95 ± 0.01 | 156.5 ± 0.4 | 163.8 ± 5.8 |
| J182013−042541 | 25.41 | +4.89 | −0.78 ± 0.01 | 806.3 ± 1.4 | 793.1 ± 27.4 |
| J184511−060146 | 26.85 | −1.36 | −0.58 ± 0.03 | 104.1 ± 0.7 | 102.9 ± 3.6 |
| J183253−042628 | 26.86 | +2.09 | −0.91 ± 0.01 | 168.0 ± 0.4 | 168.3 ± 5.9 |
| J182634−030927 | 27.27 | +4.08 | −0.71 ± 0.01 | 410.6 ± 0.9 | 408.0 ± 12.2 |
| J183847−040042 | 27.92 | +0.98 | −0.75 ± 0.00 | 747.1 ± 0.8 | 743.0 ± 26.1 |
| J183400−030340 | 28.22 | +2.48 | −1.04 ± 0.01 | 368.1 ± 0.4 | 364.3 ± 10.9 |
| J185054−050942 | 28.27 | −2.23 | −0.65 ± 0.01 | 230.9 ± 0.4 | 240.4 ± 7.2 |
| J184415−041757★ | 28.29 | −0.36 | 0.00 ± 0.02 | 1432.1 ± 7.6 | 366.5 ± 12.9 |
| J185523−053804 | 28.36 | −3.44 | −0.92 ± 0.02 | 65.3 ± 0.3 | 64.4 ± 2.4 |
| J183652−024606 | 28.81 | +1.97 | −1.14 ± 0.01 | 260.6 ± 0.5 | 248.3 ± 8.8 |
| J185744−052527★ | 28.81 | −3.87 | −0.78 ± 0.02 | 137.5 ± 0.5 | 116.9 ± 4.1 |
| J183939−030047 | 28.91 | +1.24 | −0.73 ± 0.02 | 266.3 ± 1.0 | 257.3 ± 8.4 |
| J183717−015034 | 29.68 | +2.30 | −0.85 ± 0.01 | 730.2 ± 1.4 | 716.7 ± 23.4 |
| J182900−002018 | 30.07 | +4.84 | −0.76 ± 0.01 | 217.4 ± 0.4 | 215.6 ± 7.6 |
| J184124−015255★ | 30.11 | +1.37 | −0.64 ± 0.01 | 6663.2 ± 7.5 | 3050.5 ± 98.8 |
| J183840−012957★ | 30.14 | +2.16 | −0.96 ± 0.02 | 290.8 ± 1.3 | 198.8 ± 7.0 |
| J183551−005941 | 30.27 | +3.01 | −0.95 ± 0.01 | 220.8 ± 0.4 | 219.6 ± 7.8 |
| J190014−033504 | 30.74 | −3.59 | −0.81 ± 0.01 | 187.2 ± 0.4 | 183.3 ± 5.5 |
| J184959−013256 | 31.39 | −0.38 | −0.93 ± 0.01 | 1523.4 ± 4.1 | 1459.8 ± 43.8 |
| J183838+000858 | 31.60 | +2.92 | −0.69 ± 0.01 | 196.5 ± 0.3 | 169.3 ± 5.1 |

★ Spatially blended source (see Appendix A)

**Table B1.** (Continued) Total Intensities and Spectral Indices of On-axis Targets

| Target Source (NVSS) | $\ell$ (°) | $b$ (°) | $\alpha$ | $S_{1.4\,\mathrm{GHz}}$ (mJy) | $S_{\mathrm{NVSS}}$ (mJy) |
|---|---|---|---|---|---|
| J183418+004852* | 31.70 | +4.18 | −0.71 ± 0.01 | 446.8 ± 0.7 | 442.6 ± 13.9 |
| J183931+001447 | 31.79 | +2.76 | −1.03 ± 0.02 | 84.7 ± 0.3 | 80.9 ± 2.5 |
| J183307+011535 | 31.97 | +4.65 | +0.26 ± 0.00 | 484.3 ± 0.5 | 489.5 ± 14.7 |
| J183437+010519 | 31.98 | +4.24 | −0.49 ± 0.02 | 100.3 ± 0.6 | 94.7 ± 2.9 |
| J185822−013654 | 32.28 | −2.28 | −0.78 ± 0.01 | 429.1 ± 0.7 | 406.1 ± 12.2 |
| J184704−000446 | 32.36 | +0.93 | −0.18 ± 0.02 | 428.2 ± 1.5 | 409.5 ± 13.8 |
| J190833−023000 | 32.65 | −4.95 | −0.80 ± 0.01 | 190.7 ± 0.4 | 183.8 ± 5.5 |
| J183511+014620* | 32.66 | +4.42 | −0.91 ± 0.01 | 905.5 ± 0.9 | 483.9 ± 14.5 |
| J183337+020355 | 32.74 | +4.91 | −0.91 ± 0.01 | 468.9 ± 0.5 | 458.4 ± 13.8 |
| J184821+001108* | 32.75 | +0.77 | −0.25 ± 0.03 | 300.3 ± 2.0 | 135.6 ± 4.9 |
| J185351−002508 | 32.84 | −0.73 | −0.68 ± 0.03 | 185.5 ± 1.1 | 193.6 ± 6.9 |
| J185751−004817 | 32.95 | −1.80 | −0.79 ± 0.01 | 777.3 ± 1.0 | 722.5 ± 25.4 |
| J190042−005151 | 33.22 | −2.46 | −1.10 ± 0.01 | 222.6 ± 0.6 | 218.0 ± 7.7 |
| J190407−011342 | 33.29 | −3.38 | −0.40 ± 0.01 | 291.3 ± 0.6 | 332.7 ± 11.8 |
| J185146+003532 | 33.50 | +0.19 | −0.36 ± 0.02 | 1008.9 ± 6.3 | 839.9 ± 29.6 |
| J190832−011929 | 33.70 | −4.41 | −0.37 ± 0.02 | 105.5 ± 0.3 | 116.0 ± 3.5 |
| J184755+012221 | 33.75 | +1.41 | −0.08 ± 0.03 | 94.0 ± 0.6 | 99.8 ± 3.0 |
| J185857+000727 | 33.90 | −1.61 | −0.80 ± 0.02 | 273.1 ± 1.0 | 290.2 ± 9.8 |
| J190017+000355 | 34.00 | −1.94 | −0.50 ± 0.01 | 361.3 ± 0.6 | 338.4 ± 10.2 |
| J184435+020933 | 34.07 | +2.51 | −0.60 ± 0.03 | 65.9 ± 0.4 | 62.5 ± 2.3 |
| J190831−004855 | 34.16 | −4.17 | −0.47 ± 0.04 | 45.8 ± 0.4 | 57.5 ± 2.1 |
| J191010−005622 | 34.23 | −4.60 | −0.93 ± 0.03 | 37.1 ± 0.2 | 32.6 ± 1.1 |
| J190532−000941 | 34.40 | −3.21 | −1.11 ± 0.02 | 179.7 ± 0.7 | 176.9 ± 6.3 |
| J190559+000721 | 34.70 | −3.18 | −1.13 ± 0.01 | 181.8 ± 0.5 | 192.4 ± 6.8 |
| J190655+000339* | 34.75 | −3.42 | −1.18 ± 0.06 | 84.4 ± 0.9 | 37.2 ± 1.5 |
| J190741+000038 | 34.80 | −3.61 | −1.04 ± 0.01 | 249.9 ± 0.5 | 246.4 ± 7.4 |
| J183848+040424 | 35.13 | +4.66 | −0.39 ± 0.00 | 683.3 ± 0.7 | 666.2 ± 20.0 |
| J185515+021054 | 35.31 | +0.15 | −0.06 ± 0.08 | 93.2 ± 1.9 | 84.6 ± 2.6 |
| J191133+001449 | 35.45 | −4.36 | −0.58 ± 0.02 | 117.3 ± 0.5 | 117.6 ± 4.6 |
| J190426+011036 | 35.46 | −2.36 | +0.36 ± 0.05 | 80.5 ± 0.8 | 61.7 ± 1.9 |
| J185114+025939 | 35.57 | +1.41 | −0.84 ± 0.00 | 529.3 ± 0.6 | 510.8 ± 15.3 |
| J184320+040256 | 35.62 | +3.65 | −0.82 ± 0.01 | 115.8 ± 0.3 | 112.6 ± 3.4 |
| J190944+005558* | 35.85 | −3.65 | −0.91 ± 0.01 | 262.6 ± 0.7 | 121.2 ± 4.3 |
| J191417+002421 | 35.91 | −4.90 | −0.83 ± 0.04 | 52.7 ± 0.4 | 57.7 ± 2.1 |
| J190712+012709 | 36.02 | −2.84 | −0.37 ± 0.01 | 703.8 ± 0.7 | 712.2 ± 21.4 |
| J185213+033255 | 36.18 | +1.44 | −0.57 ± 0.02 | 223.4 ± 0.8 | 234.3 ± 7.0 |
| J185837+024518 | 36.20 | −0.34 | −0.82 ± 0.04 | 98.0 ± 0.9 | 93.6 ± 3.4 |
| J184500+043812 | 36.33 | +3.54 | −0.84 ± 0.05 | 54.0 ± 0.5 | 54.3 ± 2.0 |
| J184604+043450 | 36.40 | +3.28 | −0.27 ± 0.02 | 84.7 ± 0.4 | 93.5 ± 2.8 |
| J185802+031316 | 36.55 | +0.00 | −0.15 ± 0.01 | 1011.2 ± 2.0 | 1089.6 ± 32.7 |
| J185306+044052 | 37.29 | +1.76 | −1.03 ± 0.03 | 63.9 ± 0.4 | 48.2 ± 1.5 |
| J184718+055022* | 37.67 | +3.57 | −0.67 ± 0.03 | 136.8 ± 0.7 | 101.5 ± 3.5 |
| J184438+062651 | 37.91 | +4.44 | −1.04 ± 0.02 | 180.1 ± 0.6 | 186.1 ± 5.6 |
| J191406+025549 | 38.13 | −3.70 | −0.51 ± 0.02 | 82.3 ± 0.3 | 83.1 ± 2.5 |
| J184432+064257 | 38.14 | +4.58 | −0.82 ± 0.00 | 806.5 ± 0.6 | 794.3 ± 23.8 |
| J185513+052158 | 38.14 | +1.60 | −0.76 ± 0.02 | 123.4 ± 0.6 | 111.4 ± 4.0 |
| J184919+063211 | 38.52 | +3.44 | −0.43 ± 0.06 | 23.2 ± 0.3 | 24.5 ± 0.9 |
| J191325+034308* | 38.76 | −3.18 | −0.77 ± 0.01 | 601.6 ± 1.1 | 396.9 ± 13.1 |
| J191849+030442 | 38.81 | −4.67 | −0.94 ± 0.01 | 540.2 ± 0.7 | 550.7 ± 19.3 |
| J184753+071538 | 39.00 | +4.09 | −0.45 ± 0.02 | 194.0 ± 0.6 | 196.5 ± 6.6 |
| J190343+055256 | 39.56 | −0.04 | −0.41 ± 0.04 | 513.4 ± 4.4 | 477.3 ± 14.3 |
| J190043+064546 | 40.01 | +1.03 | −0.84 ± 0.01 | 129.2 ± 0.4 | 127.4 ± 3.8 |
| J191725+044236 | 40.10 | −3.61 | −0.39 ± 0.01 | 211.2 ± 0.6 | 193.9 ± 5.8 |
| J192049+042052 | 40.17 | −4.52 | −1.01 ± 0.01 | 206.3 ± 0.5 | 216.6 ± 6.5 |
| J190734+060446 | 40.18 | −0.80 | −0.97 ± 0.01 | 283.8 ± 0.6 | 270.2 ± 8.1 |
| J191840+043932 | 40.20 | −3.91 | −1.08 ± 0.06 | 31.2 ± 0.3 | 29.8 ± 1.0 |
| J184731+090047 | 40.53 | +4.96 | −0.69 ± 0.01 | 173.9 ± 0.5 | 174.1 ± 5.7 |
| J192258+044354 | 40.76 | −4.82 | −0.49 ± 0.04 | 97.2 ± 0.7 | 104.8 ± 3.8 |

\* Spatially blended source (see Appendix A)

**Table B1.** (Continued) Total Intensities and Spectral Indices of On-axis Targets

| Target Source (NVSS) | $\ell$ (°) | $b$ (°) | $\alpha$ | $S_{1.4\,\mathrm{GHz}}$ (mJy) | $S_{\mathrm{NVSS}}$ (mJy) |
|---|---|---|---|---|---|
| J192243+045126 | 40.84 | −4.71 | −0.80 ± 0.01 | 929.7 ± 1.0 | 909.5 ± 27.3 |
| J191310+064158 | 41.37 | −1.75 | −0.79 ± 0.02 | 82.6 ± 0.4 | 80.5 ± 2.5 |
| J184951+094850 | 41.51 | +4.81 | −0.87 ± 0.02 | 99.9 ± 0.4 | 102.9 ± 3.7 |
| J191917+061942 | 41.75 | −3.27 | −0.28 ± 0.02 | 262.0 ± 1.1 | 262.3 ± 7.9 |
| J190614+084226 | 42.36 | +0.70 | −0.79 ± 0.06 | 45.8 ± 0.6 | 42.3 ± 1.4 |
| J185557+102011 | 42.66 | +3.70 | −0.81 ± 0.00 | 353.2 ± 0.3 | 353.6 ± 10.6 |
| J192233+071048 | 42.88 | −3.59 | −0.55 ± 0.01 | 129.9 ± 0.3 | 128.9 ± 3.9 |
| J190741+090717 | 42.90 | +0.57 | −0.60 ± 0.01 | 647.8 ± 1.7 | 566.8 ± 20.0 |
| J192245+073933 | 43.33 | −3.40 | −0.53 ± 0.02 | 188.6 ± 0.8 | 193.6 ± 6.8 |
| J192820+070355 | 43.46 | −4.91 | −0.84 ± 0.02 | 223.1 ± 0.7 | 245.0 ± 8.3 |
| J191906+081920 | 43.49 | −2.30 | −0.84 ± 0.02 | 84.5 ± 0.4 | 82.3 ± 2.5 |
| J185728+111021★ | 43.57 | +3.75 | −0.76 ± 0.01 | 221.9 ± 0.4 | 120.6 ± 4.3 |
| J191641+090147 | 43.84 | −1.44 | −1.01 ± 0.07 | 37.6 ± 0.5 | 32.8 ± 1.1 |
| J185952+112514 | 44.06 | +3.34 | −0.77 ± 0.03 | 120.3 ± 0.6 | 124.0 ± 4.3 |
| J190323+112905 | 44.51 | +2.60 | −1.16 ± 0.01 | 260.2 ± 0.4 | 283.6 ± 9.9 |
| J192840+084849 | 45.04 | −4.15 | −0.31 ± 0.00 | 428.0 ± 0.4 | 466.5 ± 14.0 |
| J192355+094424 | 45.31 | −2.68 | −0.91 ± 0.01 | 336.7 ± 0.5 | 343.9 ± 11.8 |
| J185923+125912 | 45.41 | +4.15 | −1.00 ± 0.00 | 2880.8 ± 1.5 | 2875.9 ± 86.3 |
| J191005+114748 | 45.54 | +1.28 | −0.88 ± 0.01 | 755.5 ± 1.3 | 752.1 ± 23.8 |
| J191000+122524 | 46.09 | +1.59 | −0.84 ± 0.01 | 208.7 ± 0.6 | 206.8 ± 7.2 |
| J192922+095808★ | 46.14 | −3.76 | −0.96 ± 0.01 | 1093.2 ± 1.2 | 1099.0 ± 35.3 |
| J191733+114215 | 46.31 | −0.38 | +0.81 ± 0.07 | 86.7 ± 1.5 | 76.0 ± 2.3 |
| J190501+132047 | 46.35 | +3.09 | −0.87 ± 0.01 | 252.7 ± 0.6 | 246.5 ± 8.6 |
| J193434+104340 | 47.43 | −4.52 | −0.06 ± 0.02 | 120.0 ± 0.4 | 149.8 ± 4.5 |
| J190247+145137 | 47.46 | +4.26 | −0.23 ± 0.01 | 195.3 ± 0.5 | 197.5 ± 7.6 |
| J193357+105642 | 47.54 | −4.28 | −0.85 ± 0.01 | 177.5 ± 0.4 | 173.4 ± 5.2 |
| J191025+140125★ | 47.56 | +2.24 | −0.91 ± 0.02 | 556.2 ± 1.9 | 249.3 ± 8.7 |
| J192540+122738 | 47.91 | −1.77 | +0.10 ± 0.01 | 180.7 ± 0.3 | 138.9 ± 4.2 |
| J190451+152148 | 48.13 | +4.05 | −0.82 ± 0.00 | 1096.2 ± 0.8 | 1084.4 ± 35.6 |
| J190414+153638 | 48.28 | +4.29 | −0.01 ± 0.01 | 288.1 ± 0.5 | 283.2 ± 8.5 |
| J192458+130033★ | 48.31 | −1.36 | −1.11 ± 0.07 | 295.5 ± 3.4 | 153.8 ± 5.5 |
| J190655+152342 | 48.39 | +3.62 | −0.56 ± 0.01 | 133.0 ± 0.3 | 125.9 ± 3.8 |
| J193335+120844 | 48.56 | −3.62 | −0.47 ± 0.03 | 65.3 ± 0.4 | 59.9 ± 1.8 |
| J190355+160147 | 48.62 | +4.55 | −0.74 ± 0.01 | 239.7 ± 0.4 | 245.8 ± 8.5 |
| J191644+150349 | 49.19 | +1.37 | −0.90 ± 0.01 | 442.4 ± 0.7 | 457.8 ± 16.1 |
| J192517+135919 | 49.21 | −0.97 | −0.85 ± 0.07 | 1174.0 ± 14.7 | 1044.3 ± 35.3 |
| J190516+163706 | 49.30 | +4.53 | −0.64 ± 0.01 | 560.0 ± 0.7 | 605.3 ± 20.3 |
| J193302+131335 | 49.44 | −2.98 | −0.67 ± 0.01 | 1043.3 ± 1.4 | 1068.5 ± 36.8 |
| J191133+161431 | 49.65 | +3.02 | −1.11 ± 0.00 | 610.9 ± 0.5 | 592.3 ± 17.8 |
| J191158+161147 | 49.66 | +2.91 | −0.23 ± 0.00 | 486.3 ± 0.4 | 394.9 ± 11.9 |
| J190901+163944 | 49.75 | +3.75 | −0.75 ± 0.00 | 521.5 ± 0.5 | 516.8 ± 18.2 |
| J191219+161628 | 49.77 | +2.87 | −0.94 ± 0.01 | 469.3 ± 0.5 | 472.0 ± 16.7 |
| J192835+142156 | 49.92 | −1.49 | −2.39 ± 0.14 | 37.6 ± 0.8 | 30.8 ± 1.4 |
| J192910+141952 | 49.96 | −1.63 | −0.87 ± 0.03 | 82.6 ± 0.6 | 93.8 ± 3.4 |
| J191649+155836 | 50.00 | +1.77 | −0.68 ± 0.01 | 321.9 ± 0.6 | 326.7 ± 10.9 |
| J191549+160834 | 50.04 | +2.06 | −0.93 ± 0.01 | 128.5 ± 0.4 | 127.2 ± 3.8 |
| J194012+125809 | 50.06 | −4.64 | −0.79 ± 0.01 | 178.4 ± 0.5 | 172.9 ± 6.0 |
| J191414+163640 | 50.28 | +2.62 | −0.39 ± 0.01 | 212.0 ± 0.4 | 218.0 ± 6.6 |
| J192439+154043 | 50.63 | −0.03 | +0.06 ± 0.01 | 432.8 ± 1.3 | 364.4 ± 10.9 |
| J193939+134604★ | 50.70 | −4.13 | −0.37 ± 0.02 | 111.5 ± 0.5 | 81.3 ± 2.5 |
| J192032+162557 | 50.82 | +1.20 | −0.74 ± 0.03 | 75.3 ± 0.5 | 80.4 ± 3.0 |
| J192203+162243 | 50.95 | +0.85 | −1.09 ± 0.03 | 289.9 ± 0.8 | 279.4 ± 9.8 |
| J193306+145624 | 50.95 | −2.17 | −0.80 ± 0.01 | 416.1 ± 1.0 | 400.8 ± 12.0 |
| J193321+150446 | 51.10 | −2.16 | −0.20 ± 0.01 | 394.3 ± 0.7 | 499.5 ± 15.0 |
| J193052+153235 | 51.22 | −1.41 | −0.26 ± 0.01 | 686.0 ± 0.8 | 640.8 ± 19.2 |

★ Spatially blended source (see Appendix A)

**Table B2.** Total Intensities and Spectral Indices of Off-axis Targets

| Target Source (NVSS) | $\ell$ (°) | $b$ (°) | $\alpha$ | $S_{1.4\,\mathrm{GHz}}$ (mJy) | $S_{\mathrm{NVSS}}$ (mJy) |
|---|---|---|---|---|---|
| J183756−112202 | 21.28 | −2.21 | −0.32 ± 0.13 | 21.6 ± 0.6 | 24.2 ± 1.2 |
| J184555−115813 | 21.64 | −4.22 | −0.68 ± 0.06 | 28.4 ± 0.3 | 31.5 ± 1.4 |
| J182535−083948 | 22.28 | +1.74 | +0.04 ± 0.10 | 19.8 ± 0.4 | 26.5 ± 0.9 |
| J183931−101336 | 22.48 | −2.03 | −0.27 ± 0.03 | 75.9 ± 0.5 | 72.0 ± 2.7 |
| J184906−111430 | 22.64 | −4.59 | −1.33 ± 0.03 | 47.5 ± 0.3 | 44.4 ± 1.4 |
| J184808−105535 | 22.82 | −4.23 | −1.32 ± 0.05 | 24.0 ± 0.2 | 22.8 ± 0.9 |
| J184541−093643 | 23.72 | −3.10 | −1.03 ± 0.01 | 160.4 ± 0.3 | 26.3 ± 0.9 |
| J185027−091037 | 24.64 | −3.96 | −0.72 ± 0.01 | 368.2 ± 0.7 | 372.5 ± 13.0 |
| J182058−050223 | 24.95 | +4.44 | −0.66 ± 0.01 | 280.5 ± 0.7 | 279.3 ± 9.8 |
| J184617−081126 | 25.05 | −2.59 | −0.24 ± 0.04 | 32.5 ± 0.3 | 39.9 ± 1.7 |
| J182644−030952 | 27.29 | +4.04 | −0.33 ± 0.06 | 41.7 ± 0.5 | 41.4 ± 1.3 |
| J183414−030119 | 28.28 | +2.44 | +0.02 ± 0.01 | 255.3 ± 0.5 | 266.4 ± 9.4 |
| J183701−015140 | 29.63 | +2.36 | +0.36 ± 0.03 | 144.8 ± 0.9 | 139.8 ± 4.2 |
| J183827−013111 | 30.10 | +2.19 | −0.28 ± 0.12 | 33.2 ± 0.9 | 23.7 ± 0.8 |
| J183603−005747 | 30.32 | +2.98 | +0.04 ± 0.01 | 110.1 ± 0.4 | 96.6 ± 2.9 |
| J183415+004451 | 31.64 | +4.16 | −0.25 ± 0.03 | 62.4 ± 0.4 | 67.7 ± 2.1 |
| J183935+001547 | 31.81 | +2.76 | −0.92 ± 0.02 | 90.0 ± 0.3 | 84.7 ± 2.6 |
| J183433+010127 | 31.92 | +4.22 | −1.57 ± 0.05 | 43.6 ± 0.4 | 39.0 ± 1.6 |
| J185807−004834 | 32.97 | −1.86 | +0.73 ± 0.04 | 83.1 ± 1.8 | 32.5 ± 1.5 |
| J190832−005319 | 34.09 | −4.21 | −0.15 ± 0.08 | 18.7 ± 0.3 | 35.8 ± 1.8 |
| J190721+012341 | 35.99 | −2.90 | −0.24 ± 0.08 | 46.1 ± 0.6 | 42.6 ± 1.7 |
| J185222+033347 | 36.21 | +1.42 | +0.02 ± 0.09 | 28.5 ± 0.6 | 34.4 ± 1.5 |
| J191833+043928 | 40.18 | −3.88 | −1.01 ± 0.03 | 39.4 ± 0.3 | 38.1 ± 1.6 |
| J190616+083858 | 42.31 | +0.67 | +0.60 ± 0.05 | 62.6 ± 0.8 | 52.5 ± 2.1 |
| J192802+070219 | 43.40 | −4.85 | −0.87 ± 0.04 | 59.7 ± 0.5 | 65.9 ± 2.0 |
| J191630+090223 | 43.83 | −1.39 | −0.96 ± 0.05 | 59.8 ± 0.6 | 54.2 ± 1.7 |
| J190319+112950 | 44.51 | +2.62 | −1.11 ± 0.01 | 246.1 ± 0.4 | 264.2 ± 9.2 |
| J190235+145023 | 47.41 | +4.30 | −0.74 ± 0.02 | 86.6 ± 0.4 | 93.1 ± 3.2 |
| J190653+152650 | 48.43 | +3.65 | −0.80 ± 0.02 | 80.6 ± 0.2 | 77.4 ± 2.4 |
| J193328+120953 | 48.56 | −3.59 | −0.70 ± 0.06 | 24.1 ± 0.3 | 25.9 ± 0.9 |
| J192030+162333 | 50.78 | +1.19 | +0.66 ± 0.11 | 21.7 ± 0.6 | 24.4 ± 0.9 |
| J192032+162429 | 50.80 | +1.19 | −0.08 ± 0.08 | 29.7 ± 0.5 | 31.3 ± 1.0 |
| J192157+162501 | 50.97 | +0.89 | +0.24 ± 0.01 | 204.8 ± 0.7 | 152.8 ± 4.6 |

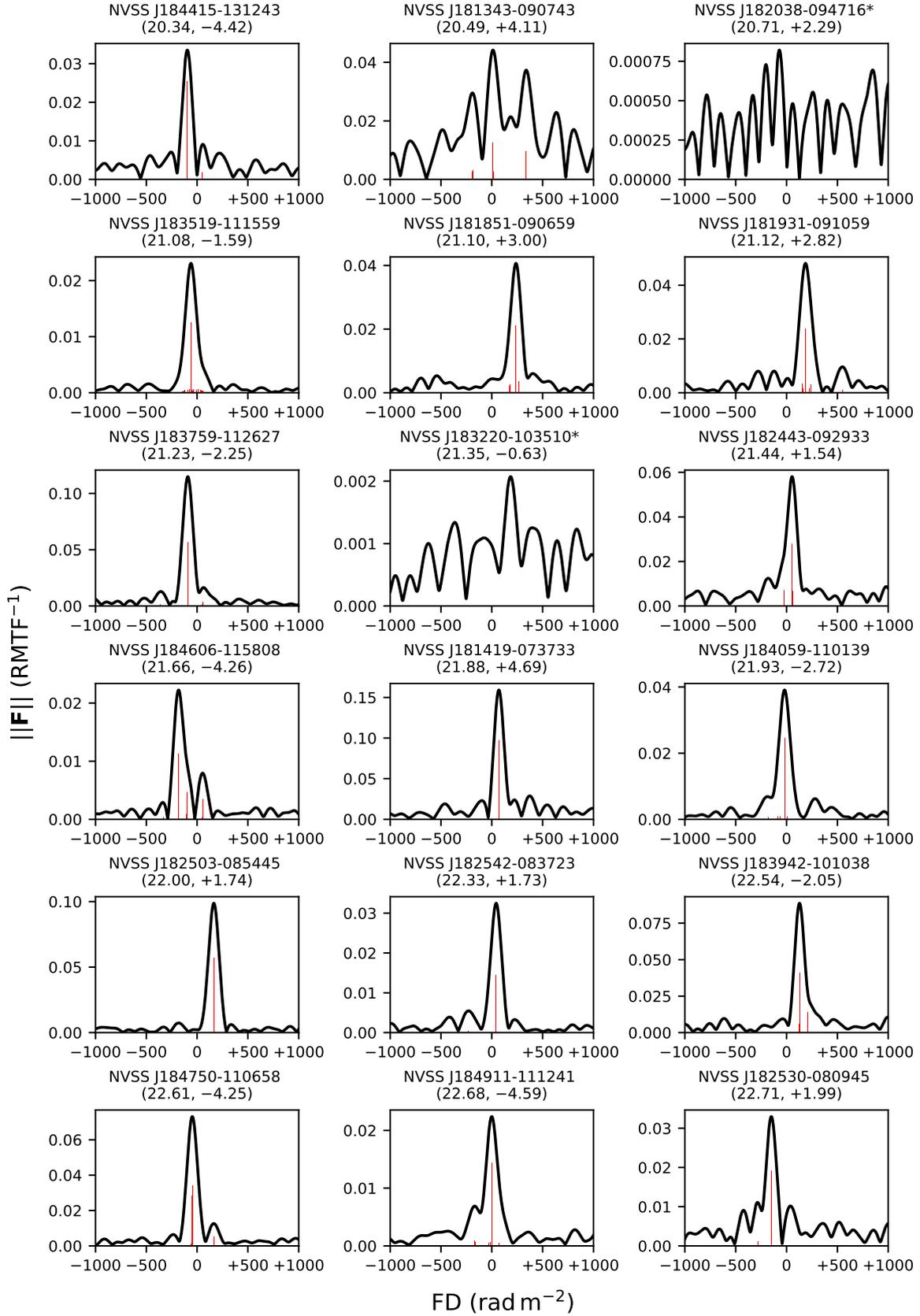

**Figure C1.** Faraday spectra of the on-axis sources, with the science targets sorted by Galactic longitude and followed by the on-axis polarisation leakage calibrator J1407+2827. The Galactic coordinates of each science target source are shown in the title of each panel in the format of $(\ell, b)$. Unpolarised sources are marked by an asterisk (*) trailing the source name. The black lines show the amplitudes of the deconvolved Faraday spectra, with the red vertical bars showing the clean components.

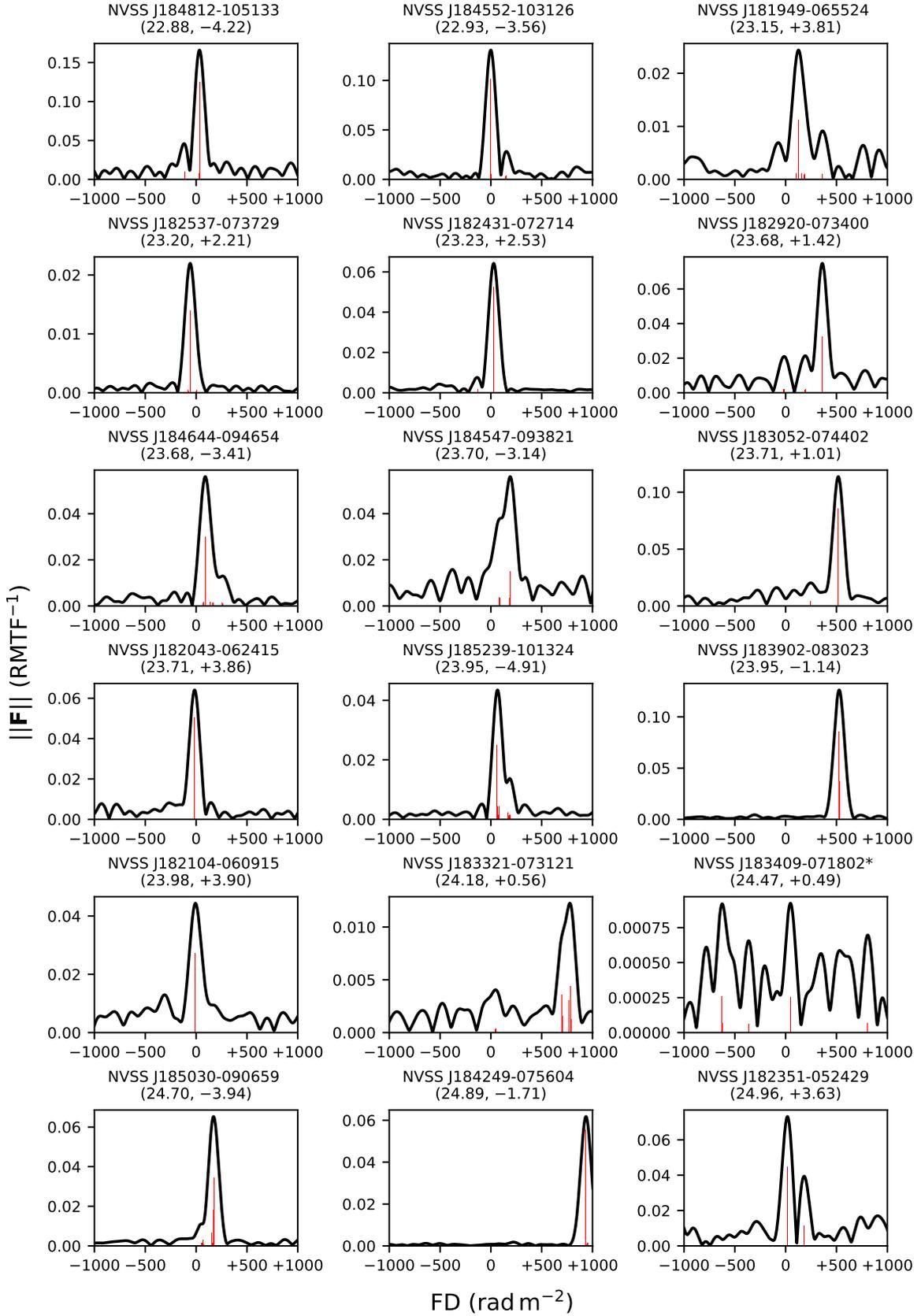

**Figure C1.** (Continued) Faraday spectra of the on-axis sources.

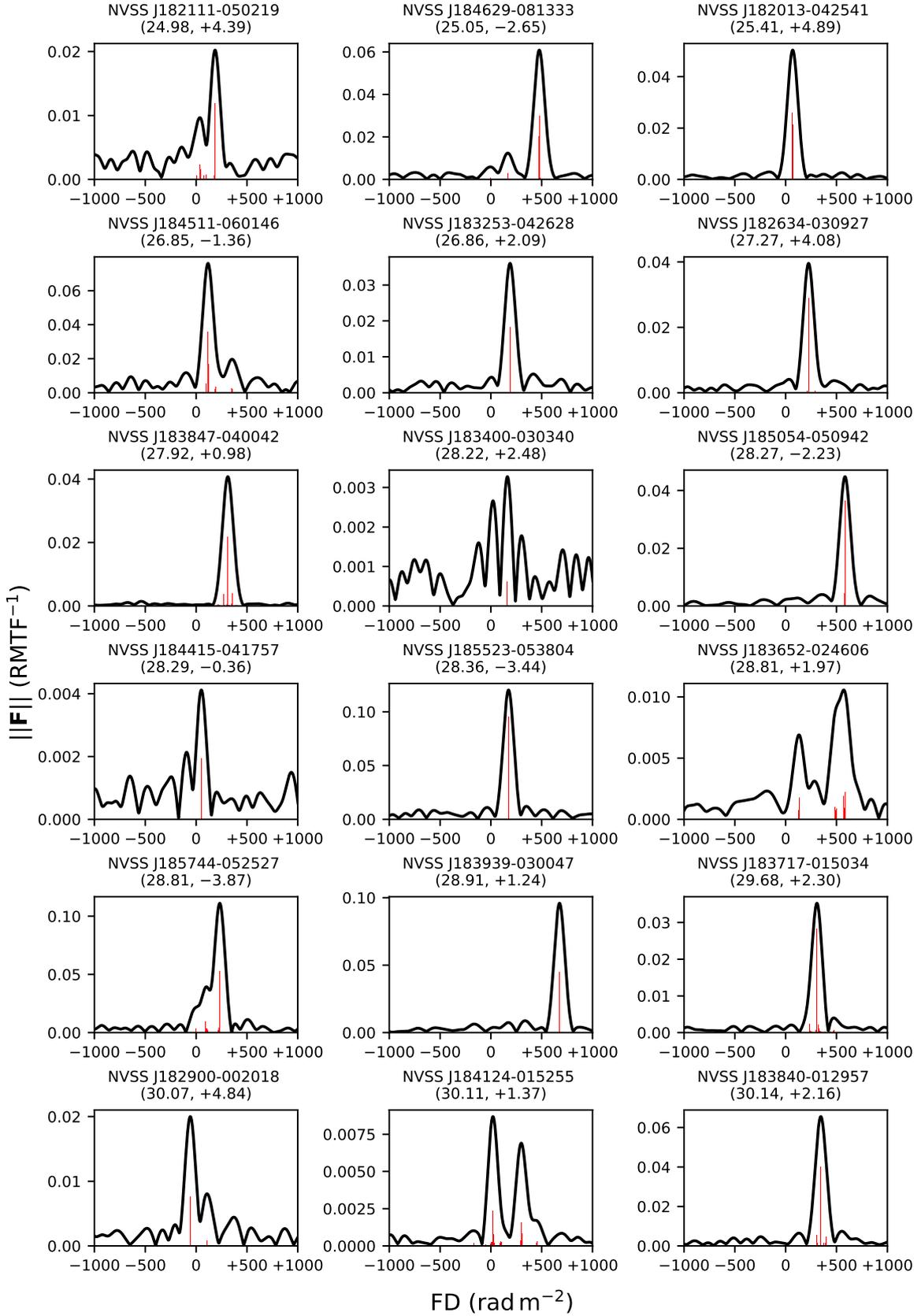

**Figure C1.** (Continued) Faraday spectra of the on-axis sources.

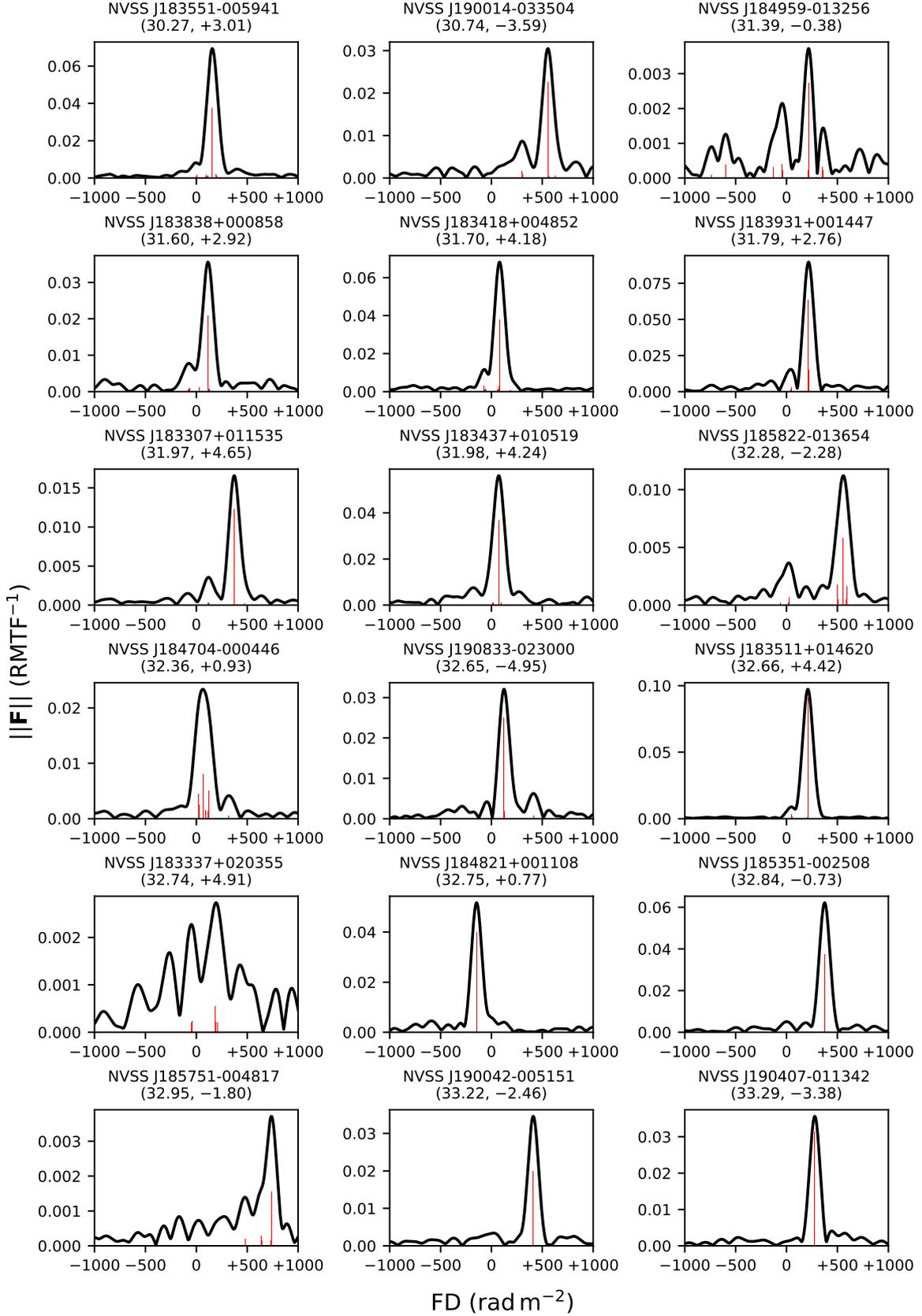

**Figure C1.** (Continued) Faraday spectra of the on-axis sources.

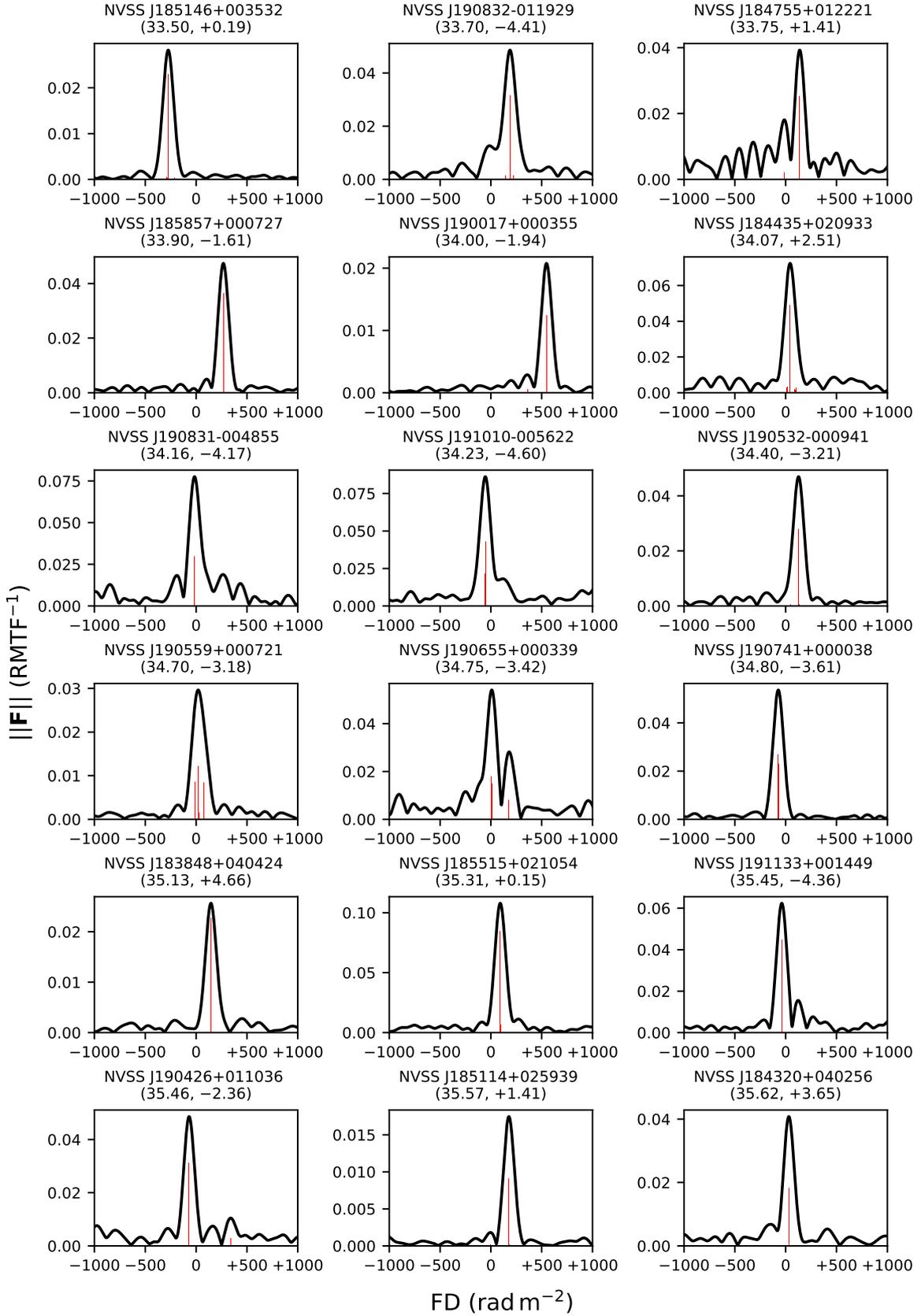

**Figure C1.** (Continued) Faraday spectra of the on-axis sources.

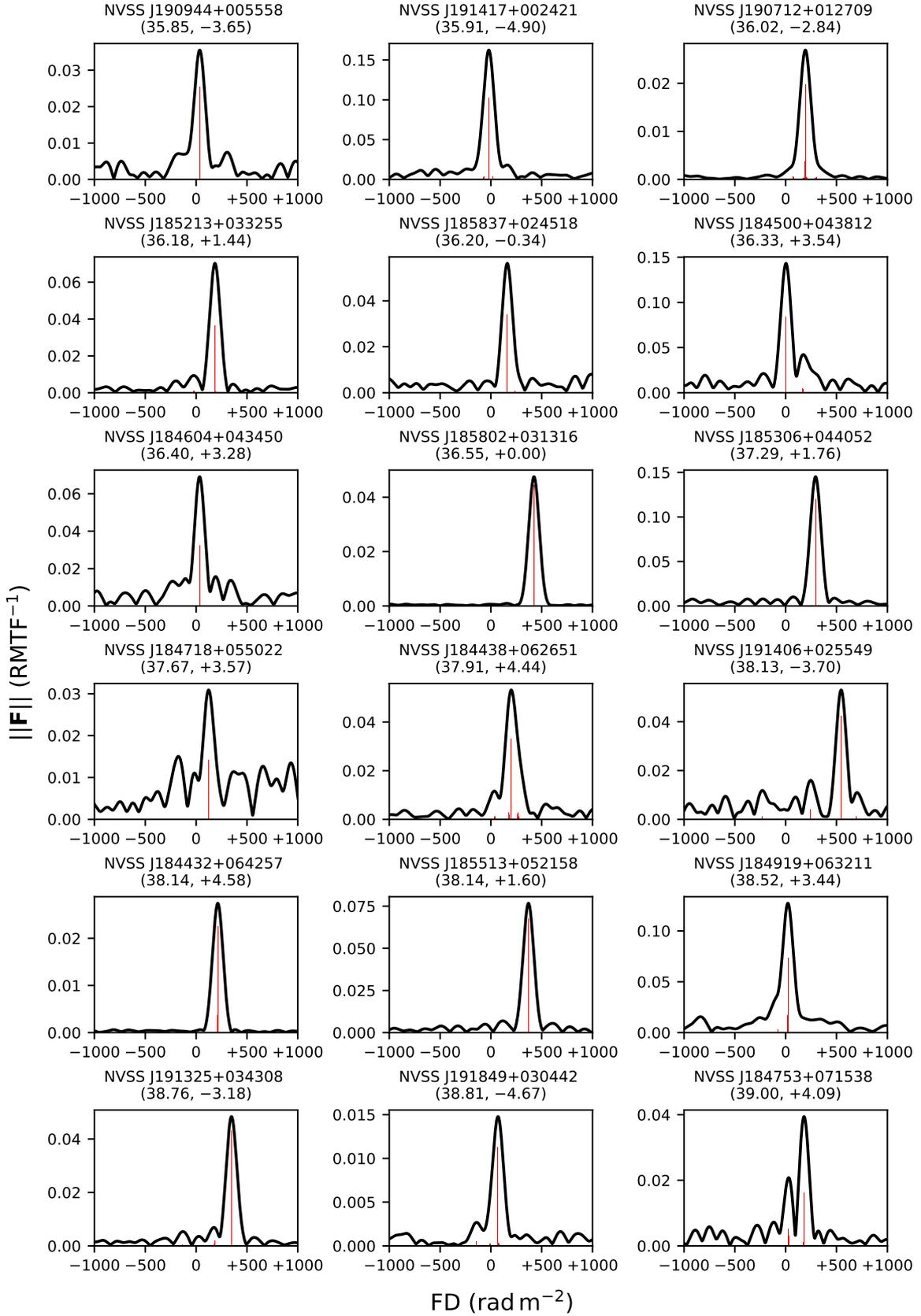

**Figure C1.** (Continued) Faraday spectra of the on-axis sources.

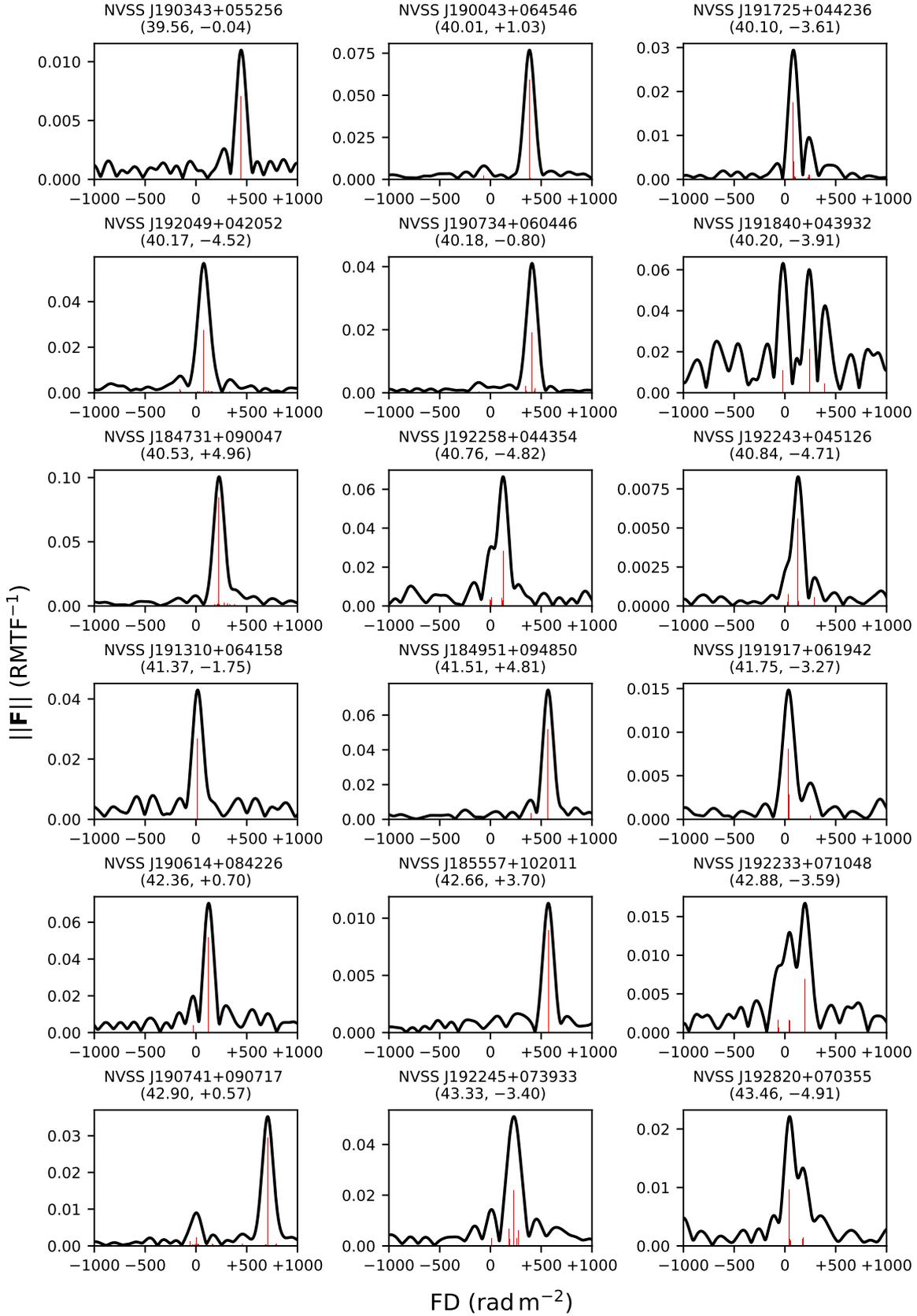

**Figure C1.** (Continued) Faraday spectra of the on-axis sources.

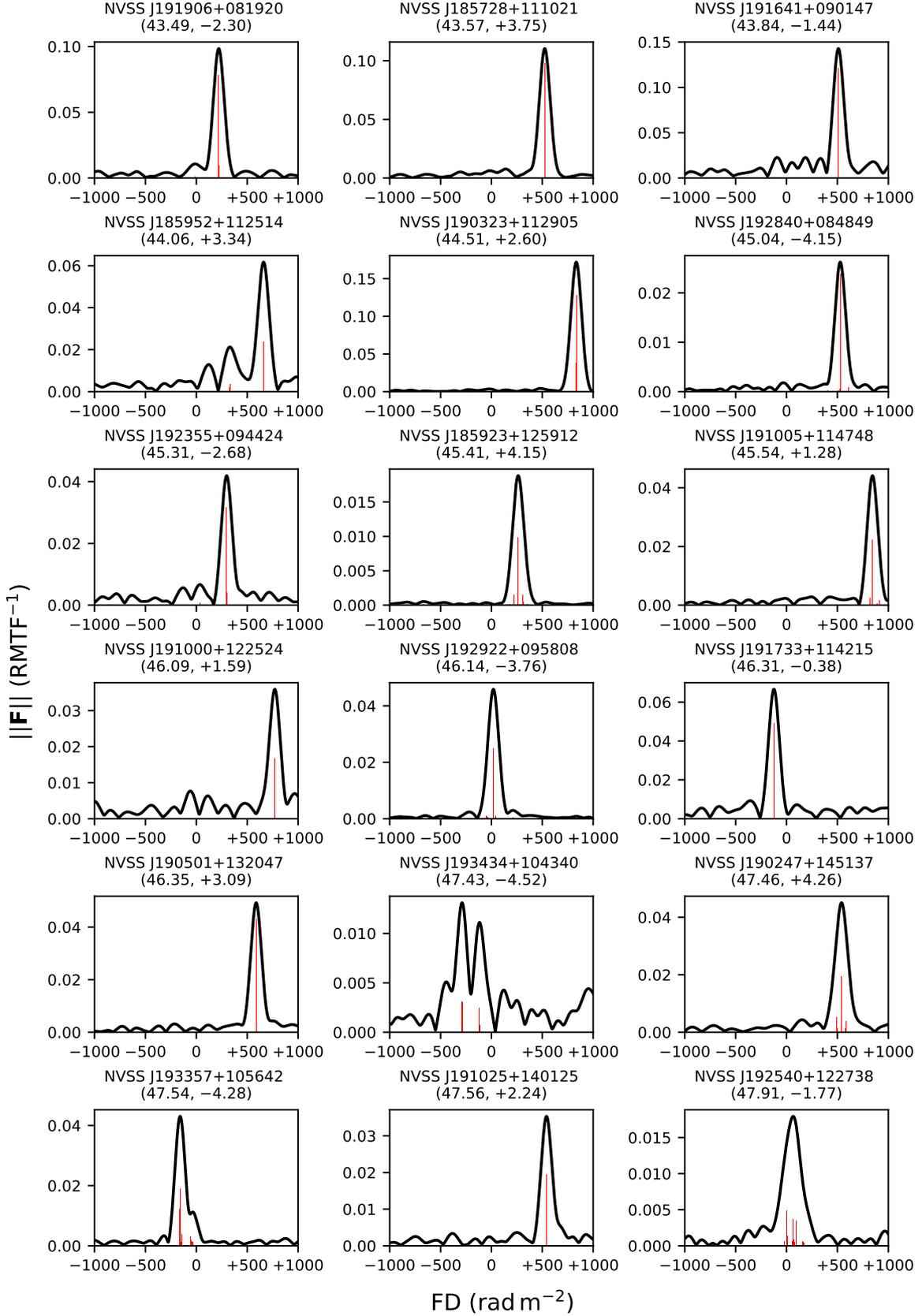

**Figure C1.** (Continued) Faraday spectra of the on-axis sources.

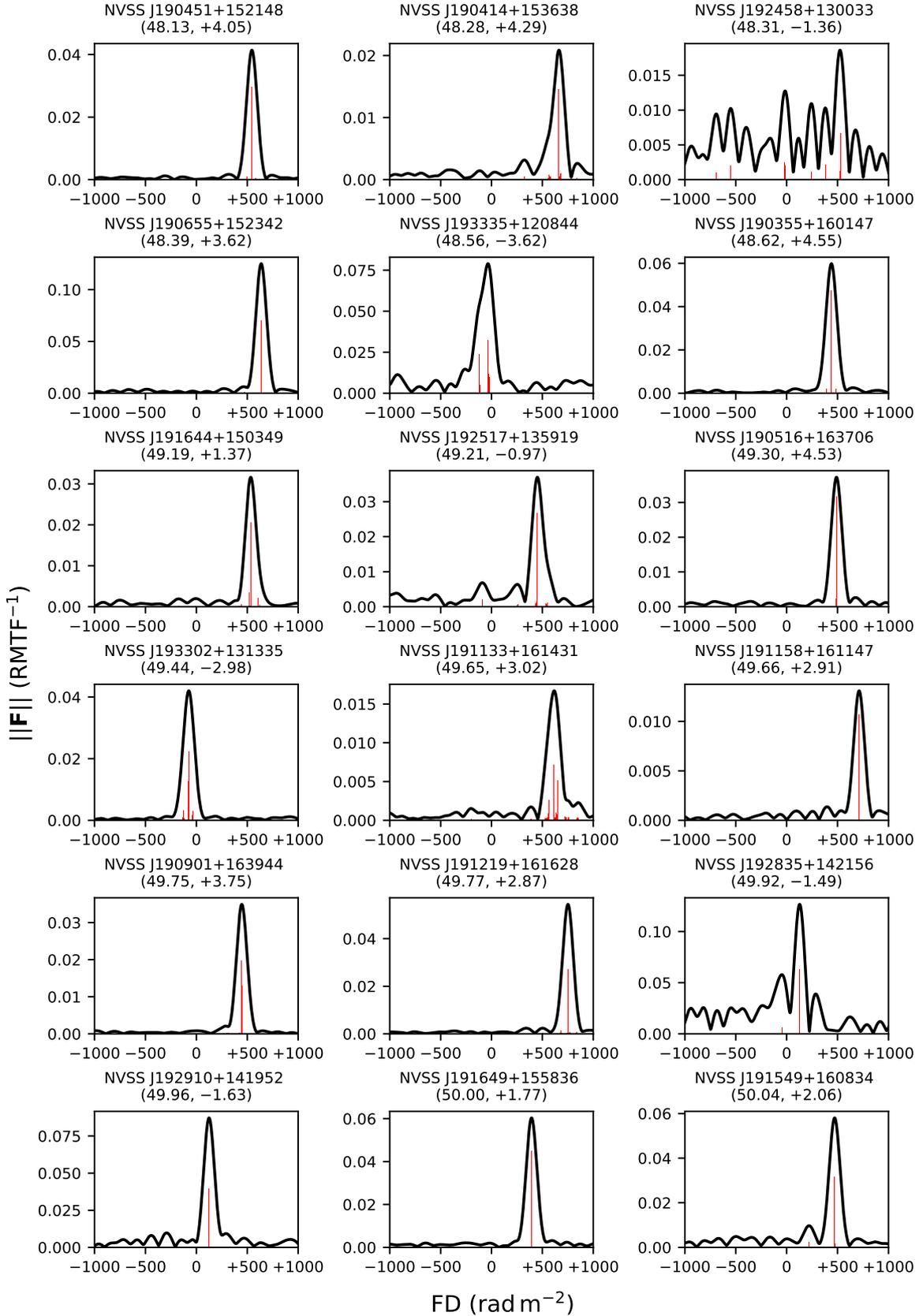

**Figure C1.** (Continued) Faraday spectra of the on-axis sources.

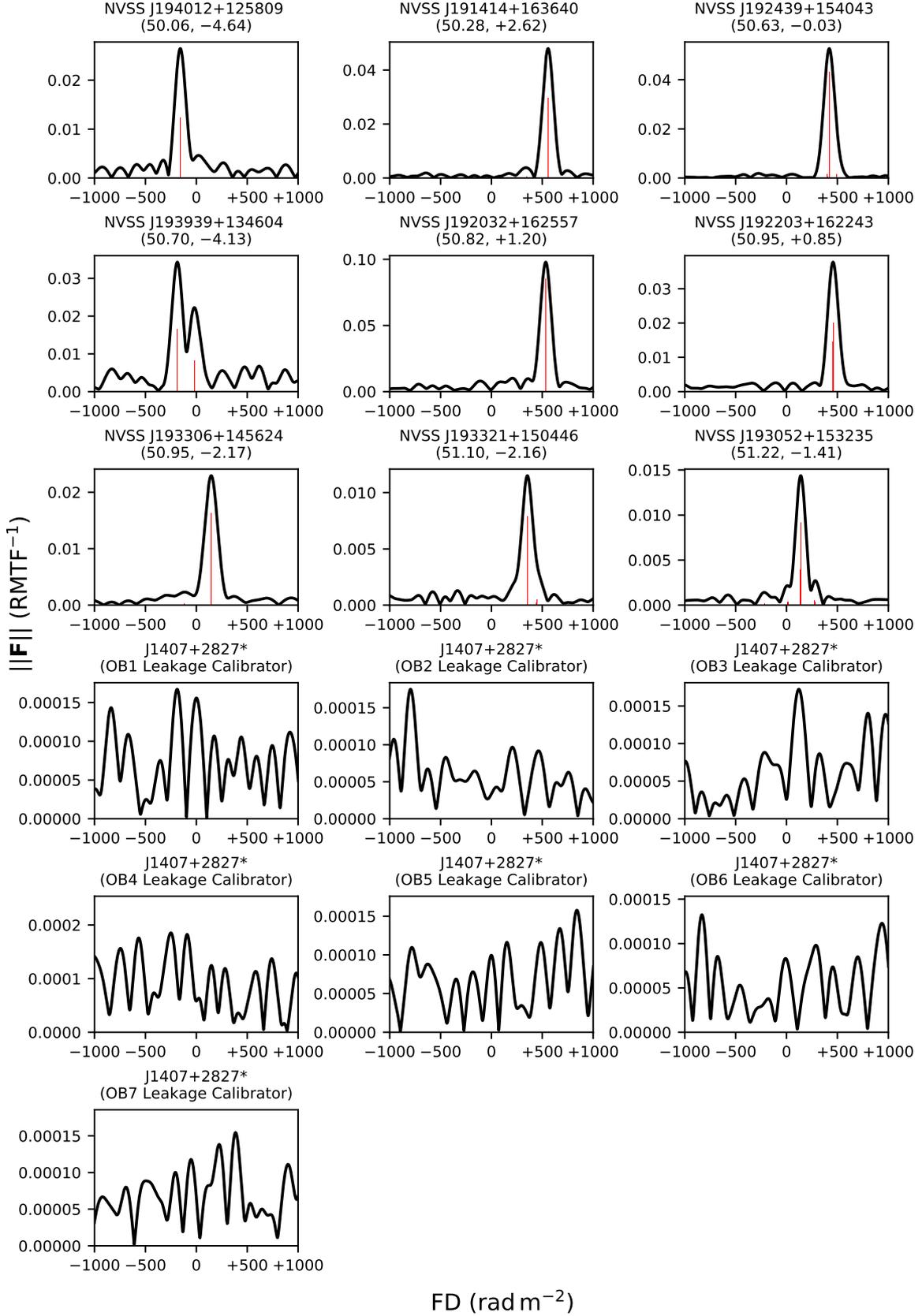

**Figure C1.** (Continued) Faraday spectra of the on-axis sources.

MNRAS, 1–23 (2020)

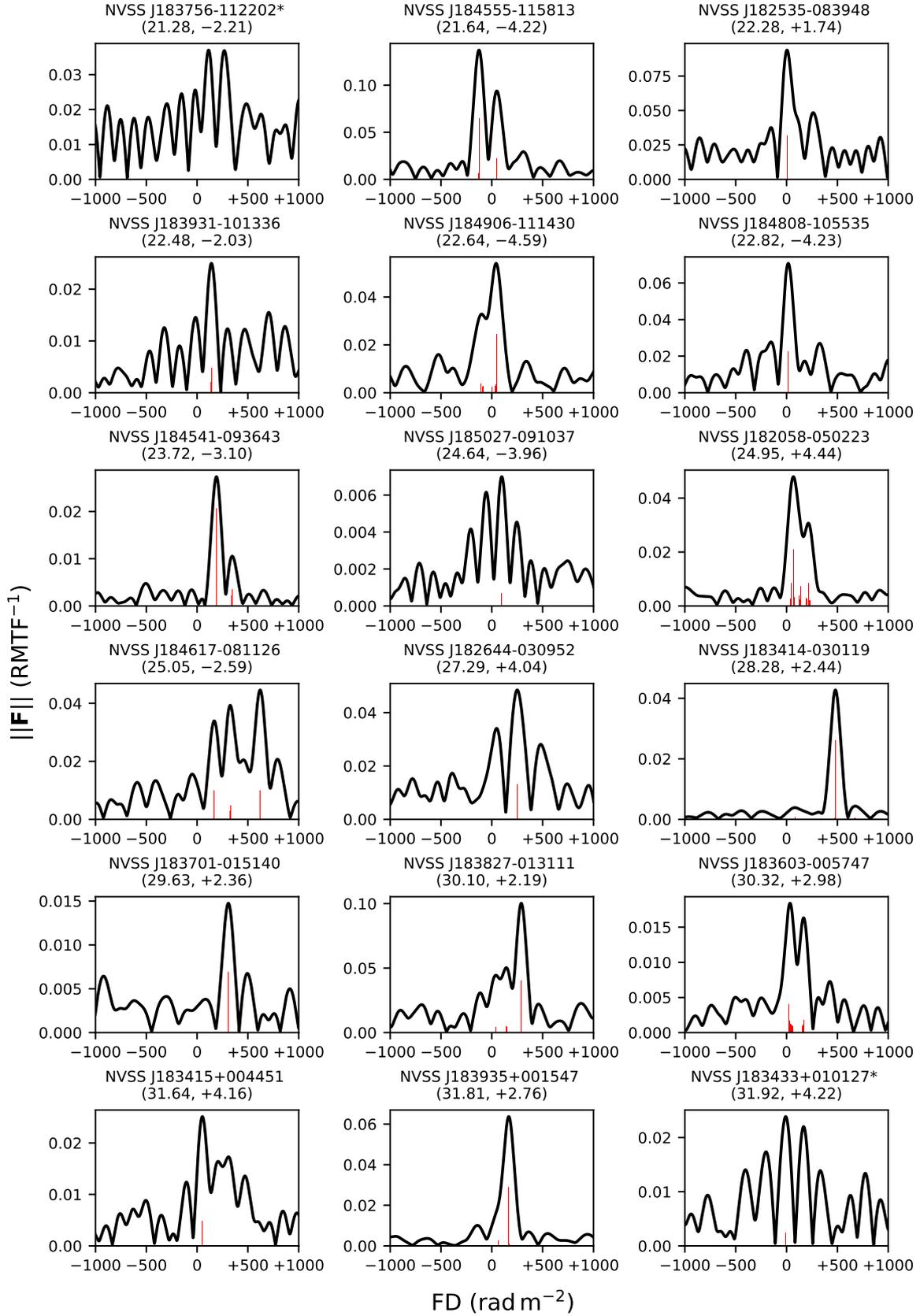

**Figure C2.** Faraday spectra of the off-axis sources, sorted by Galactic longitude. The Galactic coordinates of each source are shown in the title of each panel in the format of ($\ell$, $b$). Unpolarised sources are marked by an asterisk (*) trailing the source name. The black lines show the amplitudes of the deconvolved Faraday spectra, with the red vertical bars showing the clean components.

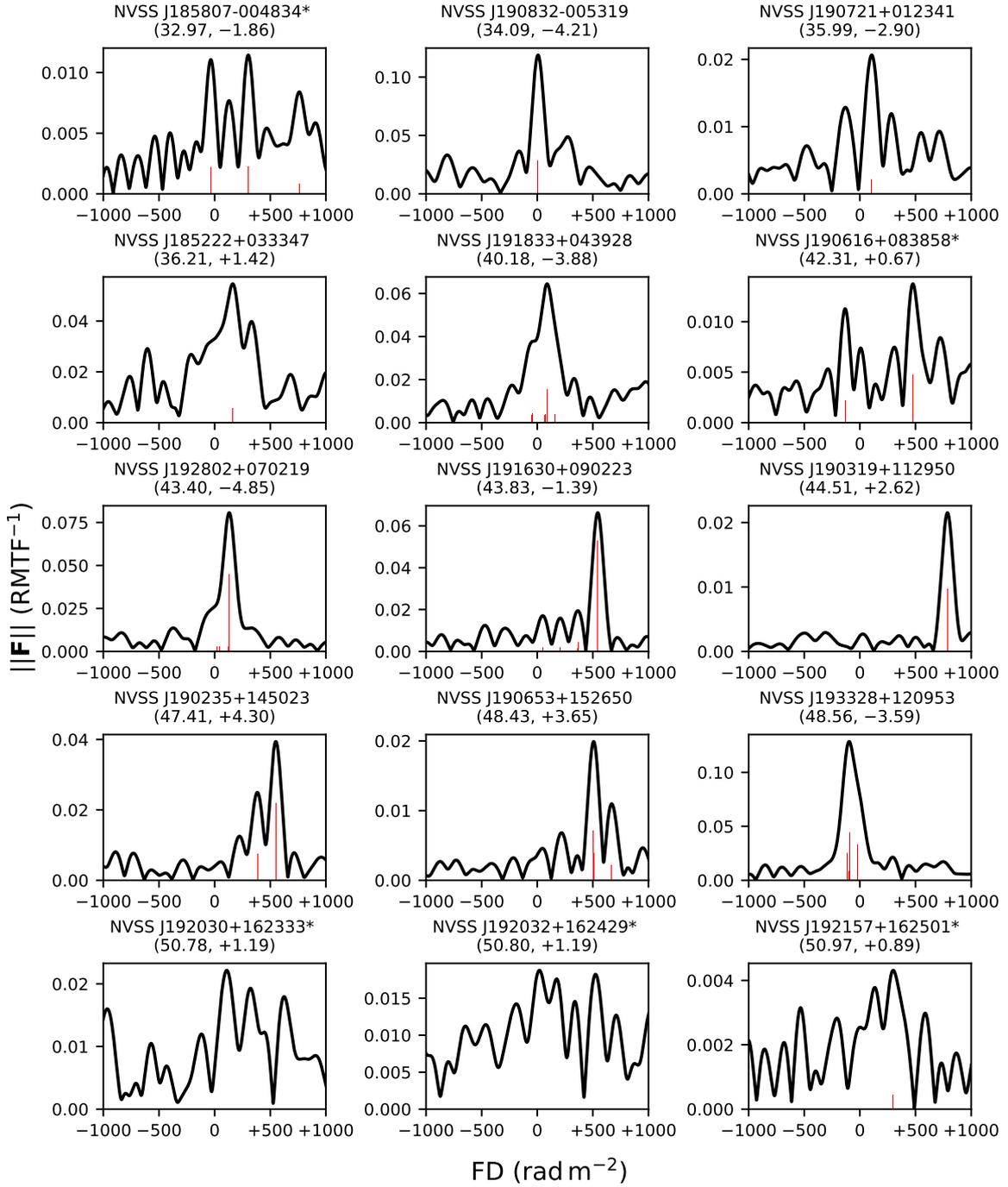

**Figure C2.** (Continued) Faraday spectra of the off-axis sources.

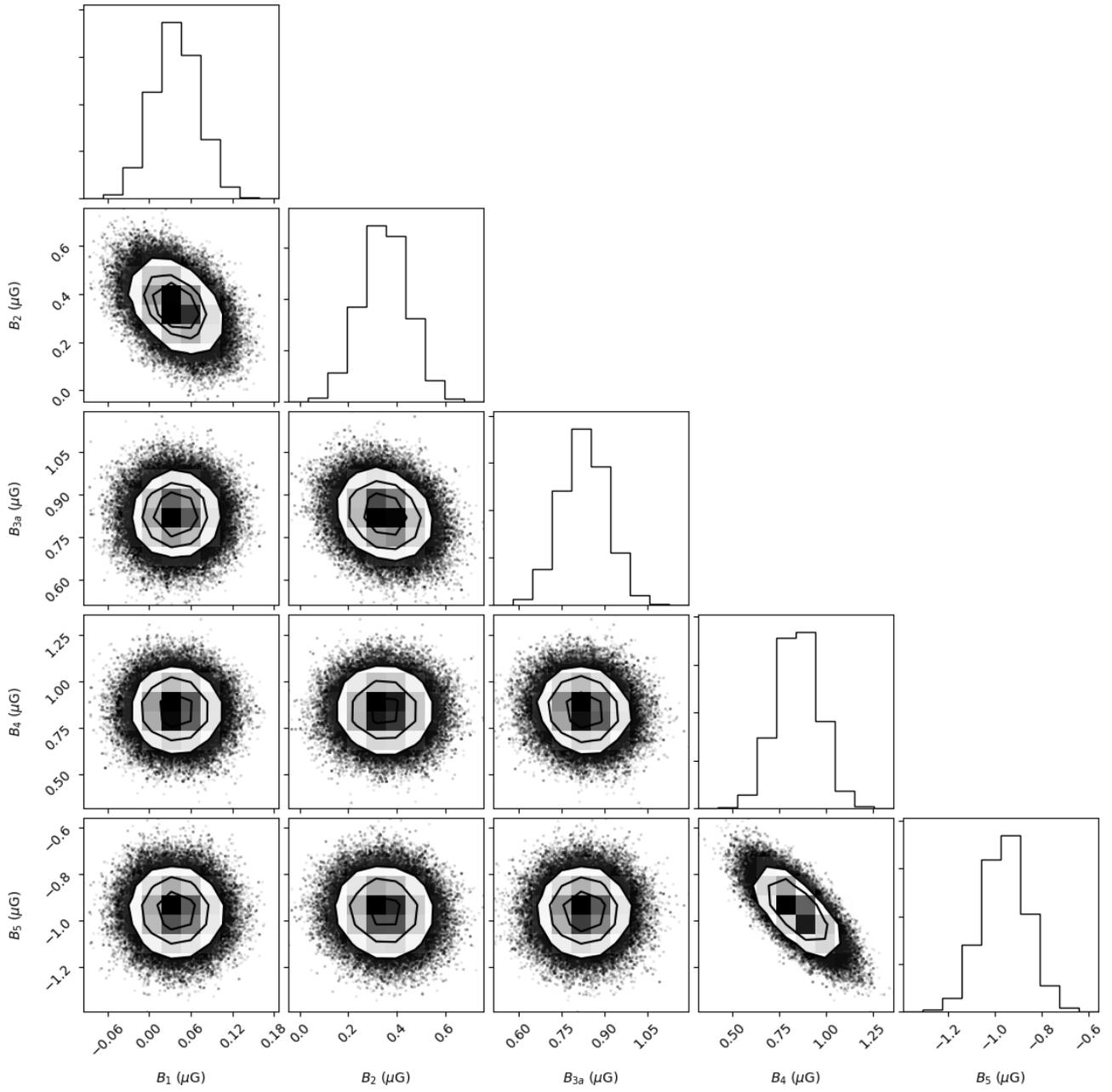

**Figure D1.** MCMC corner plot for the model Odd 3.

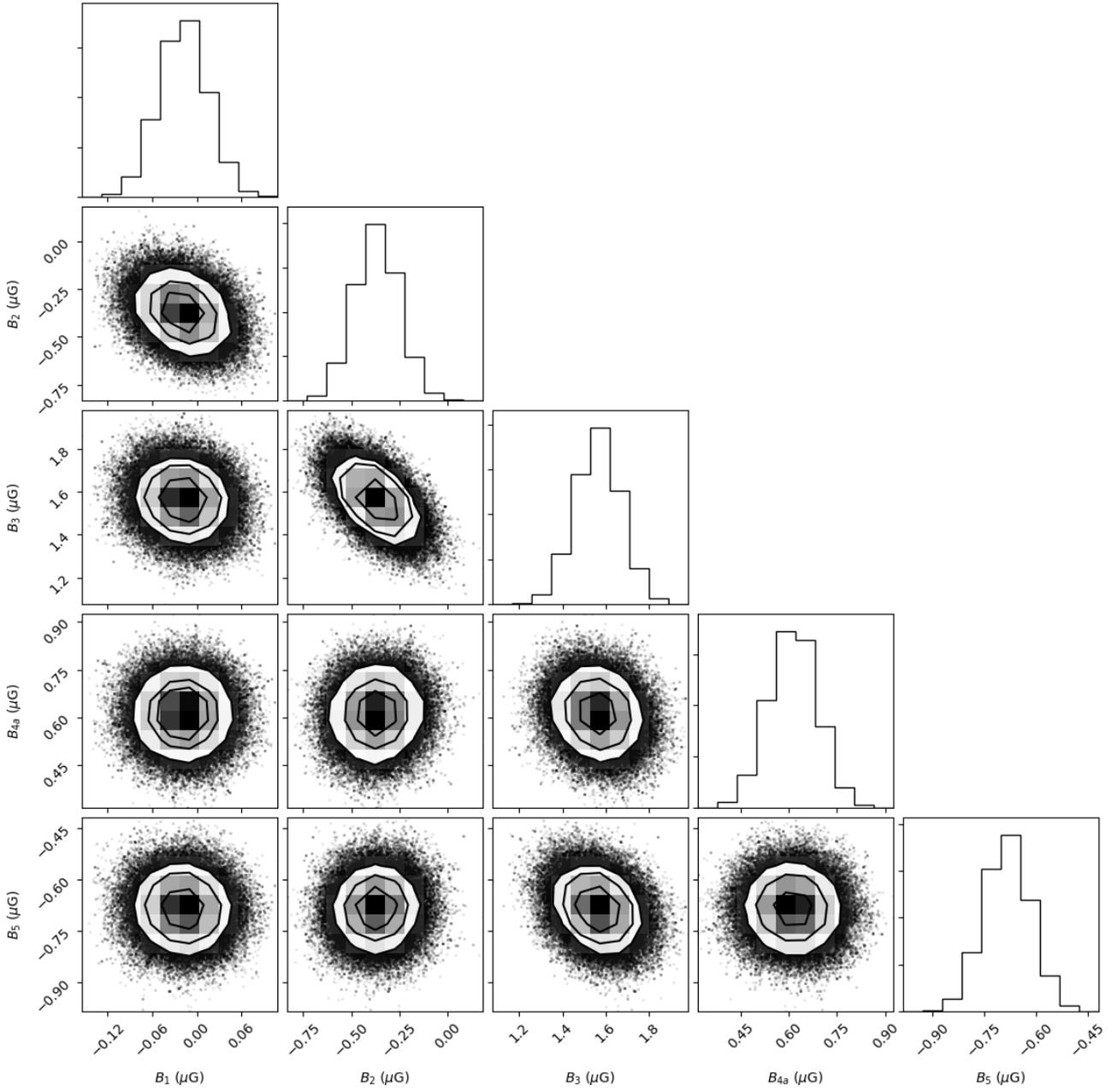

**Figure D2.** MCMC corner plot for the model Odd 4.

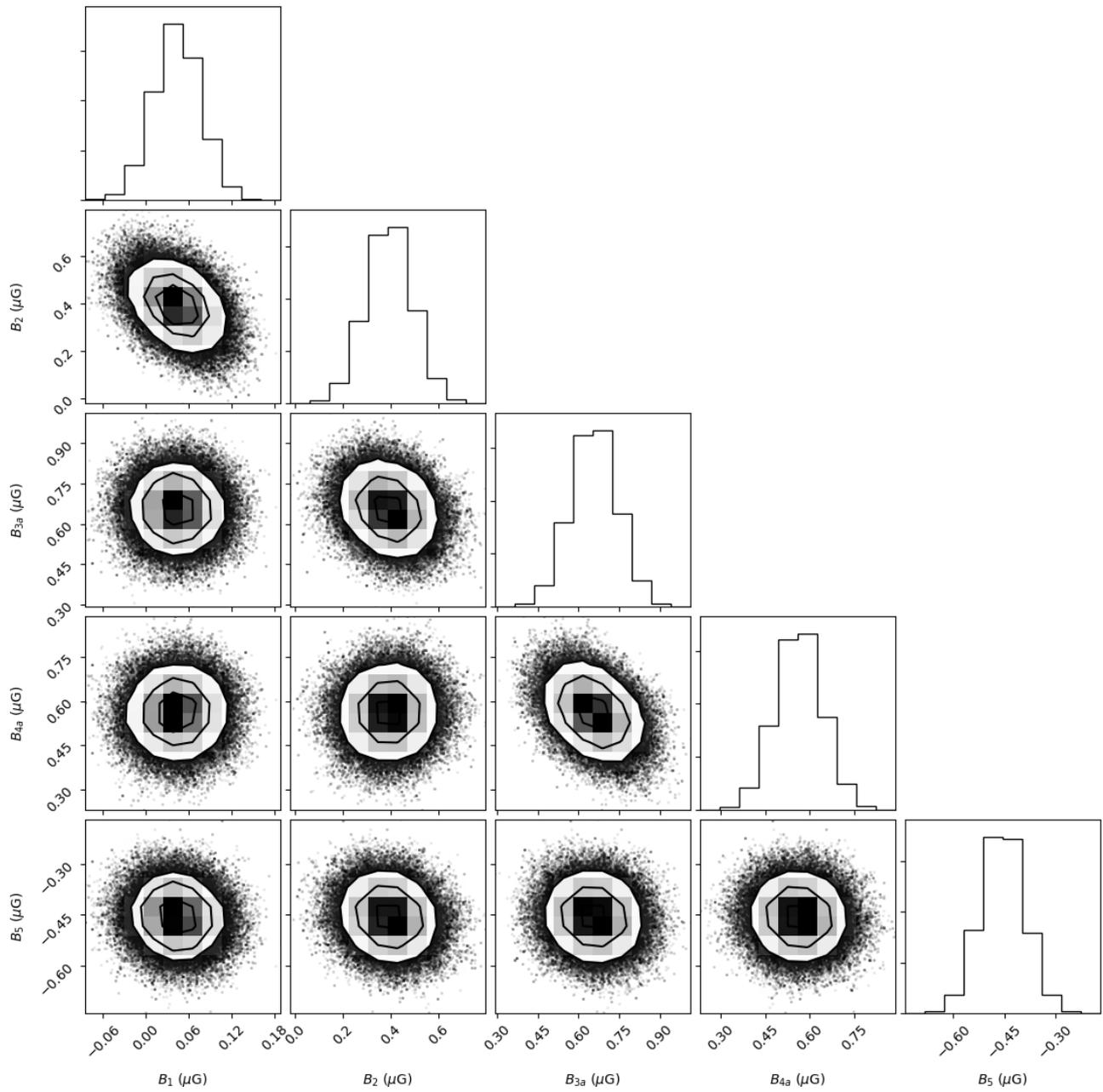

**Figure D3.** MCMC corner plot for the model Odd 3+4.

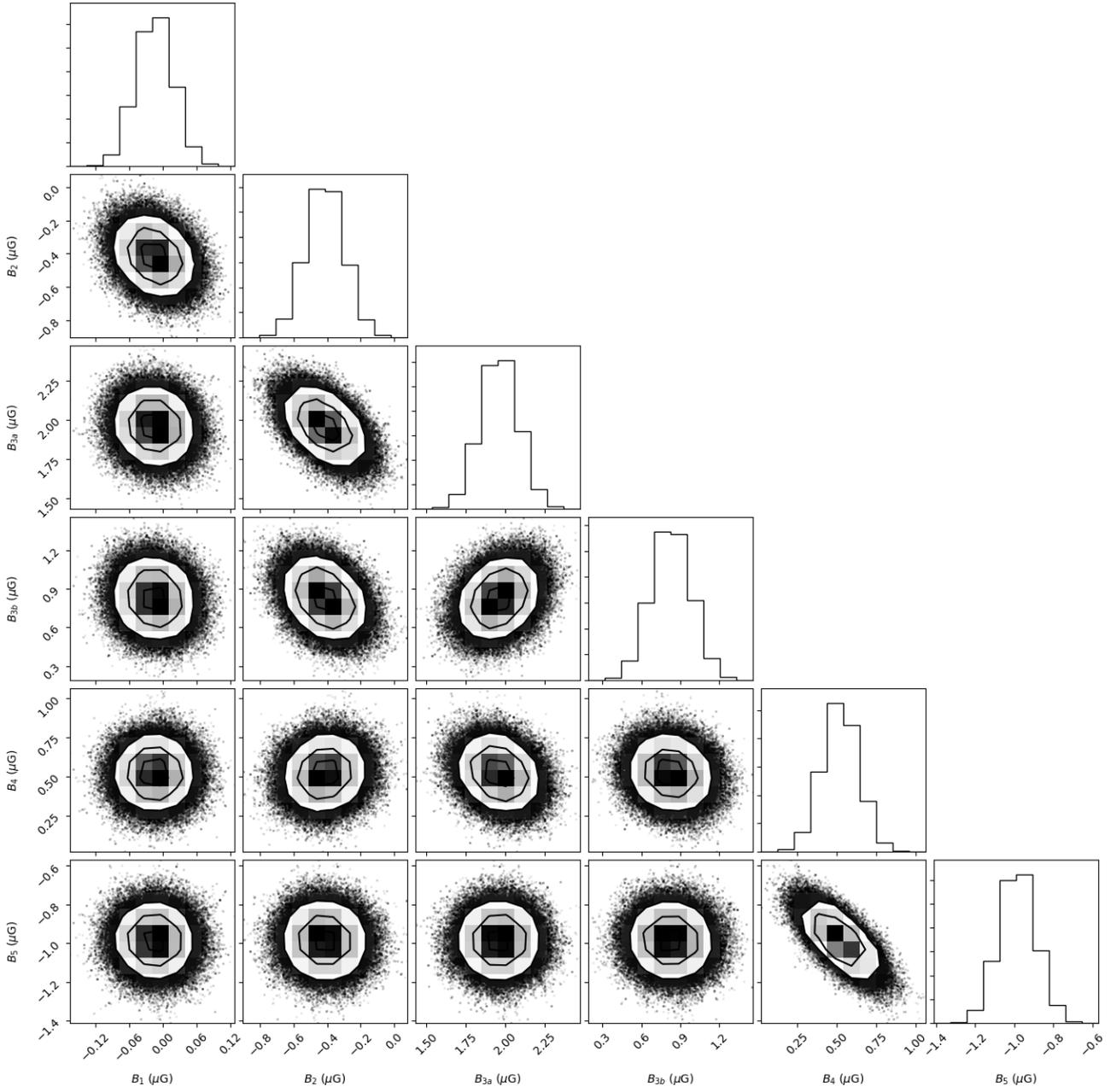

**Figure D4.** MCMC corner plot for the model Free 3.

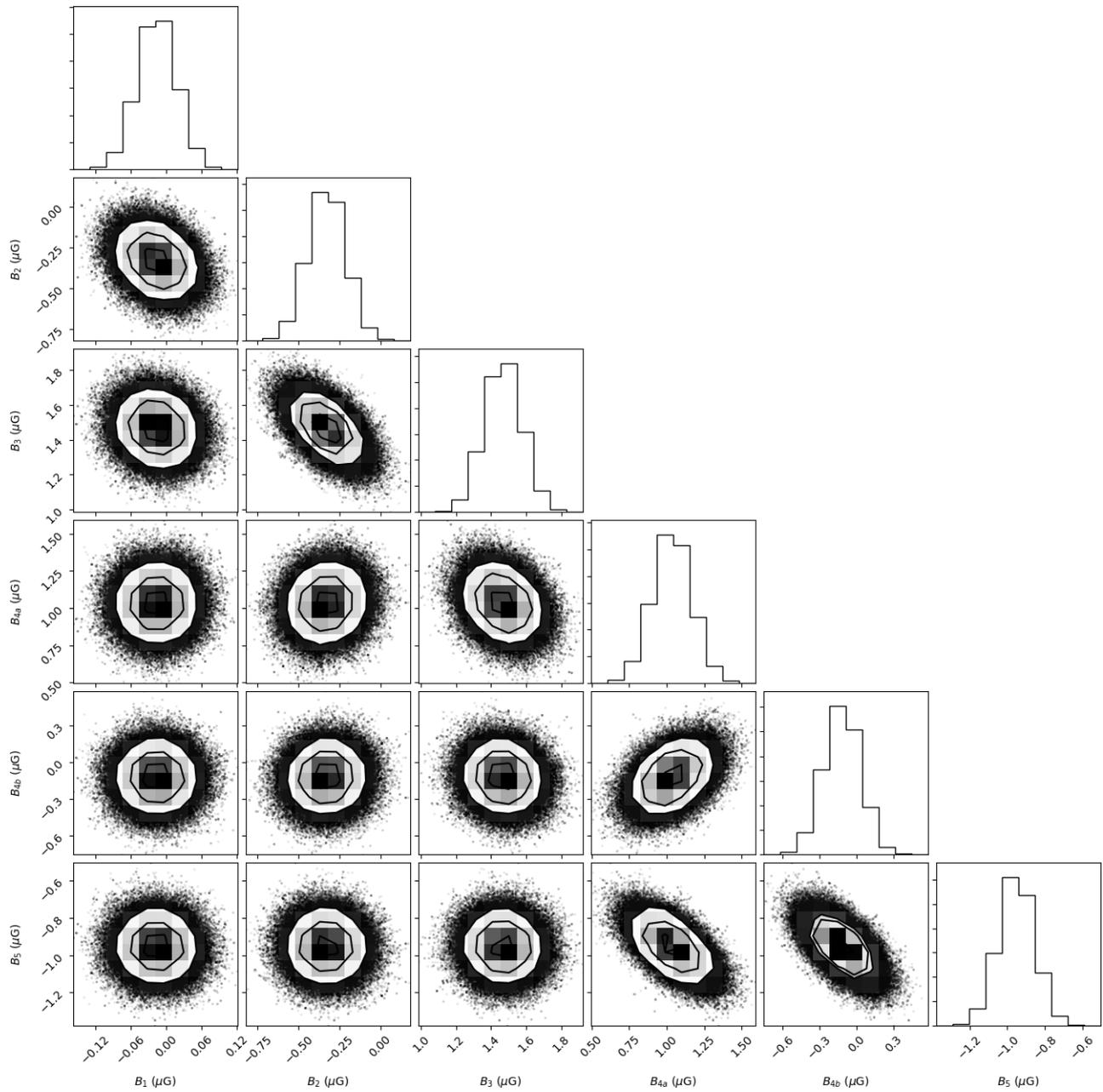

**Figure D5.** MCMC corner plot for the model Free 4.

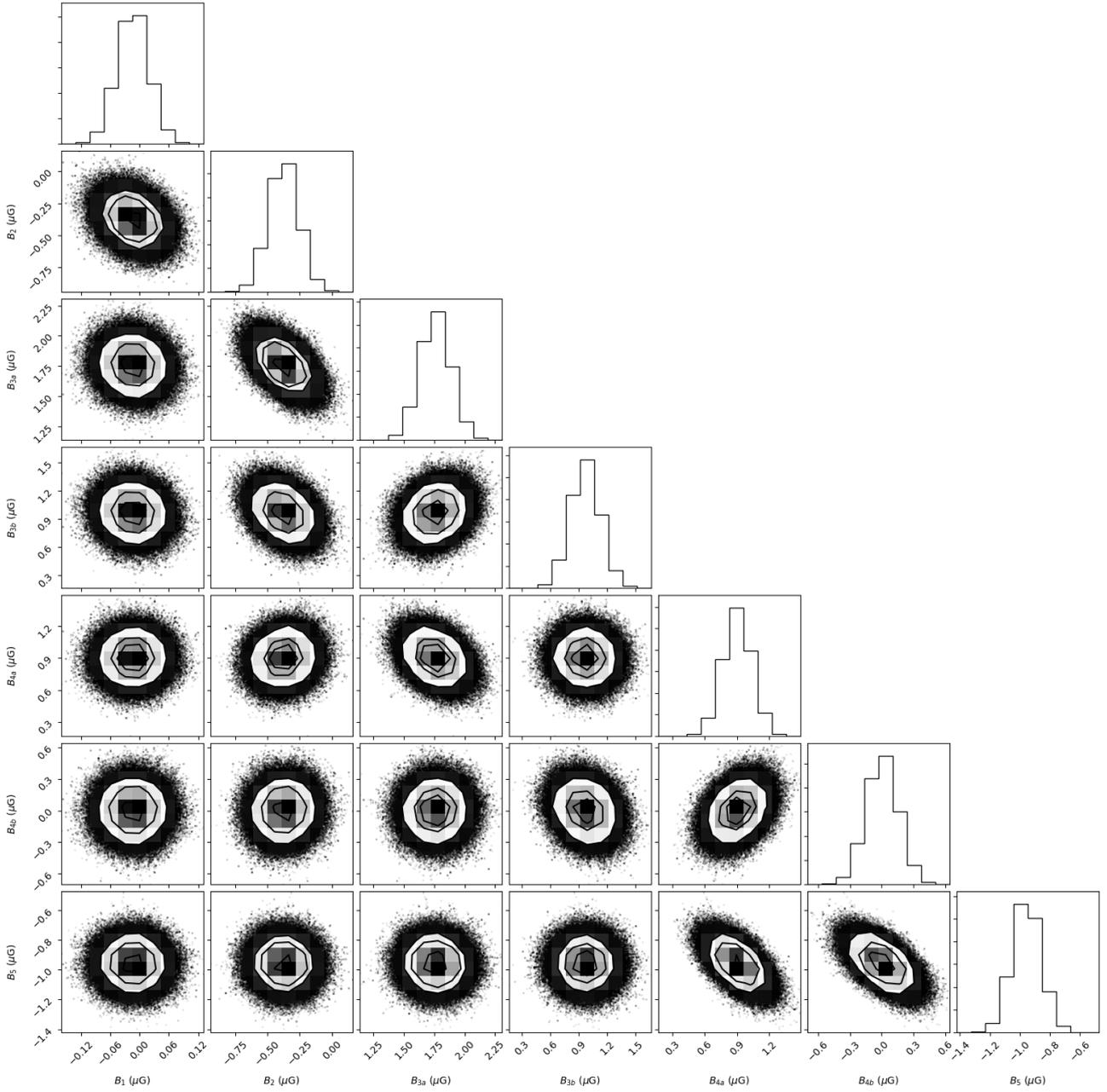

**Figure D6.** MCMC corner plot for the model Free 3+4.



\end{document}